\documentclass[fleqn,usenatbib]{mnras}

\usepackage[T1]{fontenc}

\DeclareRobustCommand{\VAN}[3]{#2}
\let\VANthebibliography\thebibliography
\def\thebibliography{\DeclareRobustCommand{\VAN}[3]{##3}\VANthebibliography}

\usepackage{graphicx}	
\usepackage{amsmath}	
\usepackage{adjustbox}
\usepackage{verbatim}
\usepackage{multirow}
\usepackage{pdflscape}
\usepackage[table,xcdraw]{xcolor}
\usepackage{placeins}
\usepackage{orcidlink}
\usepackage{afterpage}

\usepackage{newtxtext,newtxmath}

\DeclareMathAlphabet{\mathsc}{OT1}{cmr}{m}{sc}
\def\testbx{bx}%
\DeclareRobustCommand{\ion}[2]{%
\relax\ifmmode
\ifx\testbx\f@series
{\mathbf{#1\,\mathsc{#2}}}\else
{\mathrm{#1\,\mathsc{#2}}}\fi
\else\textup{#1\,{\mdseries\textsc{#2}}}%
\fi}

\newcommand{\Hei}{\ion{He}{i}}

\newcommand{\Heii}{\ion{He}{ii}}

\newcommand{\Hgi}{\ion{Hg}{i}}

\newcommand{\Oi}{[\ion{O}{i}]}
\newcommand{\Oii}{[\ion{O}{ii}]}
\newcommand{\Oiii}{[\ion{O}{iii}]}

\newcommand{\Ciii}{\ion{C}{iii}}
\newcommand{\Civ}{\ion{C}{iv}}

\newcommand{\Caxv}{[\ion{Ca}{xv}]}

\newcommand{\Mgi}{\ion{Mg}{i}}

\newcommand{\Nii}{\ion{N}{ii}}

\newcommand{\Nv}{\ion{N}{v}}

\newcommand{\Sii}{\ion{S}{ii}}
\newcommand{\Siii}{\ion{S}{iii}}

\newcommand{\NaiD}{\ion{Na}{i}~\text{D}}

\newcommand{\Fevii}{[\ion{Fe}{vii}]}

\newcommand{\Fex}{[\ion{Fe}{x}]}
\newcommand{\Fexi}{[\ion{Fe}{xi}]}

\newcommand{\Fexiv}{[\ion{Fe}{xiv}]}

\newcommand{\Neiii}{[\ion{Ne}{iii}]}
\newcommand{\Nev}{[\ion{Ne}{v}]}

\newcommand{\msun}{\mbox{M$_{\odot}$}}

\newcommand{\msol}{\mbox{M$_{\odot}$}}

\newcommand{\kms}{\mbox{$\rm{km}\,s^{-1}$}}

\newcommand{\fspectralline}[3]{[\ion{#1}{#2}]~$\uplambda#3$~\AA}

\newcommand{\sleipnir}{\textsc{SLEIPNIR}}
\newcommand{\fsf}{\textsc{FastSpecFit}}

\newcommand{\desispec}{\textsc{DESISPEC}}
\newcommand{\redrock}{\textsc{REDROCK}}

\newcommand{\galrate}{$R_\mathrm{G}=5~^{+5}_{-3}\times10^{-6}~\mathrm{galaxy}^{-1}~\mathrm{yr}^{-1}$}
\newcommand{\massrate}{$R_\mathrm{M}=1.1~^{+1.2}_{-0.6}\times10^{-16}~\mathrm{\msol^{-1}}~\mathrm{yr}^{-1}$}
\newcommand{\volrate}{$R_\mathrm{V}=2.3~^{+10.7}_{-1.6}\times10^{-8}~\mathrm{Mpc}^{-3}~\mathrm{yr}^{-1}$}

\newcommand{\SpecSample}{634551}
\newcommand{\GalaxySample}{465610}
\newcommand{\CrLSample}{208}

\newcommand{\TNSGalaxyMatches}{1219}
\newcommand{\TNSCrLMatches}{6}
\newcommand{\TDEDatabaseTotal}{370}
\newcommand{\TDEDatabaseGalaxyMatches}{11}
\newcommand{\TDEDatabaseCrLMatches}{0}
\newcommand{\SDSSLegacyGalaxyMatches}{18}
\newcommand{\SDSSLegacyCrLMatches}{5}
\newcommand{\BOSSGalaxyMatches}{7}
\newcommand{\BOSSCrLMatches}{0}
\newcommand{\MILLIQUASGalaxyMatches}{7154}
\newcommand{\MILLIQUASCrLMatches}{163}
\newcommand{\DingGalaxyMatches}{47}
\newcommand{\DingCrLMatches}{19}

\newcommand{\TNSMatchDate}{2025 August 28}

\newcommand{\WOneminusWTwo}{\textit{W}1$-$\textit{W}2}
\newcommand{\WTwominusWThree}{\textit{W}2$-$\textit{W}3}

\title[Early DESI ECLEs]{Early results in the search for extreme coronal line emitters with the Dark Energy Spectroscopic Instrument}

\author[P.~Clark~et~al.]{Peter~Clark$^{\orcidlink{0000-0002-6576-7400}}$,$^{1,2}$\thanks{E-mail:~P.S.J.Clark@soton.ac.uk}
Joseph~Callow$^{\orcidlink{0000-0002-0804-9533}}$,$^{2}$\thanks{E-mail: joecallow2809@gmail.com}
Or~Graur$^{\orcidlink{0000-0002-4391-6137}}$,$^{2,3}$
Alexei~V.~Filippenko$^{\orcidlink{0000-0003-3460-0103}}$,$^{4}$
Thomas~G.~Brink$^{\orcidlink{0000-0001-5955-2502}}$,$^{4}$
\newauthor
WeiKang~Zheng$^{\orcidlink{0000-0002-2636-6508}}$,$^{4}$
Jessica~Aguilar$^{\orcidlink{0000-0003-0822-452X}}$,$^{5}$
Steven~Ahlen$^{\orcidlink{0000-0001-6098-7247}}$,$^{6}$
Segev~BenZvi$^{\orcidlink{0000-0001-5537-4710}}$,$^{7}$
Davide~Bianchi$^{\orcidlink{0000-0001-9712-0006}}$,$^{8,9}$
\newauthor
David~Brooks$^{\orcidlink{0000-0002-8458-5047}}$,$^{10}$
Todd~Claybaugh$^{\orcidlink{0000-0002-5024-6987}}$,$^{5}$
Andrei~Cuceu$^{\orcidlink{0000-0002-2169-0595}}$,$^{5}$
Axel~de~la~Macorra$^{\orcidlink{0000-0002-1769-1640}}$,$^{11}$
Arjun~Dey$^{\orcidlink{0000-0002-4928-4003}}$,$^{12}$
\newauthor
Peter~Doel$^{\orcidlink{0000-0002-6397-4457}}$,$^{10}$
Victoria~A.~Fawcett$^{\orcidlink{0000-0003-1251-532X}}$,$^{13}$
Jaime~E.~Forero-Romero$^{\orcidlink{0000-0002-2890-3725}}$,$^{14,15}$
Enrique~Gaztañaga$^{\orcidlink{0000-0001-9632-0815}}$,$^{16,2,17}$
\newauthor
Satya~Gontcho~A~Gontcho$^{\orcidlink{0000-0003-3142-233X}}$,$^{5,18}$
Gaston~Gutierrez$^{\orcidlink{0000-0003-0825-0517}}$,$^{19}$
Mustapha~Ishak$^{\orcidlink{0000-0002-6024-466X}}$,$^{20}$
Jorge~Jimenez$^{\orcidlink{0000-0001-8528-3473}}$,$^{21}$
\newauthor
Dick~Joyce$^{\orcidlink{0000-0003-0201-5241}}$,$^{12}$
Stephanie~Juneau$^{\orcidlink{0000-0002-0000-2394}}$,$^{12}$
Theodore~Kisner$^{\orcidlink{0000-0003-3510-7134}}$,$^{5}$
Anthony~Kremin$^{\orcidlink{0000-0001-6356-7424}}$,$^{5}$
Martin~Landriau$^{\orcidlink{0000-0003-1838-8528}}$,$^{5}$
\newauthor
Laurent~Le~Guillou$^{\orcidlink{0000-0001-7178-8868}}$,$^{22}$
Marc~Manera$^{\orcidlink{0000-0003-4962-8934}}$,$^{23,21}$
Aaron~Meisner$^{\orcidlink{0000-0002-1125-7384}}$,$^{22}$
Ramon~Miquel$^{\orcidlink{0000-0002-6610-4836}}$,$^{24,21}$
\newauthor
John~Moustakas$^{\orcidlink{0000-0002-2733-4559}}$,$^{25}$
Seshadri~Nadathur$^{\orcidlink{0000-0001-9070-3102}}$,$^{2}$
Will~J.~Percival$^{\orcidlink{0000-0002-0644-5727}}$,$^{26,27,28}$
Ignasi~P\'erez-R\`afols$^{\orcidlink{0000-0001-6979-0125}}$,$^{29}$
\newauthor
Francisco~Prada$^{\orcidlink{0000-0001-7145-8674}}$,$^{30}$
Graziano~Rossi,$^{31}$
Eusebio~Sanchez$^{\orcidlink{0000-0002-9646-8198}}$,$^{32}$
David~Schlegel$^{\orcidlink{0000-0002-5042-5088}}$,$^{5}$
Michael~Schubnell$^{\orcidlink{0000-0001-9504-2059}}$,$^{33}$
\newauthor
Joseph Harry~Silber$^{\orcidlink{0000-0002-3461-0320}}$,$^{5}$
David~Sprayberry,$^{12}$
Gregory~Tarl\'{e}$^{\orcidlink{0000-0003-1704-0781}}$,$^{33}$
Benjamin~A.~Weaver,$^{12}$
Rongpu~Zhou$^{\orcidlink{0000-0001-5381-4372}}$,$^{5}$
\newauthor
and Hu~Zou$^{\orcidlink{0000-0002-6684-3997}}$$^{34}$
\\
$^{1}$ School of Physics and Astronomy, University of Southampton, Southampton, SO17 1BJ, UK\\
$^{2}$ Institute of Cosmology and Gravitation, University of Portsmouth, Portsmouth, PO1 3FX, UK \\
$^{3}$ Department of Astrophysics, American Museum of Natural History, New York, NY 10024, USA \\
$^{4}$ Department of Astronomy, University of California, Berkeley, CA 94720-3411, USA\\
$^{5}$ Lawrence Berkeley National Laboratory, 1 Cyclotron Road, Berkeley, CA 94720, USA\\
$^{6}$ Physics Department, Boston University, 590 Commonwealth Avenue, Boston, MA 02215, USA\\
$^{7}$ Department of Physics \& Astronomy, University of Rochester, 206 Bausch and Lomb Hall, P.O. Box 270171, Rochester, NY 14627-0171, USA\\
$^{8}$ Dipartimento di Fisica ``Aldo Pontremoli,'' Universit\`a degli Studi di Milano, Via Celoria 16, I-20133 Milano, Italy\\
$^{9}$ INAF-Osservatorio Astronomico di Brera, Via Brera 28, 20122 Milano, Italy\\
$^{10}$ Department of Physics \& Astronomy, University College London, Gower Street, London, WC1E 6BT, UK\\
$^{11}$ Instituto de F\'{\i}sica, Universidad Nacional Aut\'{o}noma de M\'{e}xico, Circuito de la Investigaci\'{o}n Cient\'{\i}fica, Ciudad Universitaria, Cd. de M\'{e}xico C.~P.~04510, M\'{e}xico\\
$^{12}$ National Science Foundation NOIRLab, 950 N. Cherry Ave., Tucson, AZ 85719, USA \\
$^{13}$ European Southern Observatory, Karl-Schwarzschild-Straße 2, 85748 Garching bei München, Germany\\
$^{14}$ Departamento de F\'isica, Universidad de los Andes, Cra. 1 No. 18A-10, Edificio Ip, CP 111711, Bogot\'a, Colombia\\
$^{15}$ Observatorio Astron\'omico, Universidad de los Andes, Cra. 1 No. 18A-10, Edificio H, CP 111711 Bogot\'a, Colombia\\
$^{16}$ Institut d'Estudis Espacials de Catalunya (IEEC), c/ Esteve Terradas 1, Edifici RDIT, Campus PMT-UPC, 08860 Castelldefels, Spain\\
$^{17}$ Institute of Space Sciences, ICE-CSIC, Campus UAB, Carrer de Can Magrans s/n, 08913 Bellaterra, Barcelona, Spain\\
$^{18}$ University of Virginia, Department of Astronomy, Charlottesville, VA 22904, USA\\
$^{19}$ Fermi National Accelerator Laboratory, PO Box 500, Batavia, IL 60510, USA\\
$^{20}$ Department of Physics, The University of Texas at Dallas, 800 W. Campbell Rd., Richardson, TX 75080, USA\\
$^{21}$ Institut de F\'{i}sica d’Altes Energies (IFAE), The Barcelona Institute of Science and Technology, Edifici Cn, Campus UAB, 08193, Bellaterra (Barcelona), Spain\\
$^{22}$ Sorbonne Universit\'{e}, CNRS/IN2P3, Laboratoire de Physique Nucl\'{e}aire et de Hautes Energies (LPNHE), FR-75005 Paris, France \\
$^{23}$ Departament de F\'{i}sica, Serra H\'{u}nter, Universitat Aut\`{o}noma de Barcelona, 08193 Bellaterra (Barcelona), Spain\\
$^{24}$ Instituci\'{o} Catalana de Recerca i Estudis Avan\c{c}ats, Passeig de Llu\'{\i}s Companys, 23, 08010 Barcelona, Spain\\
$^{25}$ Department of Physics and Astronomy, Siena College, 515 Loudon Road, Loudonville, NY 12211, USA\\
$^{26}$ Department of Physics and Astronomy, University of Waterloo, 200 University Ave W, Waterloo, ON N2L 3G1, Canada\\
$^{27}$ Perimeter Institute for Theoretical Physics, 31 Caroline St. North, Waterloo, ON N2L 2Y5, Canada\\
$^{28}$ Waterloo Centre for Astrophysics, University of Waterloo, 200 University Ave W, Waterloo, ON N2L 3G1, Canada\\
$^{29}$ Departament de F\'isica, EEBE, Universitat Polit\`ecnica de Catalunya, c/Eduard Maristany 10, 08930 Barcelona, Spain\\
$^{30}$ Instituto de Astrof\'{i}sica de Andaluc\'{i}a (CSIC), Glorieta de la Astronom\'{i}a, s/n, E-18008 Granada, Spain\\
$^{31}$ Department of Physics and Astronomy, Sejong University, 209 Neungdong-ro, Gwangjin-gu, Seoul 05006, Republic of Korea\\
$^{32}$ CIEMAT, Avenida Complutense 40, E-28040 Madrid, Spain\\
$^{33}$ Department of Physics, University of Michigan, 450 Church Street, Ann Arbor, MI 48109, USA\\
$^{34}$ National Astronomical Observatories, Chinese Academy of Sciences, A20 Datun Road, Chaoyang District, Beijing, 100101, P.~R.~China\\ 
}

\date{Accepted XXX. Received YYY; in original form ZZZ}

\pubyear{2025}

\begin{document}
\label{firstpage}
\pagerange{\pageref{firstpage}--\pageref{lastpage}}
\maketitle

\clearpage
\begin{abstract}
Here we present the results of our search through the Early Data Release (EDR) of the Dark Energy Spectroscopic Instrument (DESI) for extreme coronal line emitters (ECLEs) -- a rare classification of galaxies displaying strong, high-ionisation iron coronal emission lines within their spectra. With the requirement of a strong X-ray continuum to generate the coronal emission, ECLEs have been linked to both active galactic nuclei (AGNs) and tidal disruption events (TDEs). We focus our search on identifying TDE-linked ECLEs. We identify three such objects within the EDR sample, highlighting DESI's effectiveness for discovering new nuclear transients, and determine a galaxy-normalised TDE-linked ECLE rate of \galrate\ at a median redshift of $z = 0.2$ -- broadly consistent with previous work. Additionally, we identify more than 200 AGNs displaying coronal emission lines, which serve as the primary astrophysical contaminants in searches for TDE-related events. We also include an outline of the custom Python code developed for this search.
\end{abstract}




\section{Introduction}
\label{sec:Introduction}

Extreme coronal line emitters (ECLEs) are defined by the presence of prominent coronal emission lines from highly ionised iron (\Fevii, \Fex, \Fexi, \Fexiv) in their spectra. Such emission lines are rare, requiring a high-energy X-ray continuum to generate, and are most commonly seen in the solar corona (giving them their name). The two primary sources for extragalactic coronal lines are active galactic nuclei (AGNs) and tidal disruption events (TDEs). 

The actively accreting central supermassive black hole (SMBH) in an AGN can generate the requisite X-ray continuum to produce coronal lines. If present, these lines are typically seen with line strengths of a few percent that of the \fspectralline{O}{III}{5007} line \citep{nagao_2000_HighIonizationNuclearEmissionLine}.
Owing to the common use of the abbreviation ``CL'' when referring to ``changing look'' AGNs in the literature, to avoid potential confusion we abbreviate ``coronal line'' to ``CrL'' throughout this work. Thus, an AGN displaying coronal line emission is referred to as a ``CrL-AGN''. Previous work identified ECLEs as galaxies where coronal line emission is much stronger, with CrL emission line strengths exceeding 20 per cent that of \fspectralline{O}{iii}{5007} commonly applied as a diagnostic threshold \citep{wang_2012_EXTREMECORONALLINE, callow_2024_rateextremecoronal, callow_2025_rateextremecoronal}. Using this criterion, only a small fraction of CrL-AGNs are ECLEs, which indicates that AGN-linked ECLEs may either be drawn from the high end of a continuum of AGN CrL activity or reflect unique environments around some SMBHs. The use of an emission-line-strength cut to identify transient phenomena linked to the same physical processes presents challenges to which we will return.

TDEs are luminous flaring transients produced by the gravitational shredding of a star that passes too close to its galaxy's central SMBH. Around half of the star's mass remains gravitationally bound to the SMBH following the disruption and produces an accretion disc of accreting material \citep{ulmer_1999_FlaresTidalDisruption}. The mass eventually accreted is significantly less than the initial bound mass, with more material becoming unbound as the system evolves \citep{ayal_2000_TidalDisruptionSolarType}. Analysis suggests that the majority of TDEs involve SMBHs of $<10^8$~\msol. At larger SMBH masses the Roche limit (the radius within which a star will be tidally disrupted) of a solar-like star is within the SMBH's event horizon, resulting in the star being absorbed prior to disruption and so producing no observable transient \citep{hills_1975_PossiblePowerSource}. While the details of TDE emission are still debated (e.g., \citealt{piran_2015_DISKFORMATIONDISK, dai_2018_UnifiedModelTidal}, the circularisation of material into the accretion disc or collisions within the infalling material streams (or a combination of both) are the currently preferred mechanisms (e.g., \citealt{lacy_1982_NatureCentralParsec, phinney_1989_CosmicMergerMania, evans_1989_TidalDisruptionStar}). 

The first ECLEs were discovered by \cite{komossa_2008_DiscoverySuperstrongFading}, \cite{wang_2011_Transientsuperstrongcoronal}, and \cite{wang_2012_EXTREMECORONALLINE}. Spectroscopic follow-up observations of the \cite{wang_2012_EXTREMECORONALLINE} sample of seven ECLEs selected from the seventh data release of the Sloan Digital Sky Survey \citep[SDSS;][]{york_2000_SloanDigitalSky}, conducted by \cite{yang_2013_LONGTERMSPECTRALEVOLUTION}, revealed the coronal emission lines in four of the seven objects to have faded during the interim several years. This indicated that the process involved in generating the CrLs was transient in nature, as expected from a TDE. Mid-infrared (MIR) analysis conducted by \cite{dou_2016_LONGFADINGMIDINFRARED} demonstrated that those objects with fading coronal lines showed long-term declines in their MIR luminosities and changes in colour, evolving from AGN-like values at the start of observations to colours expected of non-AGN hosts. 

These analyses have been corroborated and extended by \cite{clark_2024_Longtermfollowupobservations} through additional follow-up spectroscopy obtained nearly two decades after the original SDSS spectra. Two of the \citet{wang_2012_EXTREMECORONALLINE} objects displayed persistent coronal lines, as well as optical and MIR luminosity and colour variability consistent with AGN activity. We treat these objects as AGN-ECLEs. We note that the physical division between these and other AGNs that display coronal lines is unclear, other than the elevated strength of the coronal emission. The remaining five objects display no rebrightening of their coronal lines, with the latest observations (obtained in 2021) showing that three now lack all coronal emission lines, with the remaining two (SDSS~J1241+4426 and SDSS~J1342+0530) displaying only the lowest-energy \Fevii\ lines. This spectral evolution and the long-term power-law declines in the MIR emission of these five objects were found to remain consistent with a TDE origin. We term objects such as these as ``TDE-ECLEs''.

TDEs are known to be X-ray luminous (e.g., \citealt{auchettl_2017_NewPhysicalInsights, auchettl_2018_ComparisonXRayEmission}). In the case of TDE-ECLEs, the initial disruption and subsequent accretion processes produce an intense X-ray continuum that ionises material at some (individual-system dependent) distance from the SMBH. It is this material -- unrelated to the star being disrupted -- that is responsible for both the coronal emission lines and the reprocessing of the ultraviolet (UV) -- optical flux into the long-term MIR emission seen in ECLEs. Observations have shown this emission can persist for several years following the TDE, compared to the months to years-long duration of the direct emission from the TDE seen in the optical and UV. Given that these ECLEs are driven by the reprocessing of TDE emission, they can be thought of as TDE light echoes.

Recently, direct links between TDEs and ECLEs have been made through observations of coronal emission lines developing in the spectra of live TDEs where the TDE was identified whilst the direct emission was still visible; we refer to such objects in this work collectively as CrL-TDEs. Whilst only a small number of CrL-TDEs are currently known \citep[14 at time of writing; see, e.g.,][]{fraser_2017_EPESSTOSpectroscopicClassification, neustadt_2020_TDENotTDE, onori_2022_NuclearTransient2017gge, newsome_2022_2022upjZTF22abegjtxDiscovery, wang_2024_ASASSN18apDustyTidal, hinkle_2024_CoronalLineEmitters, clark_2025_2018dyktidaldisruption}, there is already indication of significant variation in properties (e.g., the time lag between the occurrence of the TDE and the emergence of the coronal lines) which is likely to increase as more such objects are identified. \citet{clark_2025_2018dyktidaldisruption} explored the MIR properties of these objects and identified tentative evidence of a colour-luminosity relationship. Specifically, for TDEs that are observed with coronal lines, those with brighter MIR flares show more significant reddening at outburst, potentially related to material covering fractions or other environmental conditions increasing the level of MIR reprocessing. Such a relation was not seen in the AGNs explored in the study, which also displayed much smaller changes in peak MIR luminosity overall. Other work has also shown that differences in the MIR properties of TDEs and AGNs could be used as a distinguishing characteristic \citep{hinkle_2024_Midinfraredechoesambiguous,yao_2025_DistinguishingTidalDisruption}.

Here we return to the complication of using an emission-line-strength requirement in the identification of transient-linked ECLEs. As the strength of both CrL and \fspectralline{O}{iii}{5007} emission varies with time, with CrL emission generally observed to fade as \fspectralline{O}{iii}{5007} emission increases over the evolution of the transient, applying an emission-line-strength cut without spectroscopic observations covering the full evolution of the transient would bias the identification of events to only those observed early in their evolution, despite their multi-year duration and identical physical origin. As such, we do not apply an emission-line-strength cut as a requirement in our search for ECLEs in this work, preferring a more physically motivated classification. As this differs with previous work identifying ECLEs, we note for clarity that all objects here identified as the result of AGN activity are referred to as ``CrL-AGNs'', with those attributed to the result of TDEs being referred to as ``TDE-ECLEs'' regardless of emission-line strength to identify them as being classified as such based on their CrL emission rather than more traditional TDE diagnostics. All TDEs that were classified through more conventional methods and were observed to display CrL emission features (whether at the time of classification or at a later point in their follow-up monitoring) are referred to as ``CrL-TDEs'' to reflect this observational classification distinction despite being physically the same phenomena.

This work describes the results of a new systematic search for ECLEs in the early data release (EDR) of the Dark Energy Spectroscopic Instrument \citep[DESI;][]{desicollaboration_2016_DESIExperimentPart, desicollaboration_2016_DESIExperimentParta, schlafly_2023_SurveyOperationsDark, miller_2024_OpticalCorrectorDark, poppett_2024_OverviewFiberSystem} and includes an outline of our custom Python code for detecting and analysing ECLEs, the Spectroscopic Light Echo Identification Protocol Now in Realtime, \sleipnir\footnote{\url{https://github.com/Lightbulb500/SLEIPNIR}}. As described by \citet{callow_2024_rateextremecoronal, callow_2025_rateextremecoronal}, \sleipnir\ has already been used on the data from two SDSS surveys, exploring both the SDSS Legacy DR17 \citep{abdurrouf_2022_SeventeenthDataRelease} and the Baryon Oscillation Spectroscopic Survey (BOSS) LOWZ \citep{dawson_2013_BaryonOscillationSpectroscopic} surveys. As part of these analyses, \sleipnir\ successfully retrieved all seven objects from the original systematic search conducted in SDSS DR7 \citep{abazajian_2009_SeventhDataRelease} by \cite{wang_2012_EXTREMECORONALLINE}, one new TDE-ECLE in BOSS LOWZ, as well as numerous other candidates that were ultimately classified as AGN-ECLEs.

Whilst the sample size of TDE-ECLEs is still small, rates analyses are already providing important insights. Work conducted on the previous SDSS-based samples by \citet{callow_2024_rateextremecoronal, callow_2025_rateextremecoronal} finds the observed rates of TDE-ECLEs to be consistent with the general observed TDE rate, if between 10 and 40 per cent of TDEs display ECLE behaviour. Larger samples of TDE-ECLEs obtained over a range of redshifts will be needed to reduce the statistical uncertainties of such rate calculations, made challenging by the inherently low rates of TDEs generally and the wide timescales over which such behaviour can occur. The exploration of large spectroscopic samples is thus essential to improve our understanding of ECLEs.

We structure the work as follows. Section~\ref{sec:ObservationsAndData} outlines the spectroscopic and photometric datasets along with the value-added catalogue used to identify and classify our identified ECLE candidates. In Section~\ref{sec:sleipnir}, we describe our \sleipnir\ code and outline sources of potential contamination for similar searches and how these are dealt with in our survey. Section~\ref{sec:Sample_Construction_and_Classification} describes the construction, processing, and classification of the DESI EDR ECLE sample. In Section~\ref{sec:EDR_ECLEs} we move to specific descriptions of the most promising candidates identified through the search, including three objects we identify as being linked to TDEs. A sample of AGNs with coronal emission lines (CrL-AGNs) identified in the EDR search, which whilst not the focus of this work, can serve as a reference catalogue and basis for further work on such objects, is described in Appendix~\ref{sec:AGN_Candidates}. In Section~\ref{sec:Analysis}, we discuss the properties of our newly identified TDE-ECLEs and place their MIR evolution in context with the existing sample. We also determine galaxy normalised, mass normalised, and volumetric rates for TDE-linked ECLEs based on our analysis of DESI EDR, finding these to be broadly consistent with rates determined in previous work. In Section~\ref{sec:Conclusions}, we summarise our finding that DESI is an effective search engine for TDE-linked ECLEs and other coronal line galaxies, whilst highlighting the importance of follow-up spectroscopy to confirm the evolution and classification of such objects.

\section{Observations and Data Reduction}
\label{sec:ObservationsAndData}

A condensed summary of the spectroscopic datasets used here is given in Table~\ref{tab:Spectra_Summary}, with the corresponding information for the photometric datasets given in Table~\ref{tab:Photom_Basic}.

At all times, unless otherwise stated, apparent magnitudes are given as observed, with no additional corrections. In contrast, absolute magnitudes have been corrected for Milky Way extinction using the appropriate photometric extinction coefficient. Unless specified otherwise, these coefficients have been retrieved from \cite{schlafly_2011_MeasuringReddeningSloan}. To match the preferred extinction parameters of \cite{schlafly_2011_MeasuringReddeningSloan}, we apply the extinction law of \cite{fitzpatrick_1999_CorrectingEffectsInterstellar} throughout this paper and assume $R_{V} = 3.1$.

Throughout, when converting between apparent and absolute magnitudes we assume a Hubble constant H$_0 = 73$ \kms\ Mpc$^{-1}$ and adopt a standard flat cosmological model with $\Omega_M=0.27$ and $\Omega_{\Lambda}=0.73$.

\subsection{Optical spectroscopy}
\label{subsec:Optical_Spectroscopy}

\begin{table*}
\caption{Summary information for the spectral datasets used in this work.}
\label{tab:Spectra_Summary}
\begin{tabular}{lcccc}
\hline
\textbf{Telescope} & \textbf{Instrument} & \textbf{Wavelength Range}${^1}$ & \textbf{Type}${^2}$ & \textbf{No. Utilised} \\ \hline
Mayall-4\,m & DESI & 3600--9825 & Classification & \SpecSample \\
Gemini & GMOS-S & 3860--9330 & Follow-up & 2 \\ 
NOT & ALFOSC & 3500--9635 & Follow-up & 3 \\ 
Keck 10\,m & LRIS & 3135--10,255 & Follow-up & 3 \\ 
SDSS 2.5\,m & SDSS Legacy & 3585--10,400 & Archival & 2 \\ 
SDSS 2.5\,m & SDSS BOSS & 3810--9180  & Archival & 1 \\ 
\hline
\end{tabular}
\begin{flushleft}
\textit{Notes:} ${^1}$ In the observer frame. Owing to differences between instrumental configurations used for some spectra, ranges are approximate.\\
${^2}$ Classification: Spectra used to initially identify ECLE candidates. Archival: Spectra obtained significantly before the classification to identify any existing long-term variability. Follow-up: Spectra of previously identified candidates obtained after classification used to explore coronal line variability and/or improve obtain higher SNR data for individual objects of interest.
\end{flushleft}
\end{table*}

\subsubsection{Dark Energy Spectroscopic Instrument: DESI}
\label{subsubsec:DESI_Spec}

All DESI spectra included here are now publicly available through the EDR and were obtained as part of the DESI survey validation programs \citep{desicollaboration_2024_ValidationScientificProgram, desicollaboration_2024_EarlyDataRelease}.
The spectra were processed by the custom DESI spectroscopic pipeline, which includes a full suite of processing and correction steps to provide fully flux- and wavelength-calibrated spectra \citep{guy_2023_SpectroscopicDataProcessing}.

DESI itself is designed primarily as a cosmological experiment, and whilst not the focus of this work, Data Release 1 provides a range of state-of-the-art cosmological analyses, including two-point clustering measurements and validation \citep{desicollaboration_2024_DESI2024II}, baryon acoustic oscillation (BAO) measurements from galaxies and quasars \citep{desicollaboration_2025_DESI2024III} and from the Ly-$\alpha$ forest \citep{desicollaboration_2025_DESI2024IV}, as well as a full-shape study of galaxies and quasars \citep{desicollaboration_2024_DESI2024FullShape}. There are Cosmological results from the BAO measurements \citep{desicollaboration_2025_DESI2024VI, descollaboration_2025_DESIDR2results} and the full-shape analysis \citep{desicollaboration_2024_DESI2024VII}, as well as constraints on primordial non-Gaussianities.

\subsubsection{Gemini}
\label{subsubsec:GEMINI_Spec}

We obtained optical spectra of two of our TDE-ECLE candidates (see Section~\ref{sec:Sample_Construction_and_Classification} for details on candidate selection) using the Gemini Multi-Object Spectrograph \citep[GMOS;][]{hook_2004_GeminiNorthMultiObjectSpectrograph} on the 8.1~m Gemini South Telescope (Gemini-S) on Cerro Pachón, Chile as part of Gemini program GS-2021B-FT-202 (PI P. Clark). Observations were made using a combination of the B600 and R831 gratings, with the slit orientated at the parallactic angle. Data were reduced using the Data Reduction for Astronomy from Gemini Observatory North and South (\textsc{DRAGONS}) reduction package \citep{labrie_2019_DRAGONSDataReduction}, using the standard recipe for GMOS long-slit reductions. This includes bias correction, flat fielding, wavelength calibration, and flux calibration. As we did not have telluric standards for these observations, we do not attempt telluric correction of these spectra. 

\subsubsection{Nordic Optical Telescope: NOT}
\label{subsubsec:NOT_Spec_Optical}

We obtained follow-up spectra of several ECLE candidates using the Nordic Optical Telescope (NOT) and the Alhambra Faint Object Spectrograph and Camera (ALFOSC). These spectra were all obtained using a $1''$ slit and Grism~$\#$4, with the slit orientated at the parallactic angle. Data reduction was conducted using a modified version of the \textsc{pyNOT-redux} \citep{krogager_2025_PyNOTreduxDataReduction} package. Reduction included wavelength calibration to a helium comparison lamp and flux calibration using observations of a standard star obtained on the same night as the science observations.

\subsubsection{Sloan Digital Sky Survey: SDSS}
\label{subsubsec:SDSS_Spec}

Where available, we compared our ECLE candidates to archival spectra obtained by the SDSS several years prior to the classification spectra. These spectra were retrieved from SDSS DR18 \citep{almeida_2023_EighteenthDataRelease} and accessed using the Casjobs system. Owing to the differing targeting parameters of SDSS and DESI, only three of the identified ECLE candidates have corresponding SDSS spectra.

\subsubsection{Keck}
\label{subsubsec:Keck_Spec}

We acquired spectra of three of our ECLE candidates with the Low Resolution Imaging Spectrometer \citep[LRIS;][]{oke_1995_KeckLowResolutionImaging} on the 10~m Keck I telescope at the W. M. Keck Observatory. These observations utilized the $1''$ slit, the D560 dichroic, the 600/4000 grism, and the 400/8500 grating. This instrument configuration produced a combined wavelength range of $\sim 3200$--10,200~\AA\ and a spectral resolving power of $R = \lambda/\Delta\lambda \approx 900$. To minimize slit losses caused by atmospheric dispersion \citep{filippenko_1982_importanceatmosphericdifferential}, the slit was oriented at or near the parallactic angle (but note that LRIS is already equipped with an atmospheric dispersion corrector). The LRIS spectra were reduced with the \textsc{LPipe} data-reduction pipeline \citep{perley_2019_FullyAutomatedReduction}.

\subsection{Optical photometry}
\label{subsec:Optical_Photometry}

\subsubsection{Asteroid Terrestrial-impact Last Alert System: ATLAS}
\label{subsec:ATLAS_Phot}

ATLAS data were retrieved using the ATLAS forced-photometry server \citep{shingles_2021_ReleaseATLASForced}.\footnote{\url{https://fallingstar-data.com/forcedphot/}} ATLAS uses two broad-band filters: ``cyan'' (\textit{c}; approximately equivalent to \textit{g} + \textit{r}) and ``orange'' (\textit{o}; approximately equivalent to \textit{r} + \textit{i}). ATLAS observations are available over the general MJD range 57227--60724 and were processed using a modified version of \textsc{plot\_atlas\_fp.py} \citep{young_2024_plot_atlas_fppy}.

\subsubsection{DESI Legacy Imaging Surveys}
\label{subsec:Legacy_Phot}

We utilise the DESI Legacy Imaging Surveys \citep{dey_2019_OverviewDESILegacy} to retrieve processed composite \textit{grz} cutout images of the locations of candidates to provide contextual information on their morphology. 


\subsubsection{Liverpool Telescope: LT}
\label{subsec:LT_Phot}

We obtained \textit{ugriz}-photometry of several targets using the Liverpool Telescope (LT) with the IO:O instrument \citep{steele_2004_LiverpoolTelescopePerformance}. The LT~\textit{ugriz}-band photometry was measured on the preprocessed imaging following median stacking using \textsc{ysfitsutil} \citep{bach_2023_ysBachysfitsutilpyv02}. Photometric extraction was then conducted with \textsc{AutoPhOT} \citep{brennan_2022_AUTOmatedPhotometryTransients} and the Moffat point-spread function \citep[PSF;][]{moffat_1969_TheoreticalInvestigationFocal} with calibration to the SDSS magnitude system using the SDSS DR16 photometric catalogue \citep{ahumada_2020_16thDataRelease} retrieved from VizieR \citep{ochsenbein_2000_VizieRDatabaseAstronomical}.\footnote{\url{https://vizier.cds.unistra.fr/}}

\subsubsection{Zwicky Transient Facility: ZTF}
\label{subsec:ZTF_Phot}

ZTF observations were made with the \textit{gri} filters and retrieved using the ZTF Forced Photometry Service \citep[ZFPS;][]{ masci_2023_NewForcedPhotometry}. These observations cover an overall MJD range of 58301--60725, with the specific MJD range changing slightly for each candidate explored.  

\subsection{Infrared photometry}
\label{subsec:IR_Photometry}

\subsubsection{Two Micron All-Sky Survey: 2MASS}
\label{subsec_2MASS_Photometry}

To explore the behaviour of each identified object of interest well before its DESI spectrum, we retrieve near-infrared (NIR) photometry obtained by the Two Micron All-Sky Survey \citep[2MASS;][]{skrutskie_2006_TwoMicronAll} from IRSA. These data consist of \textit{JHK} photometric observations and are used to explore objects of interest in the \textit{J}--\textit{H} vs. \textit{H}--\textit{K} parameter space.

This parameter space is useful in distinguishing the source of IR emission as primarily the result of an AGN or from starlight \citep{hyland_1982_InfraredStudyQuasars, komossa_2009_NTTSpitzerChandra}. However, we note that given the wide range of NIR colours displayed by galaxies of the same spectroscopic classification, they cannot be distinguished effectively using these NIR colours alone. Additionally, given the time offset between the 2MASS and DESI observations, evolution in the NIR will not have been observed. 

\subsubsection{\textit{Wide-field Infrared Survey Explorer}: \textit{WISE}}
\label{subsec_WISE_Photometry}

Following the method described by \cite{clark_2024_Longtermfollowupobservations}, we retrieve the available MIR photometry for each candidate ECLE identified by \sleipnir. These MIR observations were obtained by the \textit{Wide-field Infrared Survey Explorer} (\textit{WISE}), from both the AllWISE \citep{wright_2010_WIDEFIELDINFRAREDSURVEY} and NEOWISE Reactivation Releases (NEOWISE-R) \citep{mainzer_2011_NEOWISEOBSERVATIONSNEAREARTH, mainzer_2014_INITIALPERFORMANCENEOWISE} from the NASA/IPAC infrared science archive (IRSA).\footnote{\url{https://irsa.ipac.caltech.edu/}}

As \textit{WISE} obtained images of each region of the sky in observing visits that comprise several observations on a roughly daily cadence in blocks spaced $\sim 6$ months apart, we process the data to give a weighted-average magnitude per observational block. Work by \cite{dou_2016_LONGFADINGMIDINFRARED} previously explored whether the known sample of ECLEs presented variability during each observation block, with no such variability detected. Thus, combining the individual observations allows for any long-term trends to be seen more easily.

We process the \textit{WISE} photometry with a custom Python script. In addition to generating the weighted per-block magnitudes, the script filters out any observation that was marked as an upper limit or was taken when the spacecraft was close to the South Atlantic Anomaly (saa\textunderscore sep $< 5.0$) or the sky position of the Moon (moon\textunderscore masked = 1). Additionally, any observations with a low frame quality or that suffered from potential ``contamination or confusion'' as flagged by the \textit{WISE} pipeline were also removed prior to further processing.

\subsection{Radio photometry}
\label{subsec:Radio_Photometry}

Significant radio emission can be used as an identifying diagnostic of AGN activity (see, e.g., \citealt{best_2005_hostgalaxiesradioloud}). To aid in distinguishing TDE sources from ongoing AGN activity, we explore the positions of our most promising candidates (see Section~\ref{sec:EDR_ECLEs}) in several radio surveys.

\subsubsection{Very Large Array Faint Images of the Radio Sky at Twenty-Centimeters (VLA FIRST)}
\label{subsec_VLASS_FIRST_Photometry}

We searched the Very Large Array Faint Images of the Radio Sky at Twenty-Centimeters (VLA FIRST) catalogue\footnote{Accessed through: \url{https://sundog.stsci.edu/cgi-bin/searchfirst}} for any recorded sources within $10''$ of our TDE candidates, with no sources returned. We additionally visually inspected $1'$ cutouts around each candidate to confirm this lack of identified source. A possible faint source is tentatively visible at the position of one candidate (see Section~\ref{sec:Raichu}), though it is close to the level of other spurious sources in the region. Given its overall faintness and lack of detection in other radio data (Section~\ref{subsec_VLASS_Photometry}), we do not consider this an indication of radio-AGN activity.

\subsubsection{Very Large Array Sky Survey: VLASS}
\label{subsec_VLASS_Photometry}

The VLASS Quick Look Epoch 1 catalogue was accessed via VizieR \citep{gordon_2021_VizieROnlineData}. We searched for any identified sources within $25''$ of each candidate position, with this search retrieving no catalogued sources.

To confirm this, and to search for variability in other epochs, we visually inspected the locations of each promising ECLE candidate in the VLASS datasets available through the VLASS Quicklook All Epochs interactive map.\footnote{Accessed through \url{https://vlass-dl.nrao.edu/vlass/HiPS/All_VLASS/Quicklook/}} No significant sources were observed in any epochs at the location of the candidates, ruling out the presence of any radio-loud AGN activity over the course of the epochs explored in the 2--4 GHz range covered by VLASS. This dataset covers an overall timespan of 2017 September (Epoch 1.1) through 2024 October (Epoch 3.2).

\subsubsection{Low-Frequency array two-metre sky survey: LoTSS}
\label{subsec_LoTSS_Photometry}

Finally, we searched for sources in the Low-Frequency array two-metre sky survey (LoTSS) DR2 catalogue \citep{shimwell_2022_LOFARTwometreSky}. Three of our candidates are within the sky coverage of LoTSS-DR2, one of which is closely associated with a detected radio source indicative of AGN activity; see Section~\ref{sec:Sandslash}.

\begin{table*}
\centering
\caption{Details of the photometric datasets used in this work.}
\begin{adjustbox}{width=0.95\textwidth}
\begin{tabular}{llll}
\hline
\textbf{Survey} & \textbf{Filters} & \textbf{MJD Range} & \textbf{Reference}\\ \hline
\textbf{Optical} & & & \\
Asteroid Terrestrial-impact Last Alert System (ATLAS) &  \textit{c}, \textit{o} $^1$ & 57227--60724 & \citet{tonry_2018_ATLASHighcadenceAllsky, shingles_2021_ReleaseATLASForced} \\
DESI Legacy Imaging Surveys & \textit{g}, \textit{r}, \textit{z} & - $^2$ & \cite{dey_2019_OverviewDESILegacy}\\
Liverpool Telescope (LT) & \textit{u}, \textit{g}, \textit{r}, \textit{i}, \textit{z} & 59792--60940 & \cite{steele_2004_LiverpoolTelescopePerformance}\\
Zwicky Transient Facility (ZTF) & \textit{g}, \textit{r}, \textit{i} & 58301--60725 & \citet{bellm_2019_ZwickyTransientFacility, masci_2023_NewForcedPhotometry}\\
\hline
\textbf{Infrared} & & & \\
The Two Micron All Sky Survey (2MASS) & \textit{J, H, K} & 50753--51838 & \cite{skrutskie_2006_TwoMicronAll}\\
AllWISE & \textit{W1, W2, W3} & 55203--55593 &\cite{wright_2010_WIDEFIELDINFRAREDSURVEY}\\
NEOWISE-R & \textit{W1, W2} & 56640--60523 &\cite{mainzer_2014_INITIALPERFORMANCENEOWISE}\\
\hline
\textbf{Radio} & & & \\
Very Large Array Faint Images of the Radio Sky at Twenty-Centimeters (VLA FIRST) & 1.4~GHz &  49473--56047 $^3$ & \citet{becker_1995_FIRSTSurveyFaint} \\
Very Large Array Sky Survey (VLASS) & 2--4~GHz & 58000--60615 $^4$ & \citet{lacy_2020_KarlJanskyVery} \\
Low-Frequency array two-metre sky survey (LoTSS) & 120--168 MHz & -$^5$ & \citet{shimwell_2022_LOFARTwometreSky} \\
\hline
\end{tabular}
\end{adjustbox}
\begin{flushleft}
\textit{Notes:} $^1$ ATLAS observations were made using two broad-band filters; \textit{c} (cyan) is approximately equivalent to \textit{g} + \textit{r} and \textit{o} (orange) is roughly \textit{r} + \textit{i}. \\
$^2$ Used for contextual imaging only.\\
$^3$ Overall MJD range. Data obtained at one epoch for each candidate.\\
$^4$ MJD range is approximate.\\
$^5$ Comprised of data obtained up to April 2021.\\
\end{flushleft}
\label{tab:Photom_Basic}
\end{table*}

\subsection{\fsf\ value-added catalogue}
\label{subsec:FSF_VAC}

In addition to the DESI spectra, we also make use of the \fsf\ \citep{moustakas_2023_FastSpecFitFastspectral} value-added catalogue (VAC) to provide additional information on the identified ECLE candidates and wider EDR sample as a whole. An outline of the parameters from \fsf\ used in this work is given in Table~\ref{tab:fastspecfit_info} of Appendix~\ref{subsec:Appendix_FSF}.

To generate a reference sample to compare our candidates to the overall population, we retrieve all DESI galaxies within the \fsf\ EDR catalogue (initial sample size: 1,397,479 galaxies) and apply the following cuts: 

\begin{itemize}
    \item $0.01 < z < 0.45$: Matches the redshift range of the main search for EDR ECLEs;
    \item ZWARN = 0: Objects that have reliable redshift measurement;
    \item SFR > 0: Objects for which a star-formation rate was measurable.
\end{itemize}

Following these cuts, our reference \fsf\ sample is reduced to 388,858 individual galaxies.

\section{Spectroscopic Light Echo Identification Protocol Now In Real-time: \sleipnir}
\label{sec:sleipnir}

\sleipnir\ is a Python-based analysis routine designed to provide rapid identification of candidate ECLEs for more detailed analysis through triage of a large number of input spectra. Its primary goal is to identify possible candidates to significantly reduce the burden of manual visual inspection, which given the rarity of ECLEs \citep{wang_2012_EXTREMECORONALLINE, clark_2024_Longtermfollowupobservations, callow_2024_rateextremecoronal, callow_2025_rateextremecoronal} is a significant barrier to their discovery.

\subsection{\sleipnir\ operation}
\label{sec:SLEIPNIR_Operation}

\subsubsection{Inputs}
\label{sec:SLEIPNIR_Inputs}

In general, as inputs to the analysis, \sleipnir\ requires a spectrum to classify, a predetermined redshift, and Milky Way extinction. In the case of the DESI analysis, \sleipnir\ also processes metadata for each spectrum to aid in classification. Such metadata include the observation date of the spectrum, DESI targeting information, and pipeline-processing flags. We discuss how these metadata are used to produce our input sample in Section~\ref{sec:DESI_Sample}.

\subsubsection{Analysis outputs}
\label{sec:SLEIPNIR_Analysis_Outputs}
As outputs, \sleipnir\ provides a rest-frame spectrum (corrected for both redshift and Milky Way extinction), a general classification of the input spectrum as an ECLE candidate or not, and an initial estimation of AGN activity based on simplistic Baldwin, Phillips, and Telervich \citep[BPT;][]{baldwin_1981_Classificationparametersemissionline} emission-line diagnostics. For each individual line analysed (see Section~\ref{sec:SLEIPNIR_Line_Selection}), the measured pseudoequivalent width (pEQW), line flux, and signal-to-noise ratio (SNR) are also returned, along with a flag for the easy identification of lines detected with high SNR (set by default to be $> 10$). For the Fe coronal lines specifically, flags to indicate if each line meets all of the detection criteria and is considered ``strong'' (see Section~\ref{sec:SLEIPNIR_Line_Selection}) are also returned. Input source-dependent metadata are also returned to aid in subsequent analysis of objects of interest. All textual outputs are returned in human and machine-readable ascii files, for easy transfer of the data to an external database of the user's choice for long-term and accessible storage.

In addition to these datafiles, \sleipnir\ also produces visual diagnostic diagrams for each of the identified candidates. These diagrams include the processed input spectrum, spectral cutouts of the CrL regions, \fspectralline{O}{iii}{5007} and H$\alpha$ complexes, BPT diagnostic diagrams, and a textual report on why the object was flagged for classification.

\subsubsection{Line selection and measurements}
\label{sec:SLEIPNIR_Line_Selection}

A limited list of spectral lines were selected for analysis with \sleipnir\ and are listed below (see Table~\ref{tab:line-list}). These are primarily the Fe coronal lines, Balmer series, and the emission features used for AGN identification through BPT diagnostics. A few additional lines produced in high-energy processes (e.g., \Caxv) or related to host-galaxy properties (e.g., the extinction tracers \NaiD\ and \Mgi) are included as well. Given their central importance to this analysis, the CrLs are explored in more detail. 

\begin{table}
\centering
\caption{Emission features used for candidate classification and analysis.}
\label{tab:line-list}
\begin{adjustbox}{width=0.95\columnwidth}
\begin{tabular}{lr|lr}
\hline
\textbf{Line} & \textbf{Wavelength (\AA)} & \textbf{Line} & \textbf{Wavelength (\AA)} \\ \hline
\textbf{Fe Coronal Lines} &  & \textbf{Other Coronal Lines} &  \\
\Fevii & 3759 & \Nev & 3347 \\
\Fevii & 5160 &  \Nev & 3427 \\
\Fevii & 5722 &  &  \\
\Fevii & 6088 & \Caxv  & 5446 \\
\Fex & 6376 &  &  \\
\Fexi & 7894 &  &  \\
\Fexiv & 5304 &  &  \\
\hline
\textbf{Other Lines} &  &  &  \\
H$\alpha$ & 6562.790 & \Oi & 6300 \\
H$\beta$ & 4861.350 & \Oi & 6363 \\
H$\gamma$ & 4340.472 & \Oii & 3728.2725 \\
H$\delta$ & 4101.734 & \Oiii & 4959 \\
 &  & \Oiii & 5007 \\
\Nii & 6548 &  &  \\
\Nii & 6584 & \Mgi & 5175 \\
 &  &  \\
\Sii & 6717 & \Neiii & 3869$^{**}$ \\
\Sii & 6731 &  &  \\
\Sii & 6724$^{*}$ & \Hei & 4478 \\
\Siii & 6313 & \Heii & 4686 \\
 &  &  \\
\NaiD\ & 5892.935$^{*}$ &  \\ \hline
\end{tabular}
\end{adjustbox}
\begin{flushleft}
\textit{Notes:} $^{*}$ Mean wavelength of a closely spaced doublet. \\
$^{**}$ This line was described incorrectly by both \citet{yang_2013_LONGTERMSPECTRALEVOLUTION} and \citet{clark_2024_Longtermfollowupobservations} as \fspectralline{Ne}{III}{3896}. The conclusions of these papers are not affected.
\end{flushleft}
\end{table}

A summary of the coronal lines used for analysis within \sleipnir\ is provided in Table~\ref{tab:crl_info}. While \sleipnir\ only uses the Fe coronal lines to identify ECLE candidates, we also check for the existence of the \Nev\ and \Caxv\ coronal lines. These lines are not used for the initial identification because they are rarely covered by the DESI spectrographs in the redshift range studied here and because of the significant step change in energy required to produce \Caxv. The use of these coronal lines in \sleipnir\ will be revisited in future studies making use of data with different wavelength coverage (e.g., the NIR where coronal lines of S and Si are present; \citealt{kynoch_2022_Multiplelocationsnearinfrared}) or when focused at higher redshifts.

All spectral lines are processed identically. Following a check to ensure the line is clear of possible edge effects by being too close to either the blue or red end of the spectrum (configured to be a conservative 50~\AA), a conversion to velocity space relative to the expected position of the spectral line is applied. The region around the spectral line is then fitted with a linear continuum which is removed prior to the measurement of each line's spectral properties. An important caveat of the parameters provided from this analysis is that \sleipnir\ does not perform line fitting. Instead, it measures the properties of lines in windows of fixed size using a direct ``area under the curve'' method. This method was chosen as a compromise between the individual accuracy of the measurements and operational speed given the large number of individual lines per spectrum. Testing has shown that the derived values are reliable, except in cases where multiple distinct features overlap (\sleipnir\ does not perform line deblending) or for very broad features that extend beyond the fixed window for measurement. These are not concerns for the narrow and isolated CrL features of most interest to our study, but can significantly affect the automated measurement of BPT emission-line ratios in the presence of broad features. As such, the BPT diagnostics provided by \sleipnir\ are treated as first-pass indications of the presence (or absence) of AGN activity, with additional line analysis performed once promising candidates have been identified.

\begin{table*}
\centering
\caption{Coronal lines used by \sleipnir.}
\label{tab:crl_info}
\begin{tabular}{lccccc}
\hline
\textbf{Ion} & \textbf{Wavelength} (\AA)$^{1}$ & \textbf{DESI $z$ min}$^{2}$ & \textbf{DESI $z$ max}$^{3}$ & \textbf{Required Energy (eV)}$^{4}$ & \textbf{Reference} $^{5}$ \\ \hline
\Fevii & 3759 & 0.0 & $>$~0.45 & 98.985~$\pm$~0.015 & \citet{gharaibeh_2011_Photoionizationmeasurementsiron} \\
\Fevii & 5160 & 0.0 & $>$~0.45 & 98.985~$\pm$~0.015 & \citet{gharaibeh_2011_Photoionizationmeasurementsiron} \\
\Fevii & 5722 & 0.0 & $>$~0.45 & 98.985~$\pm$~0.015 & \citet{gharaibeh_2011_Photoionizationmeasurementsiron} \\
\Fevii & 6088 & 0.0  & $>$~0.45 & 98.985~$\pm$~0.015 & \citet{gharaibeh_2011_Photoionizationmeasurementsiron} \\
\Fex & 6376 & 0.0 & $>$~0.45 & 233.6~$\pm$~0.4 & \citet{sugar_1985_Atomicenergylevels} \\
\Fexi & 7894 & 0.0 & 0.24 & 262.10~$\pm$~0.12 & \citet{sugar_1985_Atomicenergylevels} \\
\Fexiv & 5304 & 0.0 & $>$~0.45 & 361.0~$\pm$~0.7 & \citet{sugar_1985_Atomicenergylevels} \\
\hline
\Nev & 3347 & 0.08 & $>$~0.45 & 97.1900~$\pm$~0.0025 & \citet{kramida_1999_criticalcompilationextended} \\
\Nev & 3427 & 0.06 & $>$~0.45 & 97.1900~$\pm$~0.0025 & \citet{kramida_1999_criticalcompilationextended} \\
\hline
\Caxv & 5446 & 0.0 & $>$~0.45 & 817.2~$\pm$~0.6 & \citet{biemont_1999_IONIZATIONPOTENTIALSATOMS} \\ \hline
\end{tabular}
\begin{flushleft}
\textit{Notes:} $^{1}$ Rest wavelength as used within \sleipnir. \\
$^{2}$ Redshift at which the line becomes useful in this analysis, i.e., within the nominal wavelength range and clear of edge effects. Floored at $z = 0$. \\
$^{3}$ Redshift at which the line is too close to the edge of the nominal DESI wavelength coverage to be reliably used in this analysis. Ceilinged at the redshift limit of the study, $z = 0.45$.\\
$^{4}$ Energy required to produce this ionisation state. \\
$^{5}$ Data were obtained using the National Institute of Standards and Technology (NIST): Atomic Spectra Database \citep{kramida_2024_NISTAtomicSpectra}, with references provided to the specific per-line studies.
\end{flushleft}
\end{table*}

\subsubsection{Candidate flagging}
\label{sec:SLEIPNIR_Candidate_Flagging}

As \sleipnir\ returns measured parameters for each spectral-line region (unless too close to the edge of the observational range), additional checks are used to confirm the detection of CrLs, as outlined in Table~\ref{tab:sleipnir_CrL_detection_cuts}. These cuts are intended to remove the majority of false-positive detections from weak or contaminated features.

Each CrL that passes these cuts contributes to an individual spectrum's ``ECLE score''. The four strongest CrLs (\fspectralline{Fe}{vii}{6088}, \fspectralline{Fe}{x}{6376}, \fspectralline{Fe}{xi}{7894}, and \fspectralline{Fe}{xiv}{5304}) are each assigned an ECLE score of two. The three remaining, weaker \Fevii\ lines are each assigned an ECLE score of one. Given their relative weakness and reduced pEQW detection threshold, these lines contribute to the score only if \fspectralline{Fe}{vii}{6088} has met the detection criteria. The detection of all remaining lines is independent. As such, the maximum ECLE score a spectrum can reach is 11. In our search of DESI EDR, we set a threshold score of seven to flag a target for visual inspection. At redshifts exceeding 0.24, where the \fspectralline{Fe}{xi}{7894} line is too close to (or beyond) the red edge of the DESI spectrum, this scoring threshold is reduced to five to reflect the lower possible maximum ECLE score of nine.

As previously outlined, in contrast to the previous studies using earlier versions of \sleipnir\ \citep{callow_2024_rateextremecoronal, callow_2025_rateextremecoronal} and other such work, we do not apply the requirement that any of the Fe coronal lines have a pEQW exceeding 20 per cent that of \fspectralline{O}{III}{5007} for initial candidate identification. This threshold was motivated by analysis of the evolution of the initial ECLE sample observed in SDSS DR7 \citep{abazajian_2009_SeventhDataRelease} by \citet{wang_2012_EXTREMECORONALLINE} and was thought at the time to be a conclusive method of separating contaminating AGNs from TDE-linked objects. Follow-up spectroscopy of both previously identified ECLEs \citep{yang_2013_LONGTERMSPECTRALEVOLUTION, clark_2024_Longtermfollowupobservations} and CrL-TDEs \citep{newsome_2024_MappingInner01, clark_2025_2018dyktidaldisruption} has complicated this picture; strong \Oiii\ emission features can develop over time as the Fe CrLs fade. These features can dominate the spectrum and would thus prevent their identification as an ECLE if the relative emission-line-strength cut is applied, despite them being the same objects but at a later phase of evolution. Modelling work by \citet{mummery_2025_Galaxyscaleconsequencestidal} reveals that this short-term excitement of such \Oiii\ emission is fully consistent with the response expected from a dense circumnuclear environment following a TDE.

As such, with improvements to \sleipnir\ since previous studies (including improved spectral smoothing and handling of night-sky lines) and the size of the overall DESI EDR input sample resulting in a manageable false-positive rate, this constraint was removed. Scientifically, this provides two additional opportunities: late-time TDE-ECLEs displaying strengthened \Oiii\ emission will not be excluded from the initial samples, and a sample of CrL-AGN displaying a wide range of Fe CrL line strengths can be collected for comparison purposes.

Given the range of properties displayed by the existing samples of ECLEs, \sleipnir\ has two additional candidate selection criteria: ``Strong CrL'' flagging, and ``High line SNR'' flagging. ``Strong CrL'' flagging is triggered when any single CrL meets the detection criteria and both metrics for SNR exceed a threshold value, configured for our analysis to be ten. The motivation for this flag is to ensure that all objects with at least one CrL detected with high strength are visually inspected. The drawback for this condition is that, as it requires only one line to be detected, it is significantly more susceptible to contamination (particularly from residual sky lines) and thus has a higher overall false-positive rate. ``High line SNR'' flagging is triggered when all \Fevii\ lines are detected with both SNR metrics exceeding a threshold (set to five in this analysis) and disregarding the pEQW line-detection threshold. This flag is intended to capture any evolved objects where the higher ionisation state lines have already faded.

\begin{table*}
\caption{Each CrL must pass all of the outlined cuts to be considered ``detected''.}
\label{tab:sleipnir_CrL_detection_cuts}
\begin{adjustbox}{width=\textwidth}
\begin{tabular}{ll}
\hline
\textbf{Description} & \textbf{Values} \\ \hline
The central feature region$^{1}$ is clear from masked artefacts or spectral gaps & Central feature region contains no NaN values\\
The peak of a feature must occur within a given threshold of the expected line position & $\pm$~350~km~s$^{-1}$ \\
The targeted feature is the strongest emission feature within a region around the expected line position & $\pm$~800~km~s$^{-1}$\\
Central region of a feature must exceed a minimum threshold$^{2}$ & Peak scaled flux $> 1.05$ \\
Measured line pEQW must exceed a minimum threshold & pEQW $< -1.5$ \\
The minimum value of feature must be below a minimum threshold & Minimum scaled flux within the central feature region $\ge 0.7$ \\
Feature's peak SNR relative to the local region must exceed a given threshold & Peak relative local SNR within  region $\ge 3$ $^{3}$\\
Feature's minimum SNR relative to the local region$^{4}$ must exceed a given threshold & Minimum relative local SNR within region $\ge -1$ \\
Feature's minimum SNR relative to the error spectrum must exceed a given threshold & Minimum error spectrum derived SNR $\ge 3$ \\ \hline
\end{tabular}
\end{adjustbox}
\begin{flushleft}
\textit{Notes:} $^{1}$ Central feature region is defined as $\pm$~350~km~s$^{-1}$ of the expected line position.\\
${^2}$ By convention, emission features are defined with negative pEQWs. This threshold is relaxed by 1~\AA\ for the three weaker \Fevii\ emission lines. \\
$^{3}$ Defined as Maximum scaled flux density in the central feature region / Mean standard deviation of the local continuum regions.\\
${^4}$ Defined as Minimum scaled flux density  in the central feature region / Mean standard deviation of the local continuum regions. 
\end{flushleft}
\end{table*}

\subsection{Potential sources of false-positive or false-negative ECLE detections}
\label{sec:False_Detections}

\subsubsection{Sky lines}
\label{sec:Skylines}

Telluric emission lines or ``sky lines'' from Earth's atmosphere can present significant contamination problems toward the red end of the optical range, though several strong lines are also present in the bluer spectral regions. A list of the night-sky lines found to be most commonly problematic for ECLE identification is given in Table~\ref{tab:skylines_list}.

Whilst sky-line removal is often conducted as part of spectral processing in large surveys (as is the case in both SDSS and DESI) and in the reduction of individual spectra, the removal of these lines can be challenging with the possibility of residual features remaining. Furthermore, there are two additional factors that make these residual sky-line features particularly problematic in the search for ECLEs.

\begin{enumerate}
    \item Sky lines are narrow features produced by single line transitions. As such, if they are not removed fully, residual sky-line features can present morphologies very similar to the narrow coronal lines targeted by CrL searches.
    \item At some redshifts, multiple sky lines coincidentally align with several Fe features, compounding the potential for false-positive detections.
\end{enumerate}

Thankfully, as sky lines occur at fixed and known positions in the observer frame (see Table~\ref{tab:skylines_list}), objects with potential sky-line contamination can be reliably identified and treated with additional care before classification as an ECLE candidate. Moreover \sleipnir\ includes the option to perform additional preprocessing of the spectra to remove and interpolate across the strongest sky lines prior to the start of analysis. If required, the heavily contaminated telluric regions can be fully removed from consideration to reduce the number of false-positive detections if these regions are a particular concern, though with the knock-on effect of reducing the ECLE recovery ability of \sleipnir\ at redshifts where coronal lines are present within these regions of the observer frame. Given the size of the EDR sample and the number of overall candidates retrieved, we do not employ this option here.

\begin{table}
\centering
\caption{A list of the telluric emission lines (sky lines) identified as most problematic for ECLE identification.}
\label{tab:skylines_list}
\begin{tabular}{lcc}
\hline
\textbf{Line} & \textbf{Rest Frame Wavelength (\AA)} & \textbf{Source} \\ \hline
\Hgi & 4358.34 & Light pollution \\
\Oi & 5577.338 & Auroral emission \\
\Oi & 6300.304 & Auroral emission \\
OH & 7276.405 & OH Meinel emission \\
OH & 7630 & OH Meinel emission \\ \hline
\end{tabular}
\begin{flushleft}
\end{flushleft}
\end{table}

\subsubsection{Instrumental architecture}
\label{sec:Instrumental_Architecture}

Each telescope and instrument has a unique configuration and architecture that must be considered in its use for transient detection. These factors can include chip gaps across the spectrograph range where individual wavelength regimes will not be sampled (or sampled at lower SNR in composite observations), and spectra obtained using multiple spectrographs with differing wavelength coverage which are then combined to produce a single continuous spectrum.

These architectural differences can introduce configuration-specific artefacts into the resulting spectra that can resemble real coronal line features and lead to false-positive detections. As with night-sky lines, these artefacts occur at predictable wavelengths, so when identified they can be mitigated through pre-analysis processing steps on a per-configuration basis.

As described in detail by \citet{desicollaboration_2022_OverviewInstrumentationDark}, the DESI instrument system consists of three independent spectral ranges with regions of overlap between each to enable coaddition into continuous spectra. These spectrographs (or arms) are B (3600--5930~\AA),  R (5600--7720~\AA), and Z (7470--9800~\AA).
Coaddition of DESI spectra is not handled internally by \sleipnir; instead, we make use of the coadded files produced by the \desispec\ \citep{guy_2023_SpectroscopicDataProcessing} reduction pipeline directly. Whilst the coaddition performed by the \desispec\ pipeline is robust, residual features within the overlapping regions may be present in the final spectra. These features can resemble narrow emission lines following processing and are most common at the edges and centre of the coadded regions. As previously described with telluric emission features (Section~\ref{sec:Skylines}), when these features occur they do so at fixed positions in the observer frame, allowing any coincidental alignments with redshifted Fe lines to be identified and such candidates to be discounted as ECLE candidates. 

\subsubsection{Incorrect redshifts}
\label{sec:Incorrect_Redshifts}

\sleipnir\ relies on accurate external redshifts for proper line identifications. Small discrepancies in redshifts could potentially result in spectral features moving outside the line-detection windows. More significant redshift errors can also result in false-positive detections where unrelated lines are mistakenly placed at the location of the Fe coronal lines in the rest frame. We will return to this specific source of false-positive detection in our discussion of the DESI EDR sample itself in Section~\ref{sec:EDR_ECLEs}.

\subsubsection{Low signal-to-noise ratio}
\label{sec:Low_SNR}

At low SNRs, whether owing to operational issues or to faint sources, noise-related artefacts can remain in the spectra even following processing steps such as smoothing or rebinning. In such cases, if smoothing is too aggressive, strong noise features can be blurred to resemble real spectral features.

These cases can be reduced by applying observed brightness/magnitude cuts to the input spectral samples, though given the random nature of noise contamination, visual inspection remains necessary to remove all such cases from the final object samples. Given the size of the DESI EDR and \sleipnir's false-positive rate, these cuts have been set very conservatively (roughly equivalent to predicted source magnitudes of 24.5 in each DESI camera), though would likely need to be raised for larger studies to prevent an unfeasibly high absolute number of false positives during manual inspection.

\subsubsection{Multiple sources within a single spectrum}
\label{sec:Multiple_Sources}

Whilst the vast majority of spectra will contain only one source, care must be taken when handling cases where more than one object is in close proximity -- in particular, when these objects are at differing redshifts. In these cases the blending of the objects produces composite spectra with additional emission features that can align with the expected positions of Fe coronal lines and lead to false-positive detections. Visual inspection and object confirmation is required to deal with such cases, as the spectral features responsible for the false detections are real features and thus will look and behave in the same manner as the desired Fe features during spectral processing. One such example, however, was identified in the previous search of the BOSS LOWZ sample (DESI ID 39628342705525273 / SDSS~J2218+2334) and is described in depth by\citet{callow_2025_rateextremecoronal}. Thankfully, the number of such cases is small, with none directly encountered during the EDR search.  

\subsection{Detection efficiency}
\label{sec:Detection_Efficiency}

We determine the detection efficiency of \sleipnir\ by running it on simulated DESI EDR coronal line galaxy spectra. These were generated by planting coronal lines from known ECLEs into 10,000 random DESI spectra from the sample described in Section \ref{sec:DESI_Sample}. For each spectrum, the coronal lines from a randomly selected ECLE from the \citet{wang_2012_EXTREMECORONALLINE} sample were planted at the wavelengths of the coronal lines. For each simulated spectrum, the strengths of the coronal lines were modified using a scaling factor. This scaling factor was randomly sampled from between $0$ and a maximum value set depending on the presence of \fspectralline{O}{iii}{5007} in the base spectrum.
If this line was present, then the maximum scaling factor was set such that the strongest coronal line would have the same strength as \Oiii. If it was absent, the maximum scaling factor was set to $1$. This was motivated by the fact that none of the coronal lines in the \citet{wang_2012_EXTREMECORONALLINE} ECLEs was stronger than the \fspectralline{O}{iii}{5007} line in the spectra if it was present. \sleipnir\ reaches 50 per cent detection efficiency at an average coronal line strength of $\sim-1.3$~\AA, and has a maximum efficiency of $\sim90$ per cent. Compared to the efficiency of the algorithm on SDSS and BOSS spectra \citet{callow_2024_rateextremecoronal, callow_2025_rateextremecoronal}, it is slightly worse, which is likely due to the overall lower SNR of DESI spectra, though other factors likely also play a role (e.g., differences in galaxy properties between the sample). A comparison of efficiency between the samples is provided in Fig.~\ref{fig:DetEff}.

\section{Sample Construction and Classification}
\label{sec:Sample_Construction_and_Classification}

\begin{figure*}
    \centering
    \includegraphics[width=0.95\textwidth]{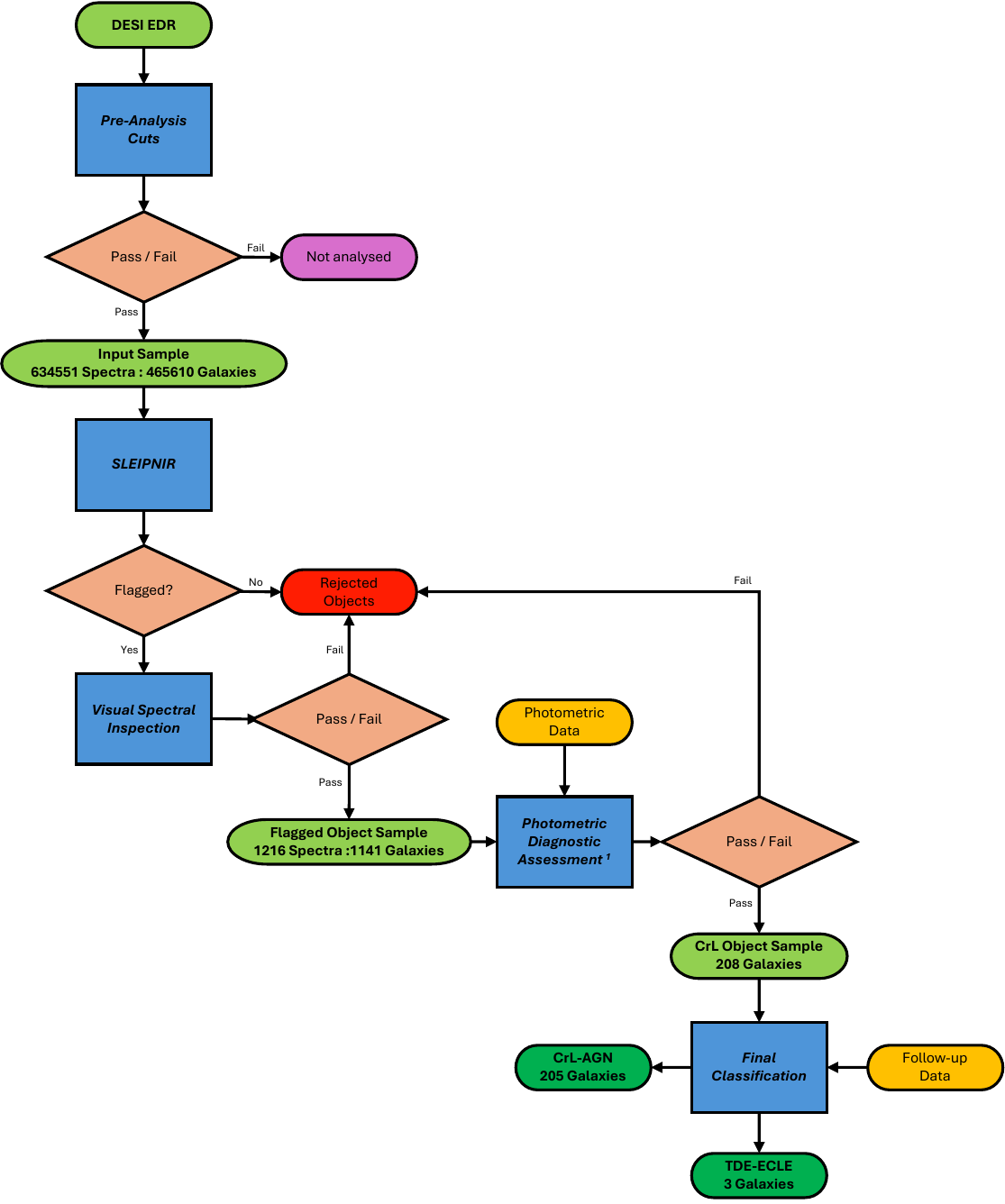}
    \caption{Summary of the workflow involved in the processing and classification of the spectra from DESI EDR.\\ Note: $^1$ Prior to this point in the workflow, individual spectra are treated independently, but after this point all spectra obtained for a single galaxy are used to provide a single classification.}
    \label{fig:DESI_EDR_Workflow}
\end{figure*}

In this section we outline the processing of the DESI EDR data and the classification of the spectra to produce our sample of TDEs and other CrL objects. A visual summary of the process is provided in Fig.~\ref{fig:DESI_EDR_Workflow}.

\subsection{Input DESI spectral sample}
\label{sec:DESI_Sample}
The DESI survey is composed of multiple survey programs with distinct (though overlapping) targeting criteria. A full description of the DESI targeting selection can be found in \citet{desicollaboration_2024_EarlyDataRelease} and the references within, whilst a summary of the relevant information for this work now follows.

The three primary DESI surveys for extragalactic targets, in order of increasing average redshift, are the Bright Galaxy Survey (BGS), the Emission Line Galaxies (ELG) sample, and the Luminous Red Galaxies (LRG) sample. Additionally, the QSO (quasistellar object) sample spans a much larger overall redshift range specifically selecting AGNs and QSOs otherwise excluded from the main targeting selections. In addition to these main programs, DESI includes several secondary-target programs with more specific science goals than those of the larger programs. Of particular interest to this work is the \textit{BGS-AGN}. This sample consists of QSOs and other AGNs at low redshifts (distribution peak $z \approx 0.5$) which were rejected by the main BGS survey's star-galaxy selection criteria, but meet photometric diagnostics indicative of an AGN rather than stellar nature. Objects within this sample have been shown to match the properties of the main DESI-QSO sample, though at significantly lower redshifts \citep{juneau_2025_IdentifyingMissingQuasars}. Owing to overlapping targeting criteria, an individual galaxy may be selected by multiple programs.

As this work is interested in detecting TDE-ECLEs and other galaxies with coronal line emission (e.g., AGNs with a range of coronal emission lines for comparison purposes), we do not restrict the study to a specific program, though we do preserve each object's original targeting program(s) for additional analysis. However, given the large number of small secondary programs, we condense these into a single ``Secondary'' category.

Selection for \sleipnir\ relies on meta information from the DESI spectroscopic pipeline, including the measured spectral redshift measured by the DESI redshift fitter \redrock\ and a basic spectroscopic classification of the object as a galaxy rather than a stellar source \cite{guy_2023_SpectroscopicDataProcessing}. The full selection parameters and reasoning are outlined in Table~\ref{tab:spectra_section}.

\begin{figure*}
    \centering
    \includegraphics[width=\textwidth]{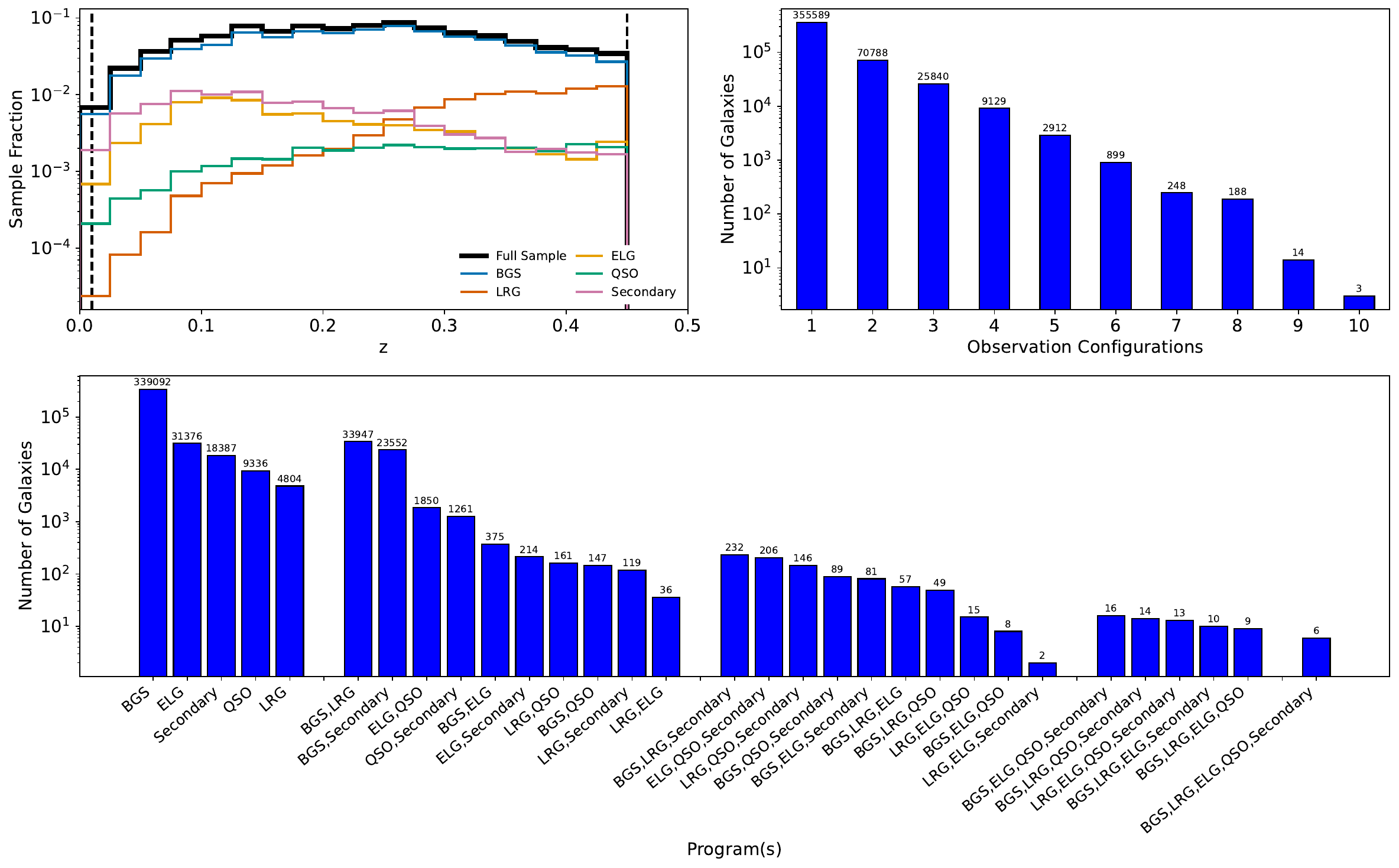}
    \caption{\textit{Upper left:} The redshift distribution of the \sleipnir\ input object sample including a per-observational-program breakdown highlighting the numeric dominance of the DESI-BGS sample. Note: Each targeted galaxy can be included in the selection for more than one program; thus, the full sample redshift distribution is not a simple numeric sum of the individual program redshift distributions. BGS selected targets dominate the sample with 85.4\% of objects.
    \textit{Upper right:} Breakdown of the number of observational configurations (and hence spectra) per galaxy in the \sleipnir\ input object. 23.63\% of the sample have been observed in multiple configurations. \textit{Bottom:} Breakdown of the DESI observational program distribution of the \sleipnir\ object input sample. 13.44\% of the \sleipnir\ input object sample are part of multiple observational programs.}
    \label{fig:Sample_Breakdown_1}
\end{figure*}

All criteria must be met for an object to be processed by the primary \sleipnir\ analysis routine. Whilst we do not select spectra from specific observational programs, the upper redshift cut ($z \le 0.45$) significantly limits the number of spectra selected from the LRG, ELG, and QSO samples, for which the peak of the redshift distributions are significantly higher than our upper cutoff. As such, our input sample is dominated numerically by the BGS sample. The upper-left panel of Fig.~\ref{fig:Sample_Breakdown_1} is a redshift distribution showing the breakdown of the input sample. Note that a single galaxy can be classified for observation as part of more than one program; as such, the full sample redshift distribution is not a simple numeric sum of the individual program redshift distributions. The bottom panel of Fig.~\ref{fig:Sample_Breakdown_1} serves as a summary of this program distribution.

Some objects were observed by DESI multiple times under potentially different observing conditions and for varying effective exposure times. This is a consequence of some objects being located in the footprint of multiple DESI observing ``tiles'' (telescope pointings with specific target configurations) and meeting the requirements for observation as part of each. As our initial selection criteria for analysis are primarily based on basic parameters of each galaxy, the analysis is conducted on a per-tile basis. As such, galaxies that have been observed in multiple tiles will have their spectra processed and analysed independently multiple times. Given the potential for time-variability in the CrL signatures, we do not pre-stack or combine the spectra of objects across different tiles; instead, observations of any galaxy in multiple tiles are compared during the candidate analysis stage. In some cases, an individual tile may be observed more than once if the first observation failed to reach the depth required for the main survey. The data in such cases are combined by the DESI pipeline; the combined data are then fed into \sleipnir. As a result of these considerations, the \sleipnir galaxy sample explored in this work comprises \GalaxySample\ unique galaxies. A breakdown showing the number of objects observed with differing numbers of tiles and configurations is given in the upper-right panel of Fig.~\ref{fig:Sample_Breakdown_1}. As such we define the ``\sleipnir\ Galaxy Sample'' as the sample of unique galaxies explored by this work (which consists of \GalaxySample\ galaxies). 

\begin{table*}
\centering
\caption{Outline of the selection criteria to construct our input spectral sample.}
\label{tab:spectra_section}
\begin{adjustbox}{width=\textwidth}
\begin{tabular}{llll}
\hline
\textbf{Selection Criterion} & \textbf{Code} & \textbf{Rationale} \\ \hline
\texttt{isTGT} & \texttt{fibermap[`OBJTYPE'] == `TGT'} & Removes fibers used for reference sky calibration etc. \\
\texttt{isProgram}  & `Reduced program' contains at least one of: `BGS', `LRG', `ELG', `QSO', `Secondary' $^1$ & Removes identified Milky Way stellar targets \\
\texttt{isGal}  & \texttt{zbest[`SPECTYPE'] in [`GALAXY', `QSO']} & Additional cut to filter potential stellar sources $^2$ \\
\texttt{isGoodFiber} & \texttt{fibermap[`COADD\_FIBERSTATUS']} == 0 & Removes any spectra from fibers with detected issues \\
\texttt{isGoodZBest} & \texttt{(zbest[`DELTACHI2'] > 25.)} \& (zbest[`ZWARN'] == 0) & Quality cut on determined redshifts \\
\texttt{inZRange} & 0.01 < \texttt{zbest['Z'] <= 0.45} & $^3$ \\
\texttt{hasGFlux}, \texttt{hasRFlux}, \texttt{hasZFlux} & fibermap['FLUX$_{X}$'] > 0.25 & $^4$ \\\hline
\end{tabular}
\end{adjustbox}
\begin{flushleft}
\textit{Notes:} $^1$ As TDEs (and thus TDE-ECLEs) occur in a wide range of galaxy types, we do not restrict our initial input sample to galaxies selected by any specific criteria. \\
$^2$ Determined by the \redrock\ redshift analysis pipeline, where the best-fitting spectral template is either a QSO or normal galaxy.\\
$^3$ Restricts analysis to objects within a redshift range where H$\alpha$ is within the DESI spectrograph range. Also further removes any remaining galactic sources. We note some classes of source (e.g., QSOs) whose redshift are better measured by specific ``afterburner'' codes such as \textsc{QUASARnet} \citep{green_2025_UsingActiveLearning} rather than \redrock. In this search we make use of \redrock-derived redshifts only. Further searches (in particular at higher redshifts where QSOs are a larger component of the sample) will make use of these additional sources of redshift determination.\\
$^4$ Cut on the observed brightness of each object as a basic cut on expected SNR. Each filter must pass the flux cut (in nanomaggies; see \href{https://www.sdss3.org/dr8/algorithms/magnitudes.php\#nmgy}{here} for a detailed explanation) corresponding to an approximate magnitude of 24. This is a relaxed cut to initially filter only extremely faint spectra. \\
\end{flushleft}
\end{table*}

\subsection{Sample classification}
\label{sec:Sample_Classification}

Of the \GalaxySample\ unique galaxies in the \sleipnir\ Galaxy Sample, \sleipnir\ flagged a total of 1141 potential candidates, 0.25 per cent of the sample. A breakdown of the flagged candidates based on the condition upon which they were flagged as a potential candidate is given in Table~\ref{tab:Sleipnir_Sample_Results}.

\begin{table*}
\caption{Breakdown of the candidates selected by \sleipnir. Values without brackets give the raw breakdown (i.e., per observation with no consideration of duplicates), whilst those in brackets are for the sample of unique objects.}
\label{tab:Sleipnir_Sample_Results}
\begin{adjustbox}{width=\textwidth}
\begin{tabular}{lcccccc}
\hline
\textbf{Flagging Characteristic} & \multicolumn{2}{c}{\textbf{Number Flagged}} & \multicolumn{2}{c}{\textbf{Percentage of \sleipnir\ Input Samples}} & \multicolumn{2}{c}{\textbf{Percentage of Flagged Samples}} \\ \hline
 & \multicolumn{1}{c}{Spectral Sample} & \multicolumn{1}{c}{Galaxy Sample} & \multicolumn{1}{c}{Spectral Sample} & \multicolumn{1}{c}{Galaxy Sample} & \multicolumn{1}{c}{Flagged Spectral Sample} & \multicolumn{1}{c}{Flagged Galaxy Sample} \\ \hline
High Scoring & 966 & 928 & 0.152\% & 0.199\% & 79.44\% & 81.33\% \\
High Scoring \& Strong Fe Feature & 38 & 29 & 0.008\% & 0.006\% & 3.13\% & 2.54\% \\
Strong Fe Feature & 131 & 129 & 0.021\% & 0.028\% & 10.77\% & 11.31\% \\
\Fevii\ SNR & 81 & 79 & 0.013\% & 0.017\% & 6.66\% & 6.92\% \\ \hline
Any & 1216 & 1141 & 0.192\% & 0.245\% & - & - \\
Input Sample Totals & \SpecSample\ & \GalaxySample\ & - & - & - & - \\ \hline
\end{tabular}
\end{adjustbox}
\end{table*}

All spectra flagged by \sleipnir\ as being a potential ECLE underwent a series of additional analysis and classification steps to remove false-positive identifications (as detailed in Section~\ref{sec:False_Detections}) and to classify these as being either TDE- or AGN-related. The sample is initially cleaned via visual inspection of the flagged spectra. Following this, classification is made through a combination of BPT diagnostics, archival NIR colour-colour classification, MIR photometric analysis, optical photometric analysis, crossmatching with existing databases, and follow-up spectra for the most interesting candidates.

\begin{figure*}
    \centering

    \includegraphics[width=\textwidth]{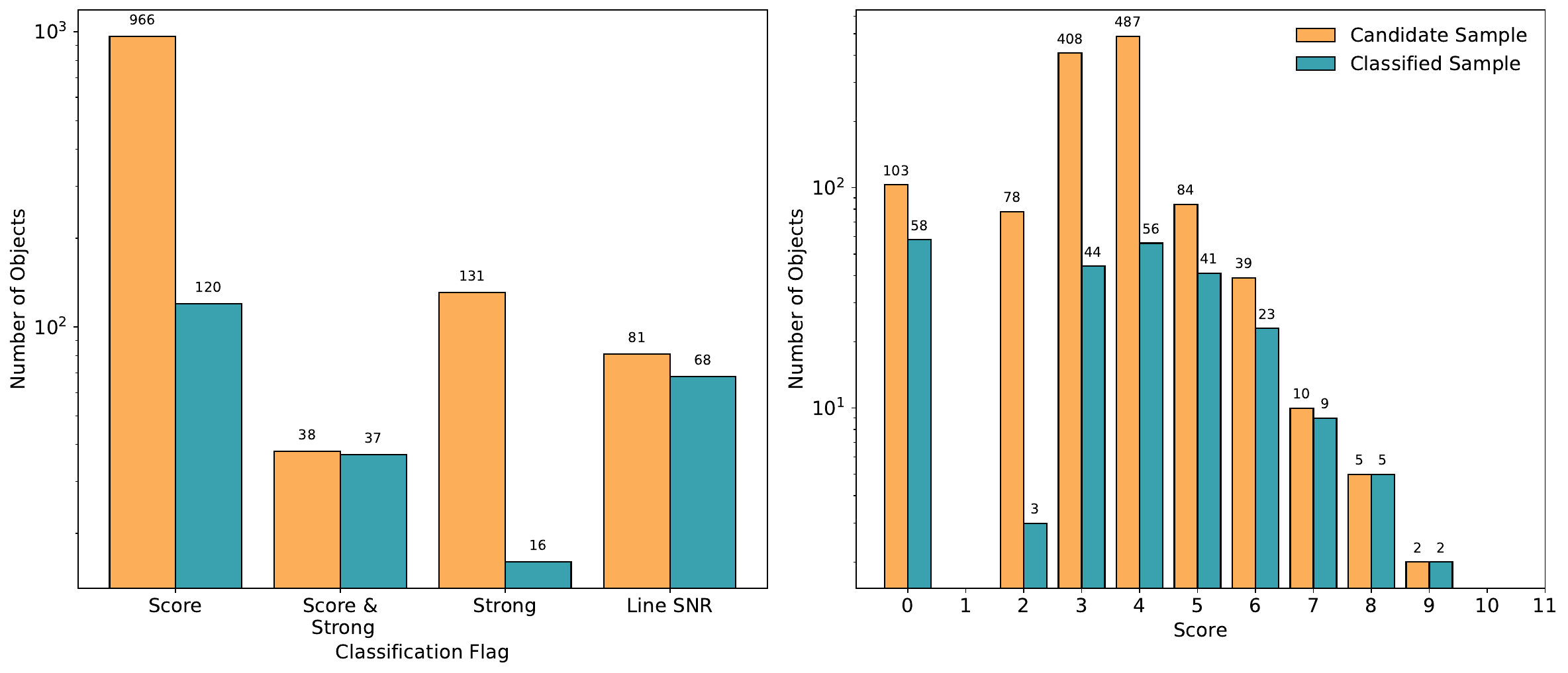}
    \caption{\textit{Left:} Comparison of the flagging criteria between the number of spectra flagged by \sleipnir\ before and following full classification. The highest fractions of false positives are seen in the spectra flagged on general score and single strong features. This is the result of a combination of low SNR, sky-line contamination, and the effect of single serendipitously located bright artefacts. \textit{Right:} Comparison of the ECLE detection score between the number of spectra flagged by \sleipnir\ before and following full classification. As expected, those objects flagged with lower scores (i.e., fewer confident Fe CrL detections) are more likely to be assessed as false positives.}
    \label{fig:Sample_Classification_Breakdown}
\end{figure*}

As noted in Section~\ref{sec:SLEIPNIR_Candidate_Flagging}, we do not apply a line-strength cut on the CrLs relative to \fspectralline{O}{III}{5007} emission in this study, and make use of other diagnostics to distinguish between TDEs and AGNs. As such, our definition of ``ECLE'' is somewhat different from that of other studies. Furthermore, given the increasing range of CrL strengths observed in AGNs (see, e.g., \citealt{clark_2024_Longtermfollowupobservations}), we do not subdivide our AGN sample further based on line strengths, deferring such analysis to future work. Instead, we divide our CrL object sample simply into TDE- and AGN-linked classifications. Fig.~\ref{fig:Sample_Classification_Breakdown} shows the distribution of candidates across the various flagging criteria and ECLE scores obtained by the initial \sleipnir\ automated flagging compared to the distributions obtained following the removal of all false-positive detections.

\begin{figure}
    \centering
    \includegraphics[width=0.95\columnwidth]{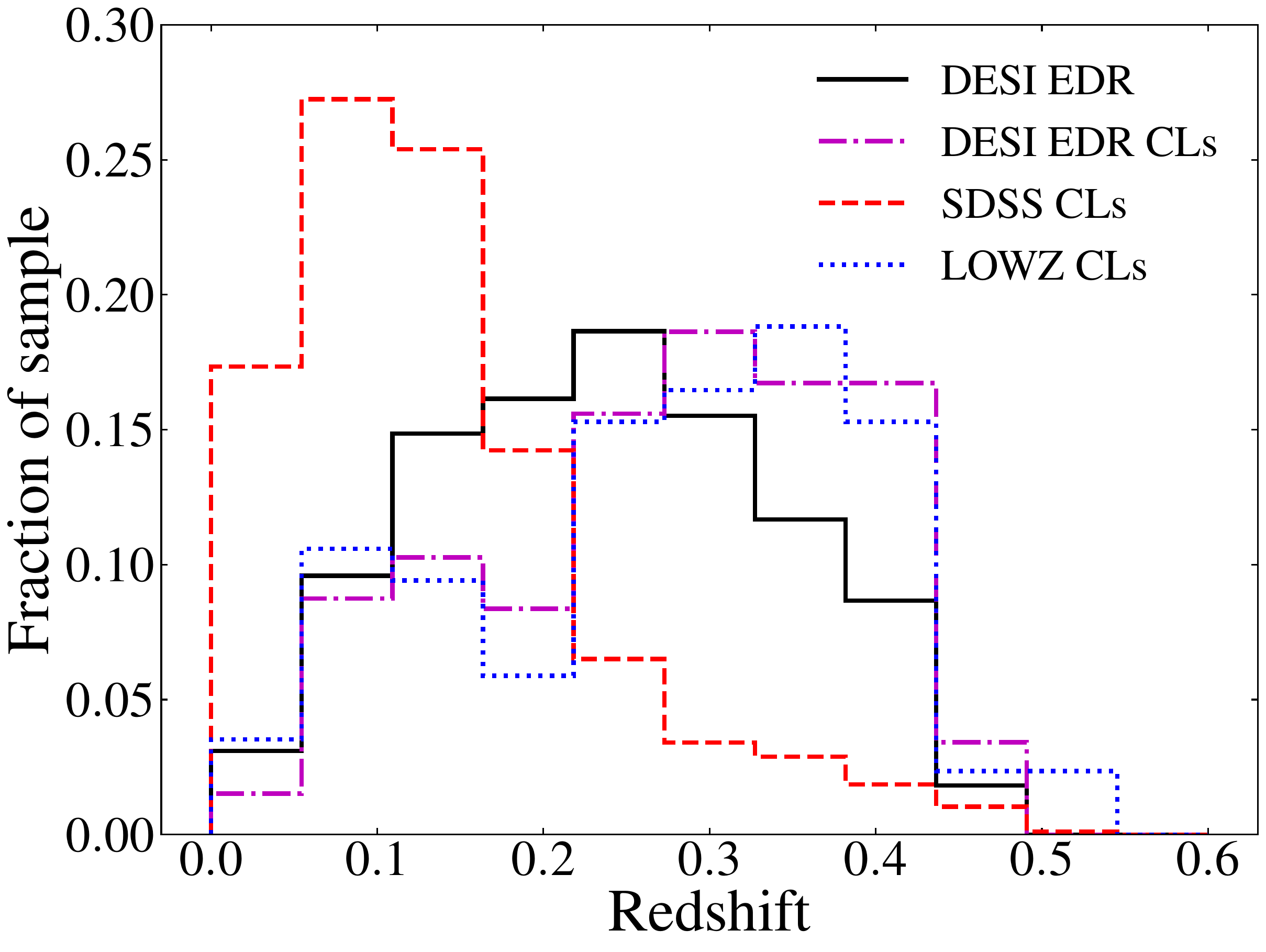}
    \caption{Comparison of the redshift distributions of the DESI input sample and the \sleipnir\ flagged sample of candidate coronal line galaxies (solid black and dot-dashed purple curves, respectively). We also include the distributions of the coronal line galaxies selected from searches of SDSS Legacy and BOSS LOWZ galaxy samples \citep[red dashed and blue dotted curves, respectively;][]{callow_2024_rateextremecoronal,callow_2025_rateextremecoronal}. The EDR coronal line galaxies are underrepresented compared to the overall EDR sample at $0.11 < z < 0.22$ and overrepresented in the range $0.28 < z < 0.44$. Note: As in \protect\cite{callow_2025_rateextremecoronal}, the comparison BOSS LOWZ sample has been filtered to remove galaxies at the narrow redshift bands most heavily contaminated by night-sky emission lines at the position of the Fe-CrLs.}
    \label{fig:zDistComp}
\end{figure}

\begin{figure}
    \centering
    \includegraphics[width=0.95\columnwidth]{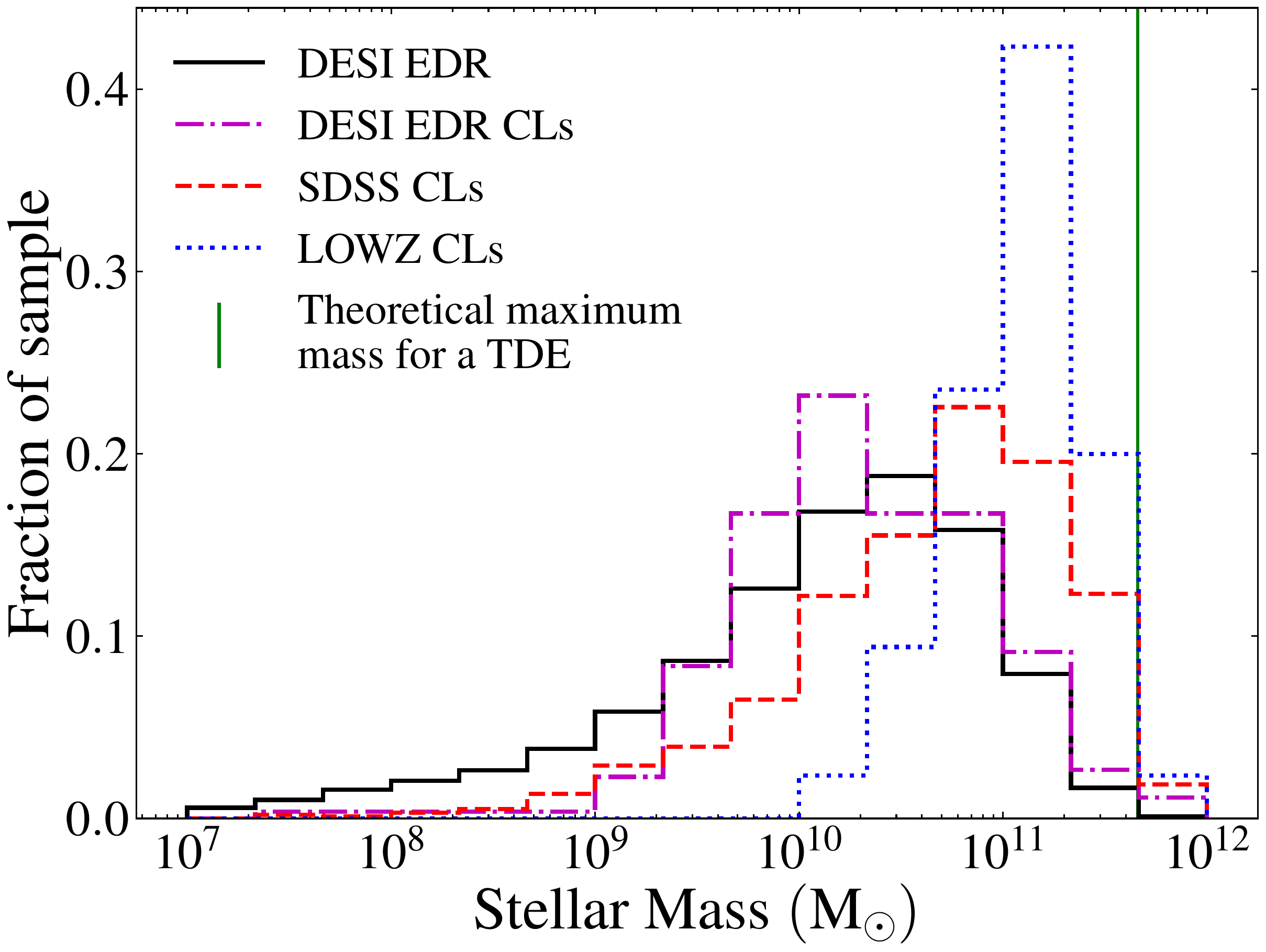}
    \caption{Comparison of the mass distributions of the DESI input sample and the \sleipnir\ flagged sample of candidate coronal line galaxies (solid black and dot-dashed purple curves, respectively). We also include the distributions of the coronal line galaxies selected from searches of SDSS Legacy and BOSS LOWZ galaxy samples \citep[red dashed and blue dotted curves, respectively;][]{callow_2024_rateextremecoronal,callow_2025_rateextremecoronal}. The vertical green line marks the theoretical galaxy stellar mass limit above which a Sun-like star would fall directly into the galaxy’s SMBH instead of being disrupted as a TDE \citep{rees_1988_Tidaldisruptionstars}. The relation between the stellar mass of a galaxy and the mass of its SMBH used in this calculation is from \citep{reines_2015_RELATIONSCENTRALBLACK}. The peak of the EDR coronal line galaxy sample is lower than the overall EDR sample, but very few galaxies with masses below $10^9~\msun$ are detected as CrL galaxies.}
    \label{fig:MassDistComp}
\end{figure}

We compare the redshift and mass distributions of the potential candidates to the overall EDR galaxy sample in Figs.~\ref{fig:zDistComp} and \ref{fig:MassDistComp}. Also included are the distributions of the coronal line galaxies detected in SDSS Legacy and BOSS LOWZ \citep{callow_2024_rateextremecoronal,callow_2025_rateextremecoronal}. Compared to the overall EDR galaxy sample, the candidate galaxies are underrepresented in the range $0.11<z<0.22$ and overrepresented at $0.28<z<0.44$. The high-redshift overrepresentation is likely due to the fact that \fspectralline{Fe}{xi}{7894} is not within the wavelength range of DESI above $z\approx0.3$, so the threshold for a galaxy to be detected by our algorithm is lowered. Therefore, galaxies with fewer coronal lines will be detected more often above this redshift. The underrepresentation covers a redshift range in which the coronal lines frequently lie at the observed wavelengths of known problematic sky lines, so are removed as part of the preprocessing of the spectra. A Kolmogorov-Smirnov test between the EDR overall galaxy and candidate galaxy samples produces a $p$-value $<0.01$, meaning we can reject the null hypothesis that these distributions are drawn from the same overall population.
Further reinforcing these reasons for the differences between the overall EDR sample and candidate galaxies is the similarity between the EDR and LOWZ coronal line galaxy distributions. The overall samples cover similar redshift ranges, so the similarities are likely a result of our detection algorithm.

Comparing the mass distributions of the same samples, we see that the peak of the EDR candidate galaxy sample is slightly lower than that of the overall EDR sample.
We also detect very few candidate galaxies with masses below $10^9~\msol$, similar to the results of the SDSS Legacy search.
A reason for this lack of lower-mass galaxies may be that they typically host lower-mass BHs, as accretion onto a SMBH is one of the main processes that can create coronal lines.
Lower-mass SMBHs are less able to sustain accretion and therefore produce coronal lines, so we are less likely to detect lower-mass galaxies with coronal lines.
We are able to reject the null hypothesis that the EDR galaxy and candidate galaxy samples are drawn from the same population with a Kolmogorov-Smirnov test that gave a $p$-value $<0.01$. We remind the reader that these distributions are constructed from the galaxies satisfying the automatic \sleipnir\ classification criteria and should not be used to determine specifics of the final object samples.

\subsection{Sample crossmatching}
\label{subsec:Crossmatching}

To further explore our sample and determine if any of the objects within are of special interest (based on existing galaxy classifications or transient activity reports) we crossmatch against six catalogues. A breakdown of these crossmatches follows, with a summary provided in Table~\ref{tab:Crossmatch_Results}. More detailed breakdown of the results of each crossmatch are included in Appendix~\ref{sec:Crossmatch_Results}.

\begin{table*}
\caption{Results summary for the crossmatch searches conducted between the DESI EDR ECLE candidate sample and numerous catalogues as described in the text.}
\centering
\label{tab:Crossmatch_Results}
\begin{tabular}{lcccc}
\hline
 & \multicolumn{2}{c}{\textbf{\sleipnir\ Galaxy Sample $^1$}} & \multicolumn{2}{c}{\textbf{Final CrL Object Sample $^2$}} \\
\textbf{Crossmatch Source} & \multicolumn{1}{c}{\textbf{Matches}} & \multicolumn{1}{c}{\textbf{Percentage Matched}} & \multicolumn{1}{c}{\textbf{Matches}} & \multicolumn{1}{c}{\textbf{Percentage Matched}} \\ \hline
TNS Classified Sources $^3$ & \TNSGalaxyMatches & 0.262\% & \TNSCrLMatches & 2.885\% \\
Literature Derived TDE Candidate Catalogue $^4$ & \TDEDatabaseGalaxyMatches & 0.002\% & \TDEDatabaseCrLMatches & 0.000\% \\
SDSS Legacy CrL Sample $^5$ & \SDSSLegacyGalaxyMatches & 0.004\% & \SDSSLegacyCrLMatches & 2.404\%\\
BOSS LOWZ CrL Sample $^6$ & \BOSSGalaxyMatches & 0.002\% & \BOSSCrLMatches & 0.000\%\\
MILLIQUAS $^7$ & \MILLIQUASGalaxyMatches & 1.536\% & \MILLIQUASCrLMatches & 78.365\% \\
Ding et al. (2025) $^8$ & \DingGalaxyMatches & 0.010\% & \DingCrLMatches & 9.135\% \\
\hline
\end{tabular}
\begin{flushleft}
\textit{Notes:} $^1$ This sample consists of the \GalaxySample\ unique galaxies present within the initial sample processed by \sleipnir. \\
$^2$ This sample consists of the \CrLSample\ unique objects identified as displaying real coronal lines following processing by \sleipnir\ and manual inspection and confirmation. \\
$^3$ Crossmatch performed on the full TNS database as of \TNSMatchDate.\\
$^4$ See Section~\ref{sec:Clark_Crossmatch} for details. \\
$^5$ Consists of the \sleipnir\ flagged objects from the SDSS Legacy Survey DR17 as detailed in \citet{callow_2024_rateextremecoronal}. \\
$^6$ Consists of the \sleipnir\ flagged objects from the BOSS LOWZ sampled as detailed by \citet{callow_2025_rateextremecoronal}. \\
$^7$ Utilises v8 of the Million Quasars (MILLIQUAS) catalogue, consisting of 1021800 individually identified AGNs from a range of sources; see \citet{flesch_2023_MillionQuasarsMilliquas} for details.\\
$^8$ Sourced from the independent search of ECLEs in DESI EDR detailed by \citet{ding_2025_ExploringLinkExtreme}.
\end{flushleft}
\end{table*}

\subsubsection{Transient Name Server: TNS}
\label{sec:TNS_Crossmatch}

The position of each candidate identified by \sleipnir\ is used to query the Transient Name Server (TNS)\footnote{\url{https://www.wis-tns.org/}} to identify any potential existing transient connection to our candidate sample. Transient matching is performed using a cone search with a radius of $3''$. This radius was chosen to be the diameter of the DESI fiber to remain focused on the nuclear regions of galaxies whilst allowing some tolerance for the presence of transients some distance from the nucleus. 

These crossmatches identified \TNSGalaxyMatches\ galaxies within the \sleipnir\ Input Galaxy Sample hosting at least one TNS recorded transient (outlined in Table~\ref{tab:Appendix_Full_TNS_Crossmatch}). Of these, \TNSCrLMatches\ galaxies are within the \sleipnir\ CrL Object Sample (as outlined in Table~\ref{tab:CrL_TNS_CrossMatch}), though none of the corresponding TNS transients has been independently classified.

\begin{table*}
\caption{Results of the crossmatch search between the \sleipnir\ CrL Object Sample and objects recorded in the full TNS database as of \TNSMatchDate.}
\label{tab:CrL_TNS_CrossMatch}
\begin{tabular}{lccc}
\hline
\textbf{DESI ID} & \textbf{CrL Classification}$^1$  & \textbf{TNS Name} & \textbf{TNS Classification} \\ \hline
39628506220463012 & CrL-AGN & AT 2018mer & None \\
39633304189666247 & CrL-AGN & AT 2019jw & None \\
39633300653869008 & CrL-AGN & AT 2021lar & None \\
39627782338120416 & CrL-AGN & AT 2021mks & None \\
39633453855017885 & CrL-AGN & AT 2021vje & None \\
39627758191511075 $^2$ & CrL-AGN & AT 2025ils & None \\ \hline
\end{tabular}
\begin{flushleft}
\textit{Notes:} $^1$ Classification of the source of the CrLs observed in this source made by this work. \\
$^2$ This transient is linked to a CrL-AGN identified as having broad CrL emission and discussed in Appendix~\ref{sec:Appendix_CrL_AGN_Sample_Information}. \\
\end{flushleft}
\end{table*}

\subsubsection{TDE literature search crossmatch}
\label{sec:Clark_Crossmatch}

In addition to a general search of the TNS, we conduct a search of objects compiled from the literature that have been identified as TDEs or related phenomena that have not necessarily been reported to the TNS.

Our database of literature objects for this crossmatching includes all objects classified as TDEs on the TNS (for general completeness), the original ECLE sample from \citet{wang_2012_EXTREMECORONALLINE}, the IR-TDE sample from \citet{masterson_2024_NewPopulationMidinfraredselected}, the TDE sample compiled by \citet{qin_2022_LinkingExtragalacticTransients}, \textit{eROSITA} X-ray selected TDEs from \citet{sazonov_2021_Firsttidaldisruption}, \citet{grotova_2025_populationtidaldisruption}, and \citet{eyles-ferris_2025_Ninetidaldisruption}. Also included in this crossmatch are those Mid-infrared Outbursts in Nearby Galaxies (MIRONGs) identified by \citet{jiang_2021_MidinfraredOutburstsNearby} and interpreted by \citet{wang_2022_MidinfraredOutburstsNearby} as being likely TDEs (noting that the MIRONG sample as a whole is likely dominated by AGNs; \citealt{dodd_2023_MidInfraredOutburstsNearby}), along with archival TDEs. In addition to objects with well-established TDE classifications, we include those which may have a TDE
-linked origin, including the samples of ambiguous nuclear transients (ANTs) from \citet{hinkle_2024_Midinfraredechoesambiguous} and \citet{wiseman_2025_systematicallyselectedsample}, as well as the quasiperiodic eruptions (QPEs) collated by \citet{jiang_2025_EmbersActiveGalactica}. In total this database currently consists of \TDEDatabaseTotal\ verified TDEs, TDE candidates, and other related nuclear transients. Given the developing nature of nuclear transient classification, this database of comparison objects is not intended to be a pure sample of conventional TDEs, rather serving as a reference for all non-AGN-related nuclear transients. 

This crossmatch identifies \TDEDatabaseGalaxyMatches\ transients with spectra in the full \sleipnir\ Input Galaxy Sample, as summarised in Table~\ref{tab:Clark_Crossmatch_Results}. Since none of these DESI spectra passed the cuts for initial visual inspection (i.e., \sleipnir\ did not identify them as potential ECLEs), all individual spectra of these galaxies were manually inspected following the crossmatch and confirmed to be absent of coronal lines. A more in-depth study of TDE host galaxies observed by DESI is planned but goes beyond the scope of this current work.

\subsubsection{SDSS crossmatches}
\label{sec:SDSS_Crossmatch}

We also perform crossmatch searches between the \sleipnir\ Input Galaxy Sample and the samples of galaxies returned as ECLE candidates by earlier versions of \sleipnir\ when analysing the SDSS samples described by \citet{callow_2024_rateextremecoronal, callow_2025_rateextremecoronal}. Crossmatches were performed against the raw forms of these SDSS samples (i.e., all objects flagged by \sleipnir\ including those later identified as false positives) for sample validation purposes. We note that the DESI spectra used in the analysis of the ECLEs identified by \citet{callow_2024_rateextremecoronal, callow_2025_rateextremecoronal} were obtained after the DESI EDR and are thus not part of the \sleipnir\ Input Galaxy Sample.

The crossmatch of the \sleipnir\ Input Galaxy Sample to the SDSS Legacy DR17 sample from \citet{callow_2024_rateextremecoronal} returned \SDSSLegacyGalaxyMatches\ objects, of which five are within the EDR CrL Object Sample. All five of the objects within the EDR CrL Object Sample have been classified as CrL-AGN prior to the crossmatch. The crossmatch of the \sleipnir\ Input Galaxy Sample to the BOSS LOWZ CrL sample from \citet{callow_2025_rateextremecoronal} returned seven objects, none of which is included in the EDR CrL Object Sample. The full results of these crossmatches are summarised in Appendix~\ref{sec:Appendix_SDSS_Crossmatches}.

\subsubsection{MILLIQUAS crossmatch}
\label{sec:MILLIQUAS_Crossmatch}

Our penultimate crossmatch compared the \sleipnir\ Input Galaxy Sample to the eighth version of the Million Quasars (MILLIQUAS) catalogue \citep{flesch_2023_MillionQuasarsMilliquas}, with the goal of further supporting or ruling out AGN classification of galaxies displaying coronal lines. We note here that as the MILLIQUAS catalogue includes classifications making use of DESI spectral data, it is not a fully independent AGN classification source in all cases.

Given the relative abundance of AGNs and the presence of coronal line AGNs, it is unsurprising that this crossmatch generated the most matches, with \MILLIQUASGalaxyMatches\ galaxies in the \sleipnir\ Input Galaxy Sample also present in the MILLIQUAS catalogue, \MILLIQUASCrLMatches\ of these within the Final CrL Object Sample. Of the latter, all were classified as CrL-AGNs.

In addition to these galaxies, 55 crossmatched AGNs were flagged by SLEIPNIR as potential ECLEs but discarded as false positives that do not display real CrL emission features at various stages of the classification process. These false-positive identifications were the result of sky-line contamination and other artefacts producing serendipitous features strong enough to flag the spectrum for inspection.

\subsubsection{Ding et al.~(2025) crossmatch}
\label{sec:Ding_Crossmatch}

A recently published independent search for ECLEs in the DESI EDR by \citet{ding_2025_ExploringLinkExtreme} identified 84 CrL galaxies which they classified as non-AGN hosts based on BPT diagnostics. Key differences between the \citet{ding_2025_ExploringLinkExtreme} analysis and the analysis presented in this work include the following:

\begin{itemize}
    \item There is no upper redshift cut on the \citet{ding_2025_ExploringLinkExtreme} sample, compared to the limit of $z < 0.45$ employed here.
    \item The \citet{ding_2025_ExploringLinkExtreme} sample applies a line-strength cut according to which at least one CrL must exceed 20 per cent of the strength of \fspectralline{O}{iii}{5007} to remove AGNs.
    \item The presence of \fspectralline{Ne}{v}{3347} and \fspectralline{Ne}{v}{3427} are included as object selection criteria for the \citet{ding_2025_ExploringLinkExtreme} sample but are not included here.
    \item Unlike the multiwavelength approach adopted here, \citet{ding_2025_ExploringLinkExtreme} classify their candidates based almost exclusively on the DESI EDR spectrum (in combination with a limited number of archival SDSS spectra).
    \item An object can be selected as an ECLE in the \citet{ding_2025_ExploringLinkExtreme} sample based purely on the presence of one emission line (e.g., \fspectralline{Fe}{xiv}{5304}) regardless of whether other lower-ionisation lines are present. In our analysis, while such objects could be flagged based on the ``Strong CrL'' criterion by \sleipnir\ and visually inspected, such objects would be rejected as unphysical artefacts if they lack any other Fe CrLs.
\end{itemize}

When a redshift cut of $z = 0.45$ is applied to the \citet{ding_2025_ExploringLinkExtreme} sample to match that of our \sleipnir-based analysis, \DingGalaxyMatches\ potential shared objects are obtained, all of which were manually reinspected by us. Of these, \DingCrLMatches\ were flagged by \sleipnir\ as potential ECLEs. From this subsample, five were rejected as being artefacts due to residual sky lines or instrumental effects. This includes DESI~J179.8036-00.5238, which was reported by \citet{ding_2025_ExploringLinkExtreme} as showing \fspectralline{Fe}{vii}{6088} in the DESI EDR spectrum but not in an archival SDSS observation, indicating potential transient activity. Following reinspection of the EDR spectrum, no other \Fevii\ emission line is observed (nor are any higher-ionisation lines). No indications of either TDE or AGN activity are observed in the object's optical or MIR photometry. Additionally, the \Fevii\ feature is not present in the DESI DR1 \citep{desicollaboration_2025_DataRelease1} spectrum of the same galaxy, which used a different version of the DESI reduction pipeline. As such, we conclude that the apparent \fspectralline{Fe}{vii}{6088} emission feature is an artefact and not an indication of CrL emission, TDE-related or otherwise. The remaining \DingCrLMatches\ objects within the subsample were found to show real CrL emission in this work but were identified as being the result of AGN activity following our classification process.

Of the remaining 23 candidates, three objects failed the initial selection for inclusion within the \sleipnir\ sample, with inspection of these revealing the apparent \Fevii\ detections to be artefacts due to either sky lines or reduction issues. The 20 remaining shared objects were included in the \sleipnir\ input sample but were not flagged as ECLE candidates for human inspection. This is the result of either their inclusion in the \citet{ding_2025_ExploringLinkExtreme} sample being exclusively due to the presence of \Nev\ emission -- which is not a selection criterion for \sleipnir\ -- or because the Fe CrL emission in the spectrum was rejected because \sleipnir's line selections indicate the feature as erroneous. Manual inspection of these spectra confirms these assessments. Owing to the outlined issues with sample construction, the \citet{ding_2025_ExploringLinkExtreme} analysis and its conclusions should be interpreted with care.

However, we note that galaxies displaying \Nev\ emission lines but no (or extremely weak) \Fevii\ emission features (e.g., DESI~J172.1834+51.8532 / DESI spectroscopic ID: 39633274758236748) are of interest for further study. Given the very similar energy requirements to produce the two ions (see Table~\ref{tab:crl_info}), these lines would be expected to occur together. The lack of \Fevii\ emission features in galaxies displaying \Nev\ offers a potentially sensitive diagnostic of the strength of the X-ray ionising continuum in such galaxies.

\section{DESI EDR ECLEs}
\label{sec:EDR_ECLEs}

Following our multistage classification, our sample of galaxies displaying verified CrLs in their EDR spectra consists of \CrLSample\ individual objects. We first focus on the three objects we ultimately classify as being linked to TDEs (Section~\ref{sec:TDE_Candidates}), before discussing six additional objects that were identified during our analysis as initially being ambiguous during their classification as either TDE- or AGN-related, though are ultimately classified as AGN activity or other contaminants (Section~\ref{sec:AGN_Candidates}).

\begin{figure}
    \centering
    \includegraphics[width=\columnwidth]{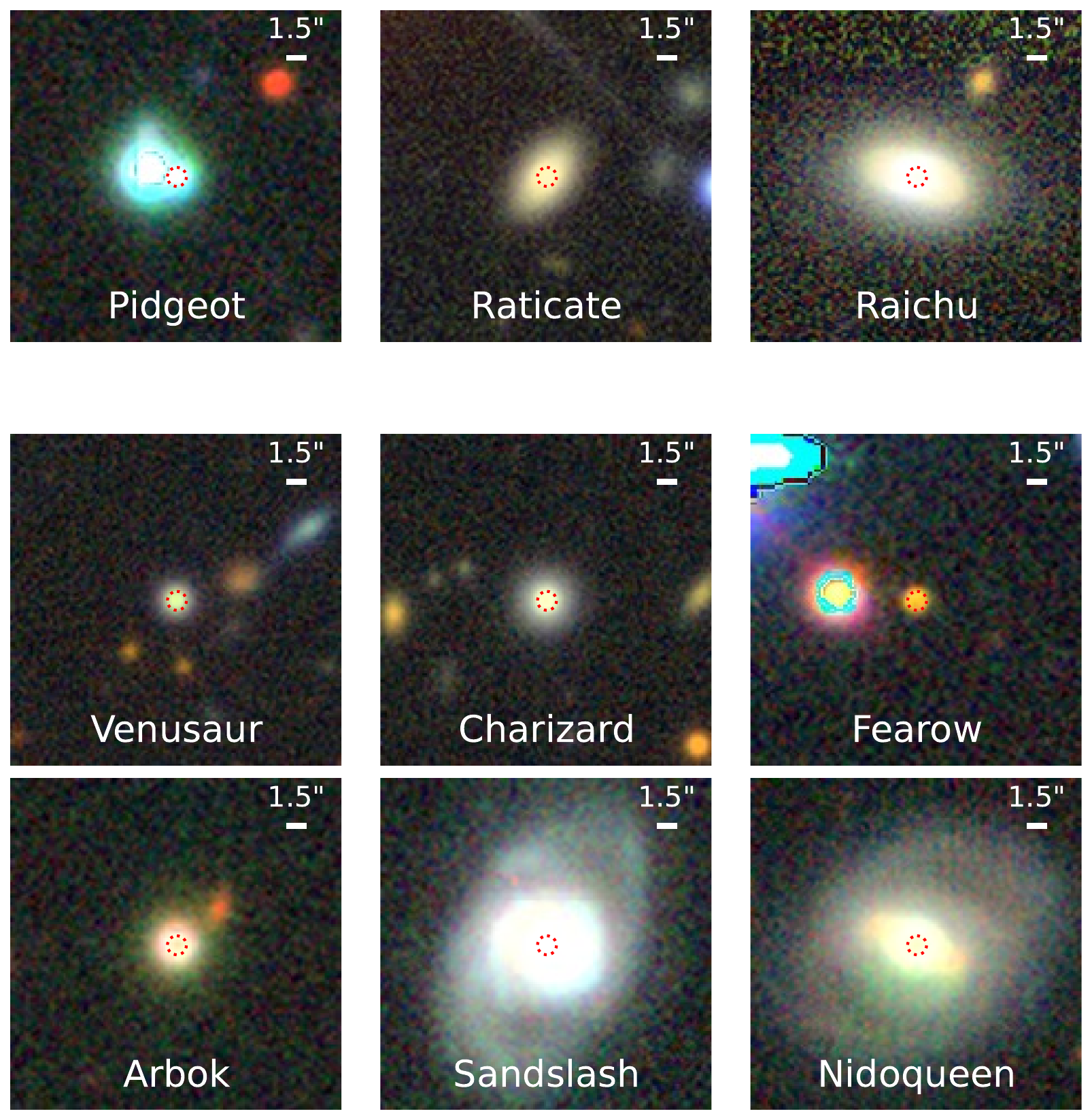}
    \caption{DESI Legacy Survey \textit{grz} composite images of each of the EDR ECLE candidates. \textit{Top row:} Candidates identified as being TDE-related following full classification. \textit{Middle and lower rows:} Candidates identified as being AGN-related following full classification. Dashed circle indicates the location of the DESI fiber used to obtain each object's spectrum.}
    \label{fig:Legacy_Image_Grid}
\end{figure}

\begin{figure*}
    \centering
    \includegraphics[width=\textwidth]{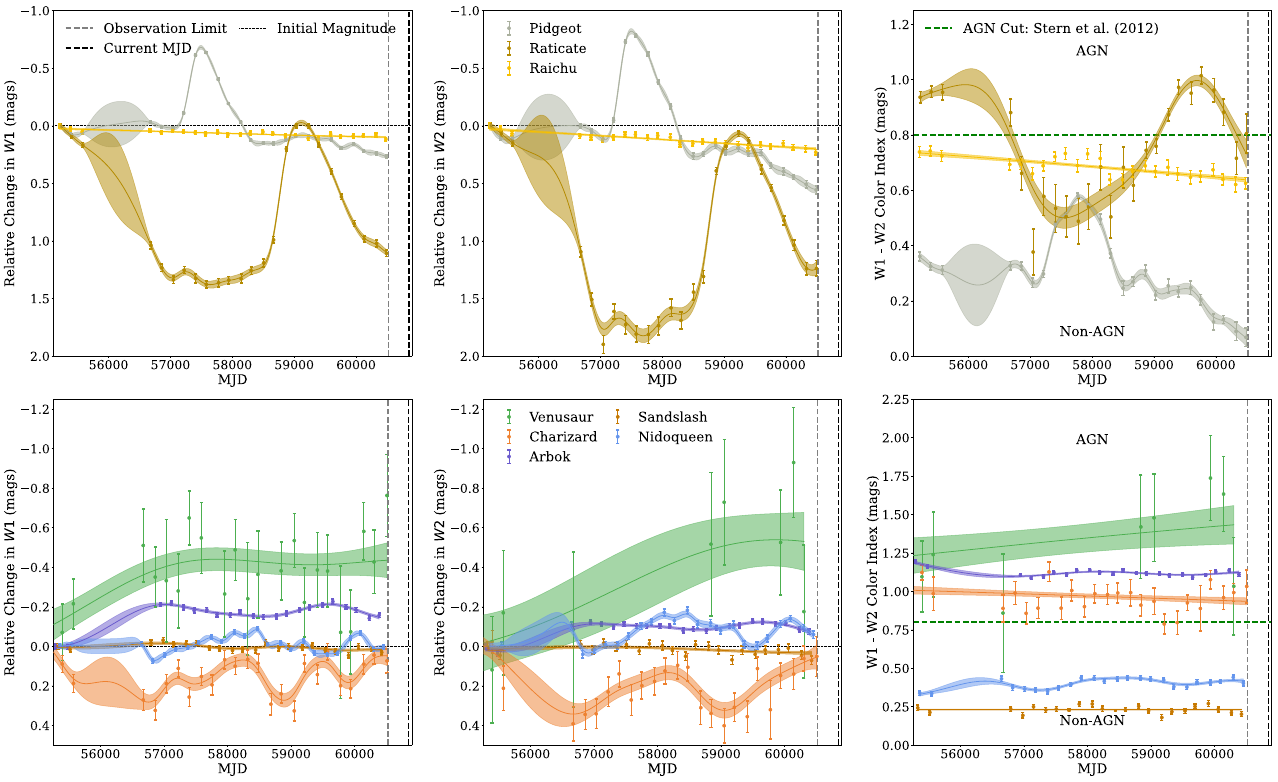}
   \caption{MIR evolution of the EDR-ECLE candidates. \textit{Top row:} Candidates linked to TDE activity following full classification. \textit{Bottom row:} Candidates linked to AGN activity following full classification. {\it Left panels:} Relative change in \textit{W1} compared to observed \textit{W1} peak. 
    {\it Middle panels:} Relative change in \textit{W2} compared to observed \textit{W2} peak. 
    {\it Right panels:} \textit{W1}--\textit{W2} colour evolution. The dashed horizontal line shown is the AGN/non-AGN dividing line from \protect\citet{stern_2012_MIDINFRAREDSELECTIONACTIVE}. Colours above this line are typical of AGNs. In all panels, fits displayed are obtained through Gaussian processes, with the shaded regions indicating the 1$\sigma$ fitting uncertainties. 
    Grey dashed vertical lines indicate the latest possible dates of \textit{WISE} observations. Black vertical lines indicate the current MJD at the time of manuscript compilation to provide a visual reference of the timescales involved.}
    \label{fig:Merged_MIR}
\end{figure*}

\begin{figure}
    \centering
    \includegraphics[width=\columnwidth]{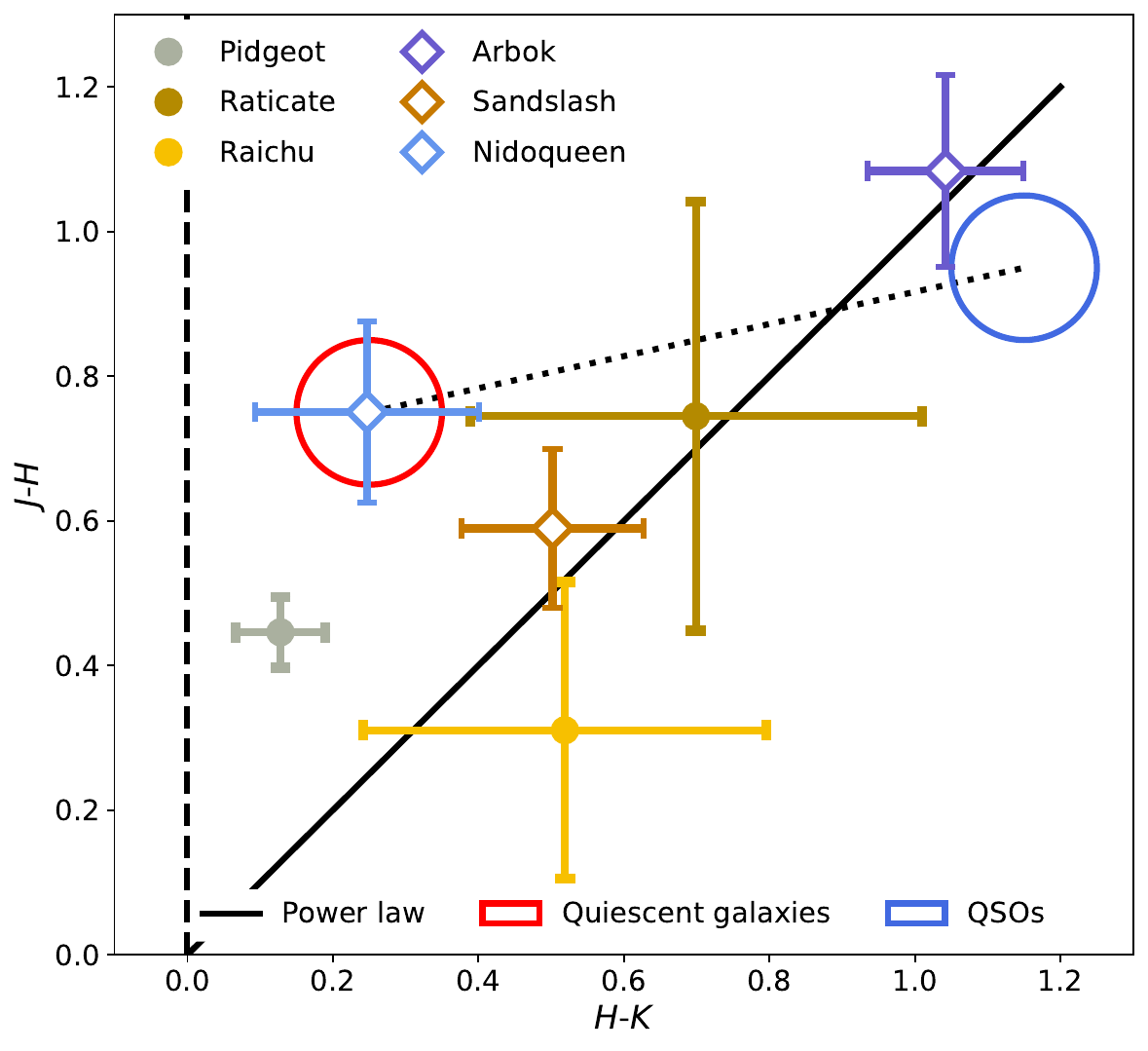}
    \caption{2MASS photometric diagnostic plot for the DESI EDR TDE-ECLE candidates. Filled markers indicate candidates linked to TDEs following overall classification. Hollow markers indicate candidates which have been ruled out as being TDE linked following overall classification. NIR classification based on \citet{hyland_1982_InfraredStudyQuasars} and \citet{komossa_2009_NTTSpitzerChandra}. As described by \citet{clark_2024_Longtermfollowupobservations}, whilst NIR classification is a useful assistive diagnostic, given both the time difference between the NIR observations and the wide range of colours displayed by galaxies of the same optical spectroscopic classification, it cannot be used to identify TDE-related activity on its own. The candidates Venusaur, Charizard, and Fearow are not included because they were not detected in the 2MASS survey.}
    \label{fig:NIR_Phot_Diagnostic}
\end{figure}

We give summary information for all of these objects in Table~\ref{tab:edr_candidate_information}, along with providing contextual \textit{grz} optical imaging from the DESI Legacy Surveys for each in Fig.~\ref{fig:Legacy_Image_Grid}, their MIR evolution in Fig.~\ref{fig:Merged_MIR}, and archival NIR colour information in Fig.\ref{fig:NIR_Phot_Diagnostic}. Note that the DESI spectra used in the follow-up analysis of the original \citet{wang_2012_EXTREMECORONALLINE} ECLE sample as part of \citet{clark_2024_Longtermfollowupobservations} were obtained following the cutoff for inclusion in EDR and are thus not part of our EDR sample.

\begin{table*}
\caption{EDR TDE Candidate Information}
\label{tab:edr_candidate_information}
\begin{tabular}{lcccccc}
\hline
\textbf{DESI Target ID} & \textbf{Internal Name $^{1}$} & \textbf{RA (J2000)} & \textbf{Dec. (J2000)} & \textbf{$z$} $^{2}$ & \textbf{E(B$-$V) (mag)} \\ \hline
\multicolumn{6}{c}{\textbf{TDE-Linked Candidates}} \\ \hline
39633332819985805 & Pidgeot & 243.37986 & 55.75283 & 0.193 & 0.008 \\
39627794400938039 & Raticate & 216.58918 & 0.14369 & 0.115 & 0.033 \\
39627884763023878 & Raichu & 204.39900 & 3.87760 & 0.057 & 0.026 \\ \hline
\multicolumn{6}{c}{\textbf{AGN-Linked Candidates}} \\ \hline
39627700695996439 & Venusaur & 30.46915 & -3.58470 & 0.160 & 0.026 \\
39627887837451274 & Charizard & 27.95254 & 4.19518 & 0.079 & 0.037 \\
39633118667214268 & Fearow & 236.47605 & 42.29324 & 0.152 $^{3}$ & 0.031 \\
39633066745924399 & Arbok & 115.57303 & 39.43668 & 0.416 & 0.056 \\
39633255741260888 & Sandslash & 172.66753 & 50.61793 & 0.058 & 0.013 \\
39633290029695741 & Nidoqueen & 212.95561 & 52.76762 & 0.074 & 0.010 \\ \hline
\end{tabular}
\begin{flushleft}
\textit{Notes:} $^1$ To prevent confusion between objects with similar DESI target IDs, each promising candidate was assigned an internal name derived from the popular Pokémon media franchise. We refer to the objects by these internal names within the following text and plots for conciseness.\\
$^{2}$ Redshift as determined by the \redrock\ pipeline. Uncertainties on \textbf{$z$} as provided by \redrock\ are on the order of a few ×10$^{-6}$.\\
$^{3}$ Erroneous determination; true redshift is $\sim 3.928$. See Section~\ref{sec:Fearow} for details.
\end{flushleft}
\end{table*}

\subsection{TDE-linked candidates}
\label{sec:TDE_Candidates}

\subsubsection{DESI 39627794400938039 : Pidgeot}
\label{sec:Pidgeot}

\begin{figure*}
\centering
\includegraphics[width=\textwidth]{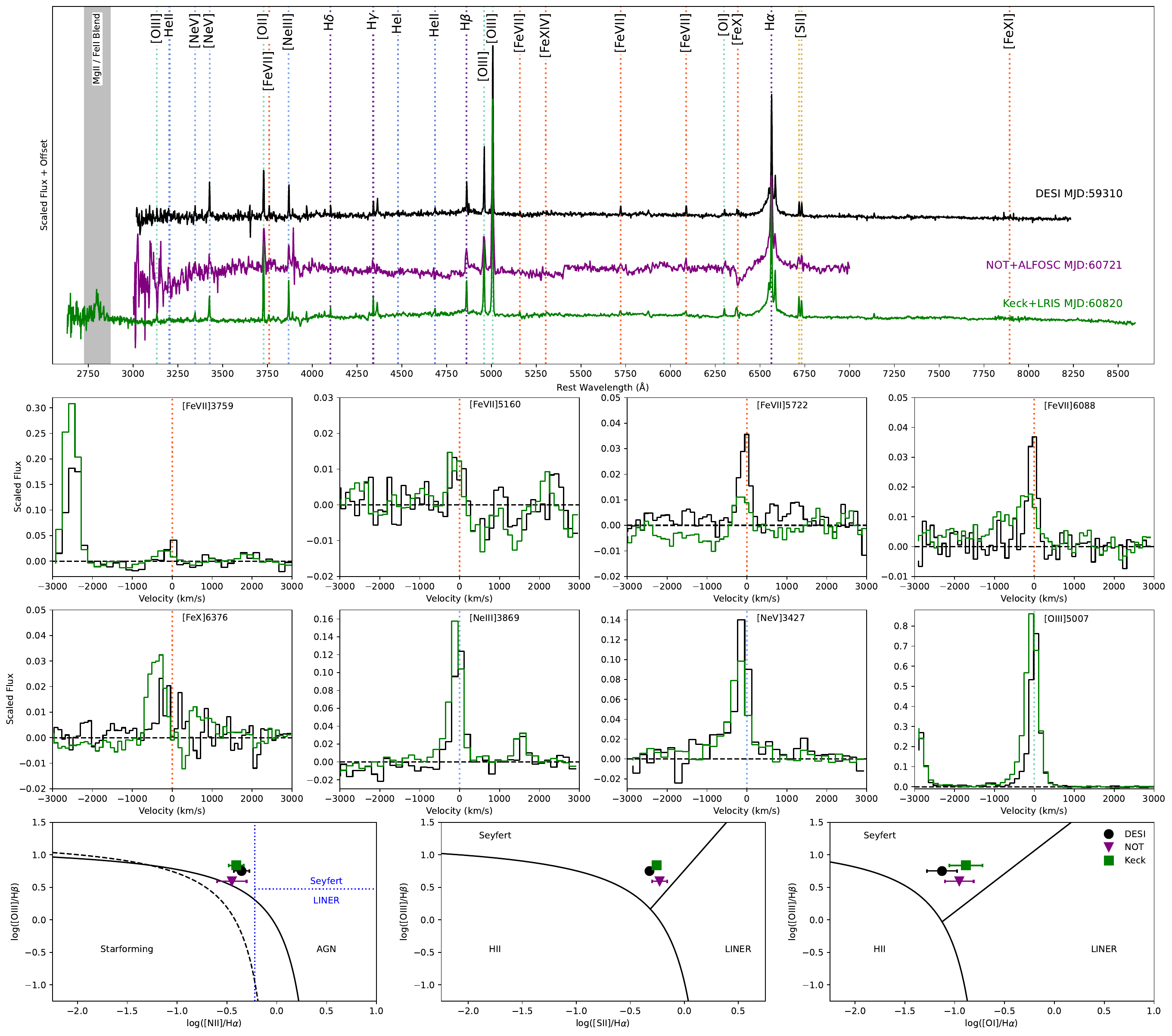}
\caption{Spectra and BPT diagrams of Pidgeot. All spectra have been rebinned to 2~\AA\ bin$^{-1}$ but have not been smoothed. \textit{Top row:} Spectral evolution of Pidgeot. The NOT+ALFOSC spectrum has prominent telluric absorption near 6400~\AA.
\textit{Second and third rows:} Comparison between the spectral features of most interest between the DESI (black) and Keck+LRIS (green) spectra. Clear reductions in \fspectralline{Fe}{vii}{5722}, \fspectralline{Fe}{vii}{6088}, and \fspectralline{Ne}{v}{3427} line strength are observed, in contrast to increasing \fspectralline{O}{iii}{5007} and \fspectralline{Ne}{iii}{3869} strength. Whilst some level of \fspectralline{Fe}{x}{6376} emission in the DESI spectrum is possible,  telluric contamination and instrumentation artefacts prevent its confirmation. The emission feature observed just blueward of \fspectralline{Fe}{x}{6376} in the Keck+LRIS spectrum is likely to be \Oi $\lambda$6364~\AA. \textit{Bottom row:} BPT diagnostic diagrams for all spectral epochs.}
\label{fig:Pidgeot_Spec_Evolution}
\end{figure*}

\begin{figure*}
    \centering
    \includegraphics[width=\textwidth]{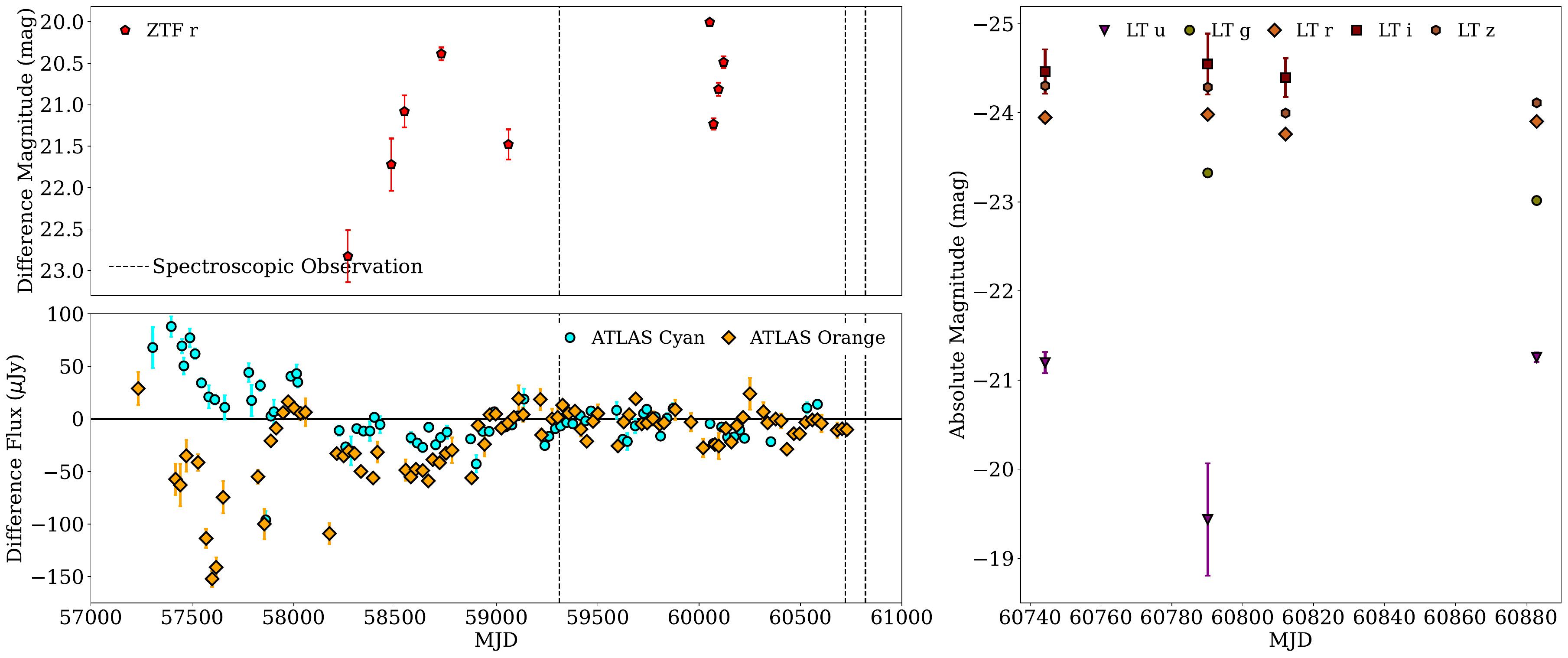}
    \caption{Photometric evolution of Pidgeot. \textit{Top left:} ZTF forced photometry difference magnitude light curve. \textit{Bottom left:} ATLAS forced photometry light curve presented in flux space. \textit{Right:} LT absolute magnitude light curve.}
    \label{fig:Pidgeot_Phot_Evolution}
\end{figure*}

Pidgeot was identified by \sleipnir\ as an ECLE candidate through a combination of multiple Fe CrL detections (\Fevii\ and \Fex) and a high-SNR \fspectralline{Fe}{vii}{5722} detection.  

Pidgeot was observed in two DESI tiles (298 and 310) with both observations meeting the \sleipnir\ flagging criteria, though only the tile 298 spectrum satisfied the high-SNR \fspectralline{Fe}{vii}{5722} requirement. We note that whilst the \fspectralline{Fe}{vii}{6088} line is close to a problematic sky line, its peak is sufficiently separated to provide a conclusive detection,  supported by the clear and unambiguous presence of other \Fevii\ emission lines. \fspectralline{Fe}{x}{6376} is close to the centre of the overlap region of the \textit{r} and \textit{z} cameras, whilst also being affected by telluric absorption and other line contamination, affecting the quality of its line profile and making a definitive detection of this line impossible. Pidgeot is located on a line of sight close to an unrelated foreground Milky Way star (see Fig.~\ref{fig:Legacy_Image_Grid}), confirmed by an additional Keck+LRIS spectrum obtained at the same time as our primary follow-up observation. As the two DESI spectra of Pidgeot were obtained just four days apart, we use the higher-SNR spectrum for our analysis.

We obtained optical follow-up spectra of Pidgeot with both NOT+ALFOSC and Keck+LRIS as shown in Fig.~\ref{fig:Pidgeot_Spec_Evolution}. The Keck spectrum has a significantly higher SNR, but both follow-up spectra show a reduction in the relative line strengths of the \fspectralline{Fe}{vii}{5722} and \fspectralline{Fe}{vii}{6088} lines, accompanied by a reduction in the strength of \fspectralline{Ne}{v}{3427} of very similar ionisation energy. Additionally, the Keck+LRIS spectrum indicates an increase in \Oiii\ and \fspectralline{Ne}{iii}{3869} emission, similar to the behaviour observed in other TDE-linked CrL galaxies \citep[e.g.,][]{clark_2024_Longtermfollowupobservations, clark_2025_2018dyktidaldisruption}.

2MASS NIR photometry does not indicate the presence of an AGN, nor is Pidgeot present within the MILLIQUAS database. However, emission-line diagnostics from the spectra indicate Seyfert-like activity, including the presence of broad components to both the H$\alpha$ and H$\beta$ lines, which are also present throughout follow-up observations (see the bottom row of Fig.~\ref{fig:Pidgeot_Spec_Evolution}). These BPT diagnostics (along with those reported for the other candidates) are not based on the automated measurements performed by \sleipnir\ during initial analysis, and instead reflect measurements made following emission-line fitting using the Image Reduction and Analysis Facility (IRAF) \textsc{splot} routine \citep{tody_1986_IRAFDataReduction, tody_1993_IRAFNineties, fitzpatrick_2024_ModernizingIRAFSupport}.

ZTF optical photometry (whilst limited) does not show the presence of any outburst behaviour, nor do our LT observations display significant evolution. ATLAS forced photometry shows a potential outburst in the \textit{cyan} band close to the start of the survey (from survey start until MJD~58000), though this is not matched by a corresponding outburst in the \textit{orange} band, which instead exhibits erratic negative difference flux relative to the baseline. We thus attribute this ``activity'' to observational artifacts near the start of the survey.

The MIR behaviour of Pidgeot is the most interesting aspect of its evolution (Figs.~\ref{fig:AllWISE_Colours} and \ref{fig:Merged_MIR}). AllWISE colour diagnostics classify it as a spiral galaxy that does not host an AGN. NEOWISE-R data indicate the galaxy is a non-AGN host that displayed a significant TDE-like MIR outburst, starting between MJD~57042 and 57215, reaching a peak at $\sim$~MJD~57575. We note that the star in close proximity to Pidgeot affects the underlying measurements through overlapping PSFs at the resolution of WISE. However, the observed outburst can conclusively be linked to the host galaxy of Pidgeot through inspection of the measured centroid position of the source which moves to the position of Pidgeot's host galaxy during the outburst before drifting back to between both objects as the outburst fades. Additionally, as noted, the AllWISE colour-colour diagnostics are indicative of a star-forming galaxy rather than a stellar source.

Whilst CrL-TDE outbursts that are superficially very similar have been seen in the past (e.g., AT~2017gge and AT~2018dyk among others; \citealt{clark_2025_2018dyktidaldisruption}), the overall evolution of Pidgeot is more complex. In previously studied transients, where the quiescent behaviour was observed both pre- and post-outburst, the brightness and colours match in both phases (see AT~2018dyk and AT~2019azh in fig.~7 of \citealt{clark_2025_2018dyktidaldisruption}). We note the limited number of objects for which this assessment is possible owing to the lifespan of the WISE spacecraft and the multiyear durations of the majority of these outbursts. This is not the case in Pidgeot, with the post-flare quiescent \textit{W1} state observed to be approximately 0.4~mag dimmer than the pre-outburst quiescent state. Additionally, the \textit{W2} luminosity displayed a 1300-day-long plateau (MJD~58300--59600) before declining an additional $\sim$~0.2~mag until the most recent observations. The colour of Pidgeot has also continued to trend blueward following the outburst (where it displayed the usual reddening behaviour typical of TDE MIR outbursts; \citealt{clark_2025_2018dyktidaldisruption}) with it now being $\sim$~0.2~mag bluer in the WISE bands than pre-outburst.

This may be the result of the TDE responsible for the outburst occurring in a galaxy already displaying long-term variability, specifically a galaxy showing a long-term reduction in MIR luminosity whilst trending toward bluer MIR colours. Alternatively, this long-term decline may be at least partially due to the contamination of the observations with flux from the nearby star. We will return to the MIR evolution of Pidgeot in Section~\ref{sec:MIR_Comp} to place it in context with the existing nuclear transient population.

Pidgeot is within the sky-coverage region of LoTSS-DR2, and is not associated with a radio source in any of the radio surveys examined in this work, ruling out radio-loud AGN activity. We conclude that Pidgeot is best explained as a TDE with a significant MIR outburst, likely occurring within a gas-rich environment.

\subsubsection{DESI 39627794400938039 : Raticate}
\label{sec:Raticate}

\begin{figure*}
    \centering
    \includegraphics[width=\textwidth]{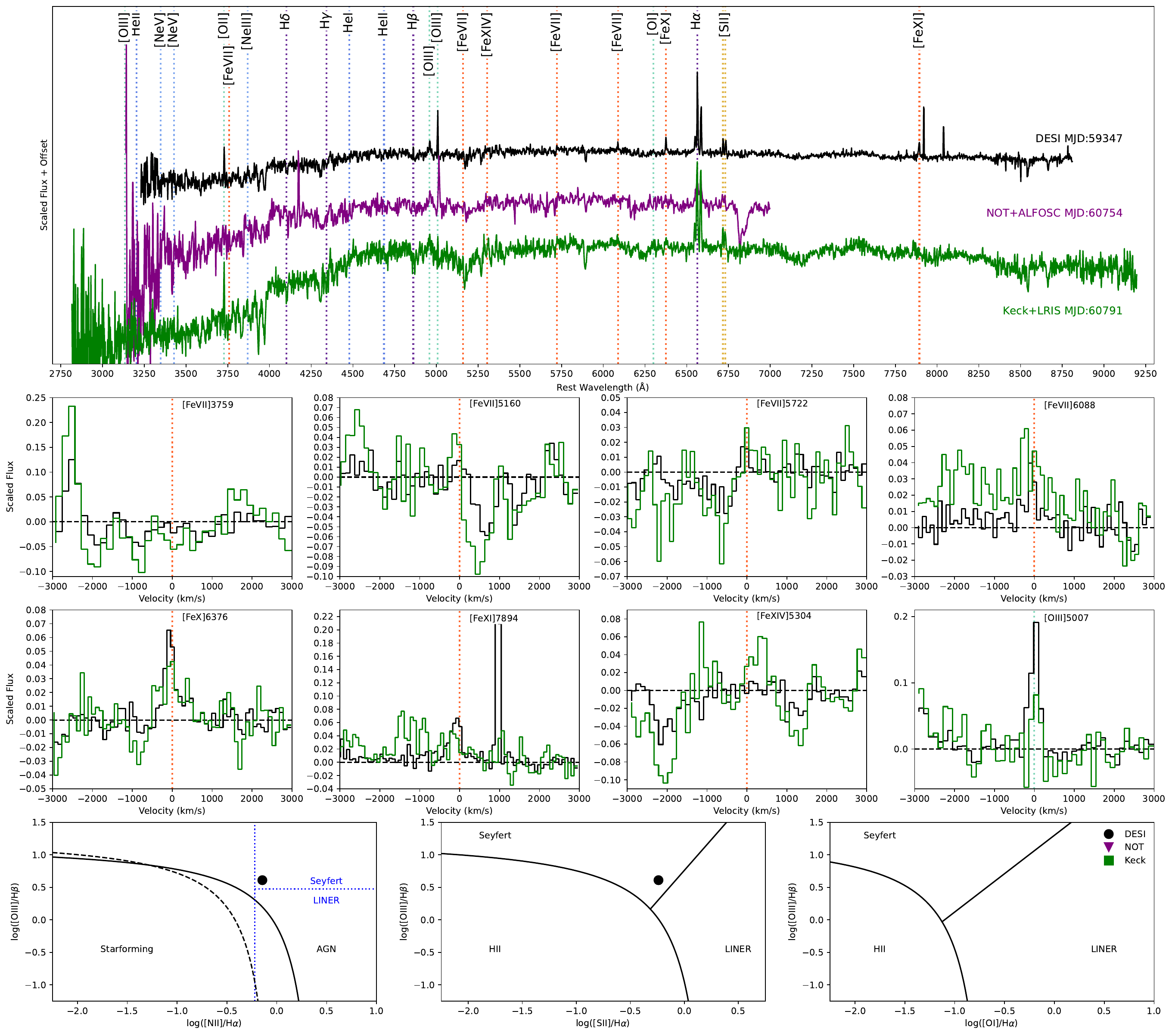}
    \caption{Spectra and BPT diagrams of Raticate. All spectra have been rebinned to 2~\AA\ bin$^{-1}$ but have not been smoothed.
    \textit{Top row:} Spectral evolution of Raticate. The NOT+ALFOSC spectrum has prominent telluric absorption near 6850~\AA. \textit{Second and third rows:} Comparison between the spectral features of most interest between the DESI (black) and Keck+LRIS (green) spectra. \textit{Bottom row:} BPT diagnostic diagrams for the DESI spectrum. The NOT+ALFSOC spectrum is not included in these subplots owing to a wavelength-calibration issue. Values for the Keck spectrum are also not included given the nondetection of H$\beta$, and $\Oi$/H$\alpha$ was not measured in the DESI spectrum because of the nondetection of \fspectralline{O}{i}{6300}.}
    \label{fig:Raticate_Spec_Evolution}
\end{figure*}

\begin{figure*}
    \centering
    \includegraphics[width=\textwidth]{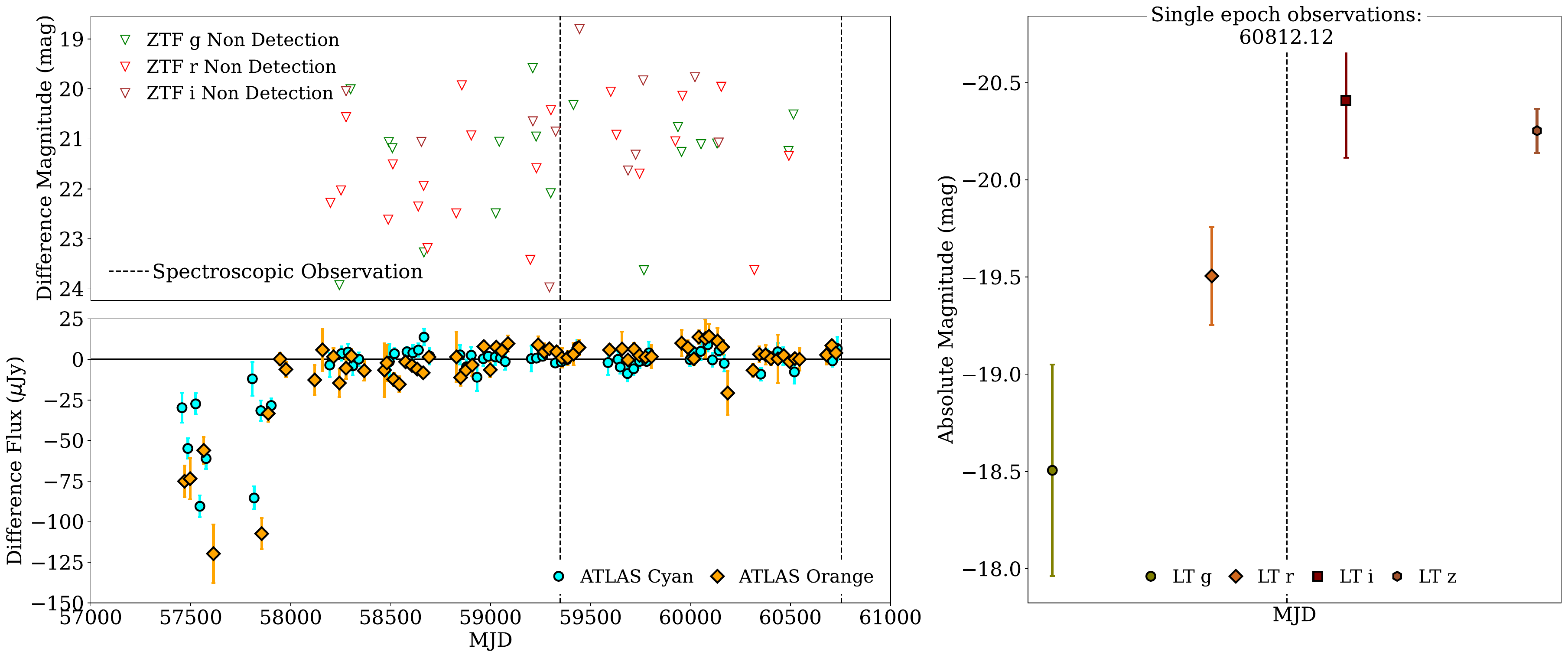}
    \caption{Photometric evolution of Raticate. \textit{Top left:} ZTF forced photometry difference magnitude light curve. \textit{Bottom left:} ATLAS forced photometry light curve presented in flux space. \textit{Right:} LT absolute magnitude light curve.}
    \label{fig:Raticate_Phot_Evolution}
\end{figure*}

Raticate was identified by \sleipnir\ as an ECLE candidate by the presence of a high-SNR \Fexi\ line, along with the detection of \Fex. Visual inspection confirmed these features, as well as the presence of weak \Fevii\ lines that did not meet the automatic pEQW detection threshold.

Sky-line contamination near \fspectralline{O}{iii}{5007} affected the automatic emission-line diagnostic determination. Inspection of this  region indicates that whilst the blue wing of the feature may be somewhat contaminated, the central core is sufficiently separated from the sky-line position in the DESI spectrum to confirm the lack of particularly strong \fspectralline{O}{iii}{5007} emission, further supported by the strength of the uncontaminated \fspectralline{O}{iii}{4959} emission. These emission-line properties disfavour the presence of a significant AGN in Raticate's host galaxy, a statement further supported by this galaxy's absence within the MILLIQUAS AGN database. However, we note that the BPT diagnostics measured from the DESI spectrum indicate Seyfert-like activity driven by the weakness of the H$\beta$ emission. Our follow-up spectra, whilst unable to provide additional BPT diagnostic measurements owing to a lack of H$\beta$ emission detection (and a slight wavelength-calibration issue with the NOT+ALFOSC spectrum), do show the complete fading of \Fex\ and \Fexi\ emission and a reduction in \Fevii, suggesting transient activity.

Photometrically, no transients were reported close to Raticate's position, and the host galaxy displayed no optical outbursts prior to the DESI observation. Pre-outburst NIR photometry from 2MASS is too uncertain to provide clear classification.

MIR photometry from AllWISE indicates Seyfert AGN-like MIR colours, though the most interesting aspect of Raticate's behaviour is seen in the NEOWISE-R dataset, where two distinct MIR outbursts are observed. The first was in its active state during the AllWISE observations (seen in both the \textit{W1} and \textit{W2} light curves and in the AGN-like \WOneminusWTwo\ colour), though the start and true peak of this outburst occurred prior to the beginning of observations. Raticate then faded between the AllWISE and NEOWISE-R surveys, reaching apparent quiescence at $\sim$~MJD~57000 with an accompanying shift to a bluer \WOneminusWTwo\ colour. This low state persisted until $\sim$~MJD~58000, with brightening and reddening then seen with a significant outburst at $\sim$~MJD~58750 reaching a luminosity peak at $\sim$~MJD~59250 and a delayed peak redness at $\sim$~MJD~59750. The MIR light curve then declined toward the previous quiescent state.

Possible explanations for this MIR behaviour include the occurrence of two distinct TDEs, each responsible for one of the outbursts; a repeating partial TDE; or an AGN repeatedly switching on and off. Additionally, whilst Raticate is not within the sky-coverage region of LoTSS-DR2, it is not associated with a radio source in either VLA-FIRST or VLASS, indicating a lack of radio-loud AGN activity. We investigate the MIR outburst behaviour and compare to other nuclear transients in Section~\ref{sec:MIR_Comp}. For the purposes of this analysis, we conclude that Raticate is the result of the occurrence of at least one TDE.

\subsubsection{DESI 39627884763023878 : Raichu}
\label{sec:Raichu}

\begin{figure*}
    \centering
    \includegraphics[width=\textwidth]{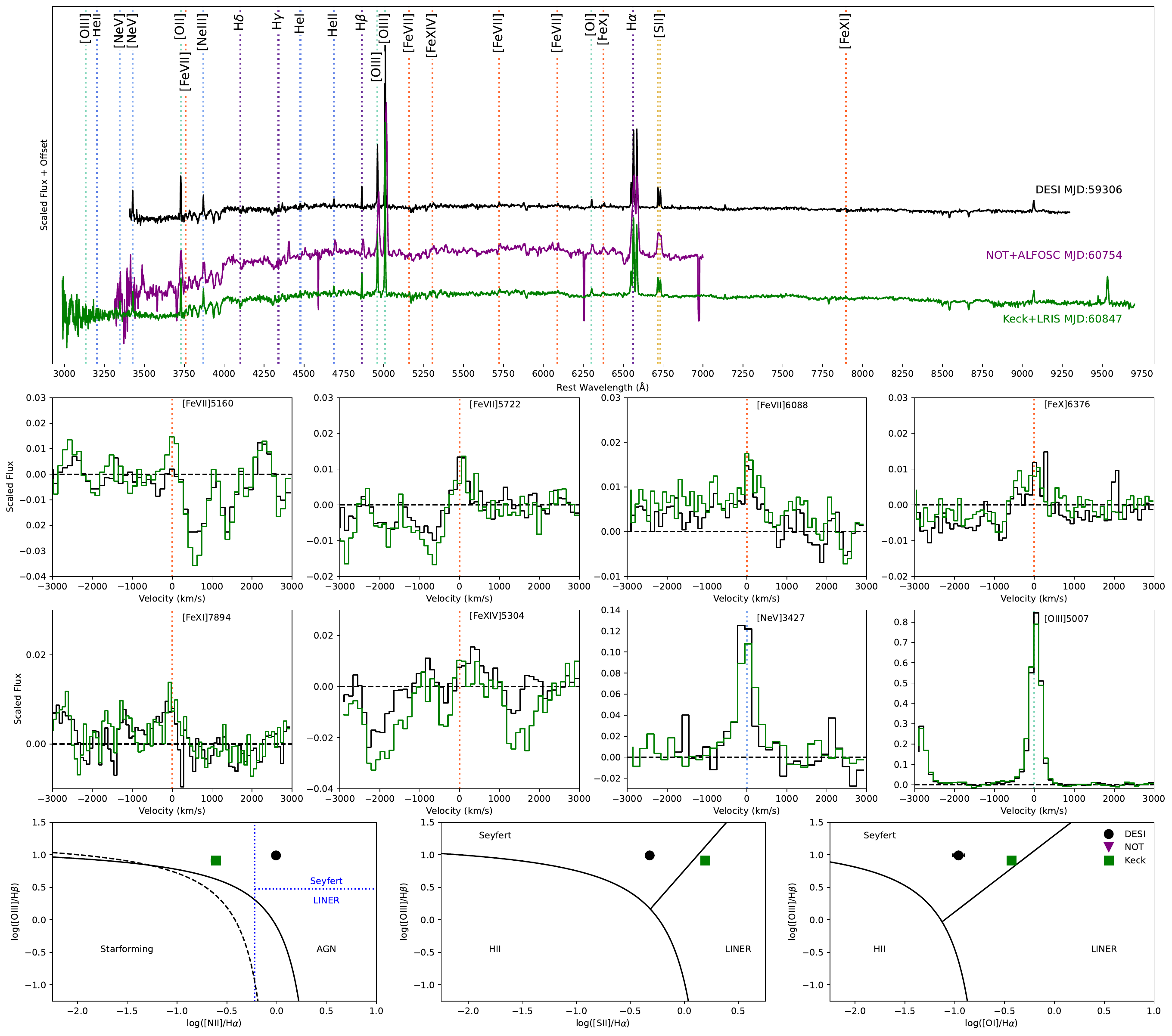}
    \caption{Spectra and BPT diagrams of Raichu. All spectra have been rebinned to 2~\AA\ bin$^{-1}$ but have not been smoothed.
    \textit{Top row:} Spectral evolution of Raichu. The sharp dips in the NOT+ALFOSC spectrum are artefacts.
    \textit{Second and third rows:} Comparison between the spectral features of most interest between the DESI (black) and Keck+LRIS (green) spectra. \textit{Bottom row:} BPT diagnostic diagrams for the DESI and Keck+LRIS spectra. The NOT+ALFSOC spectrum is not included in these subplots owing to a wavelength-calibration issue.}
    \label{fig:Raichu_Spec_Evolution}
\end{figure*}

\begin{figure*}
    \centering
    \includegraphics[width=\textwidth]{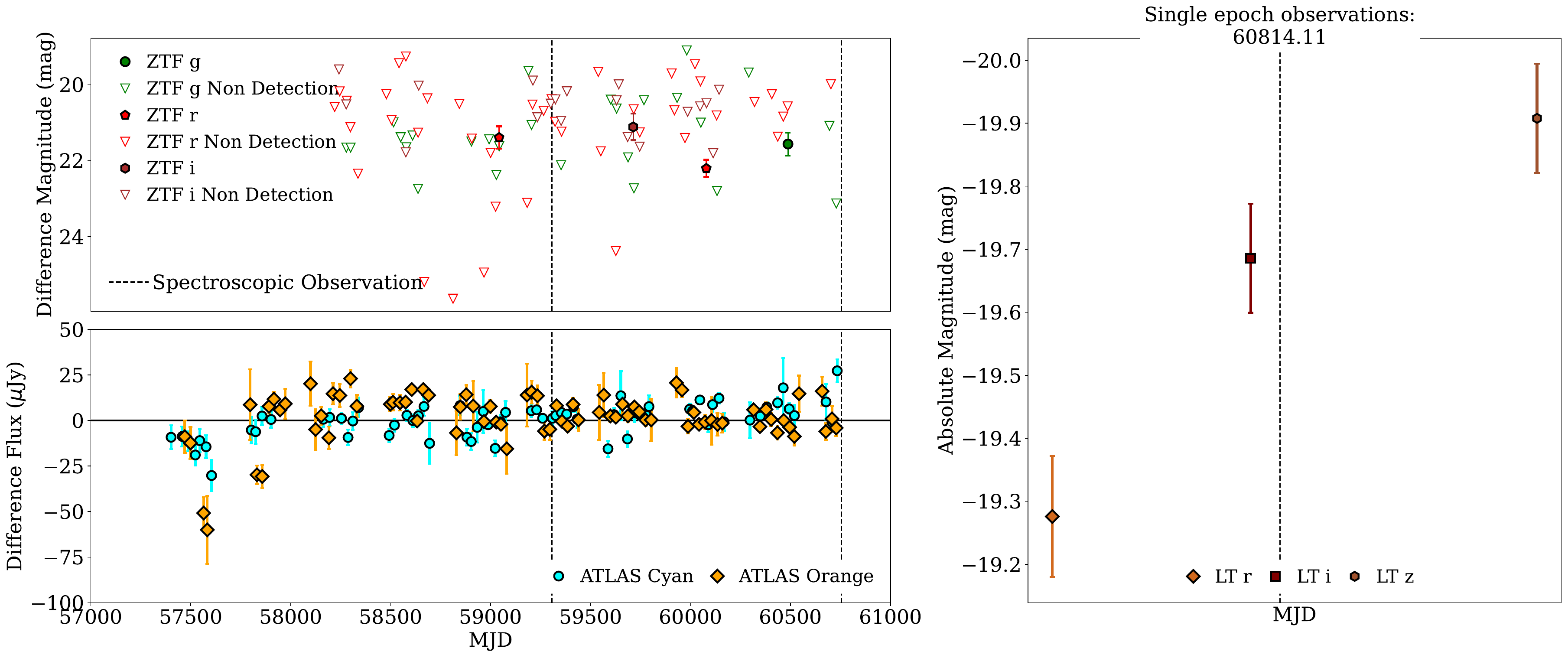}
    \caption{Photometric evolution of Raichu. \textit{Top left:} ZTF forced photometry difference magnitude light curve. \textit{Bottom left:} ATLAS forced photometry light curve presented in flux space. \textit{Right:} LT absolute magnitude light curve.}
    \label{fig:Raichu_Phot_Evolution}
\end{figure*}

Raichu was identified by \sleipnir\ as an ECLE candidate based on the weak presence of \Fevii\ lines. Visual inspection of the spectrum also shows the presence of a weak \Fex\ feature and a potential \Fexi\ feature (though this is contaminated by sky-line emission). Optical emission-line diagnostics indicate the presence of a narrow-line AGN (in particular, there are strong \Oiii\ lines). Prior NIR photometry from the 2MASS survey provides an unclear classification based on the scheme of \citet{hyland_1982_InfraredStudyQuasars}; see Fig.~\ref{fig:NIR_Phot_Diagnostic}. Optical photometry from ATLAS and ZTF shows stable fluxes over the course of observations without indication of TDE-like outbursts or AGN-like photometric variability.

Raichu is not present within the MILLIQUAS AGN catalogue, with AllWISE \WOneminusWTwo\ vs. \WTwominusWThree\ photometry obtained years prior to the DESI spectrum showing that the galaxy did not meet the AGN classification criteria of either \citet{stern_2012_MIDINFRAREDSELECTIONACTIVE} or \citet{mateos_2012_UsingBrightUltrahard}, though it occupies a somewhat ambiguous region of parameter space per the \citet{wright_2010_WIDEFIELDINFRAREDSURVEY} colour diagnostics, including both AGN and non-AGN galaxies (see Fig.~\ref{fig:AllWISE_Colours}). Raichu is not within the sky-coverage region of LoTSS-DR2 and is not associated with a catalogued source in either VLA-FIRST or VLASS. As described in Section~\ref{subsec_VLASS_FIRST_Photometry}, a possible faint source is visually observable at its location in the VLA-FIRST cutout image. This source is of similar strength to other spurious noise features in the cutout, which in combination with Raichu's nondetection in other VLASS radio observations leads us to conclude that radio-loud AGN activity is not present.

Most interestingly, Raichu displays a long-term decline in both the \textit{W1} and \textit{W2} bands, with an associated blueward trend in its \WOneminusWTwo\ colour, moving further from the red colour indices associated with AGN activity. This behaviour has been seen as one of the key trends of the SDSS TDE-ECLE sample identified by \citet{wang_2012_EXTREMECORONALLINE}, with \citet{dou_2016_LONGFADINGMIDINFRARED} and \citet{clark_2024_Longtermfollowupobservations} finding very similar behaviour observable in all the TDE-ECLEs (in particular SDSS~J1241+4426, which has also displayed long-lived -- though fading -- \Fevii\ emission). Crucially, this behaviour was not observed in the AGN-linked objects, which instead displayed repeated brightening and fading episodes. We further explore this behaviour in Section~\ref{sec:MIR_Comp}.

Additionally, strong \Oiii\ emission (and the accompanying BPT diagnostics) more typically seen in AGNs has been observed in several other TDE-ECLEs and CrL-TDEs at late times, including SDSS~J134244.41+053056.1 \citep{clark_2024_Longtermfollowupobservations} and AT~2018dyk \citep{clark_2025_2018dyktidaldisruption}. The development of these features in such objects can be explained by the delayed response of material at greater distances from the SMBH. Additionally, \citet{mummery_2025_Galaxyscaleconsequencestidal} find a clear link between the development of \Oiii\ emission features in the aftermath of a TDE and high densities of circumnuclear material, but our follow-up spectra do not show significant alterations in \Oiii\ emission strength compared to the DESI spectrum. Variation is observed around the H$\alpha$ complex, though this may be the result of the spectra probing somewhat different regions of the host galaxy (which is observed to be not spherically symmetric, see Fig.~\ref{fig:Legacy_Image_Grid}). Fe emission remains unchanged between the DESI and follow-up spectra. However, the uncertain timescales involved preclude this from being used to disfavour a TDE interpretation, with other TDE-ECLEs also known to display CrL emission for years during extended MIR declines, a behaviour in particular shared by the most similar of the previously identified TDE-ECLEs, SDSS~J1241+4426 \citep{clark_2024_Longtermfollowupobservations}. As such, for this analysis, and primarily on the basis of its MIR behaviour (which has been the most powerful follow-up classification diagnostic for these objects) and its CrL emission, we conclude that Raichu is best explained as a TDE-ECLE, where the originating TDE occurred well before the start of available optical and MIR observations. As with all objects in the limited sample of CrL transients, we will continue to monitor its further long-term evolution and will revisit its classification in the future if required.

\subsection{AGN-linked candidates}
\label{sec:AGN_Candidates}

We now describe the six candidate objects that were identified during the analysis process as potentially the result of TDEs, but were ultimately the result of AGN activity.

\subsubsection{DESI 39627700695996439 : Venusaur}
\label{sec:Venusaur}

Venusaur was identified as a potential ECLE candidate by an earlier development version of \sleipnir, and does not meet the standard selection criteria used in the ``live'' version of the code and is thus not included in the formal EDR CrL Object Sample. We discuss the object here, as owing to its early identification we obtained several follow-up datasets, including both photometry and spectra.

\begin{figure*}
    \centering
    \includegraphics[width=\textwidth]{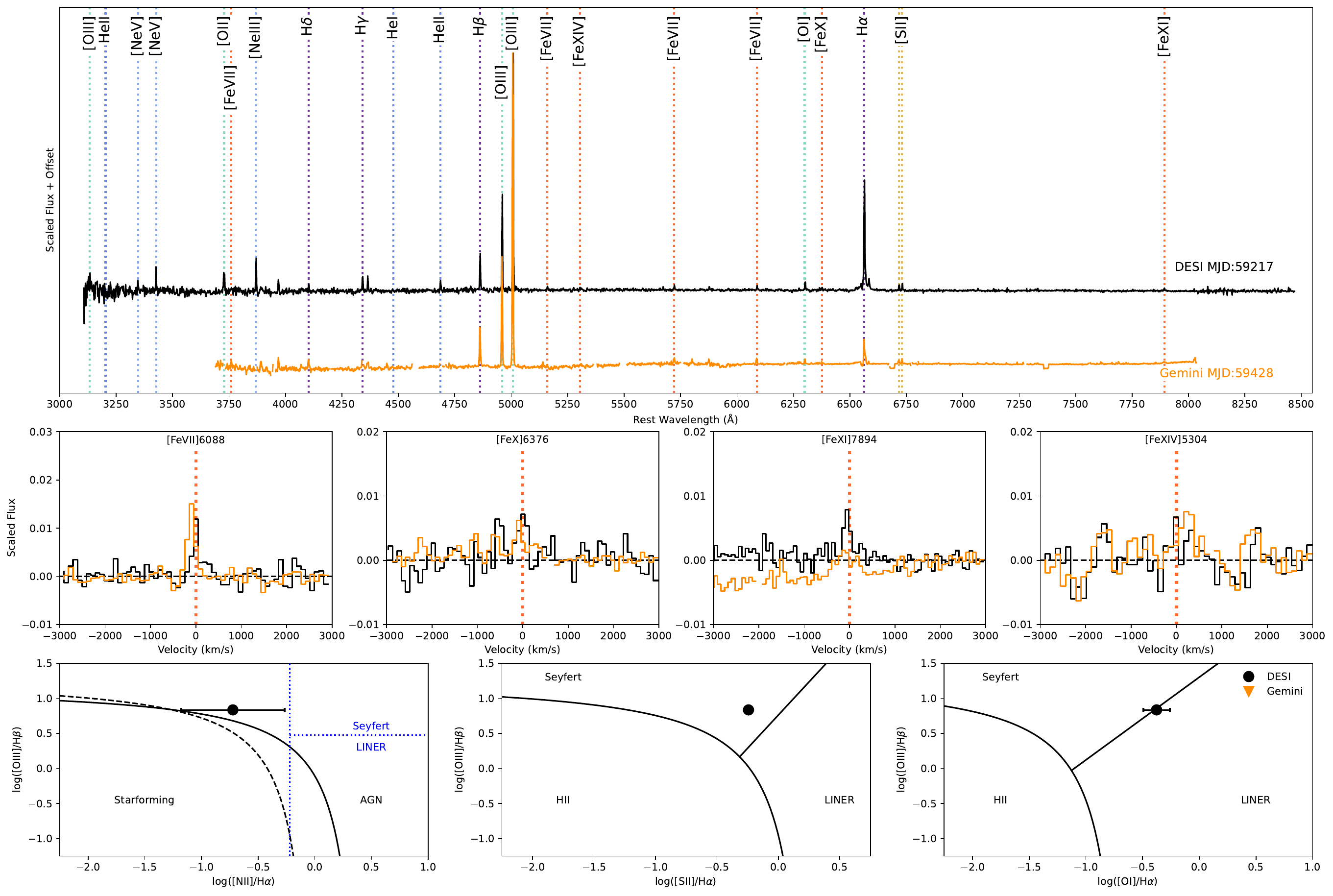}
    \caption{Spectra and BPT diagrams of Venusaur. All spectra have been rebinned to 2~\AA\ bin$^{-1}$ but have not been smoothed. \textit{Top row:} Spectra of Venusaur from both DESI and Gemini+GMOS. Prominent emission lines are labelled (coloured by element). Spectra have been normalised relative to the flux of the strongest feature and then offset from each other for ease of interpretation. \textit{Middle row:}  Comparison between the strongest Fe CrLs in velocity space, following a calibration matching each spectrum's mean flux at 2000--3000~\kms. Fe emission is unchanged between the two spectra, though not possible to confirm for \Fexi\ which suffered from an artificial offset in the Gemini+GMOS spectrum owing to issues with flux calibration. \textit{Bottom row:} BPT diagnostic diagrams for the DESI spectrum. They have not been produced for the Gemini+GMOS spectrum owing to the significant atmospheric contamination of the H$\alpha$ feature.}
    \label{fig:Venusaur_Spec_Evolution}
\end{figure*}

Venusaur was identified as a potential TDE-ECLE candidate based on the presence of \Fevii\ and \Fexiv\ emission lines (see Fig.~\ref{fig:Venusaur_Spec_Evolution}). The \fspectralline{Fe}{vii}{6088} line now fails the current line-strength requirements, with the other \Fevii\ lines are also too weak to trigger the spectrum for visual inspection through the back-up flagging criteria. 

\begin{figure*}
    \centering
    \includegraphics[width=\textwidth]{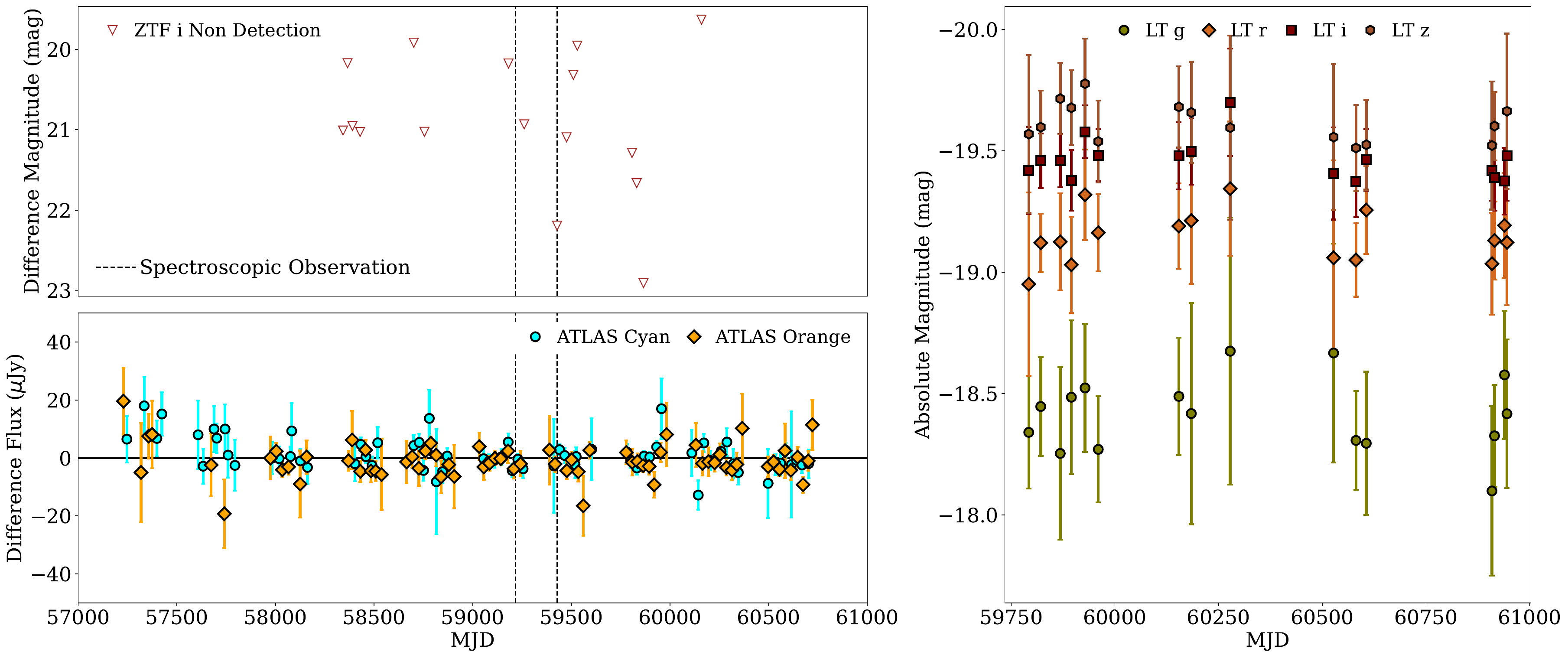}
    \caption{Photometric evolution of Venusaur. \textit{Top left:} ZTF forced photometry difference magnitude light curve. \textit{Bottom left:} ATLAS forced photometry light curve presented in flux space. \textit{Right:} LT absolute magnitude light curve.}
    \label{fig:Venusaur_Phot_Evolution}
\end{figure*}

Following its initial identification, we triggered follow-up spectroscopy  with Gemini+GMOS and optical photometry with LT+IO:O. Both sets of  observations revealed no significant changes, with the Fe CrLs well recovered in the Gemini+GMOS spectrum and remaining unchanged. Likewise, follow-up photometry with the LT (later supplemented with analysis of ATLAS and ZTF observations; Fig.~\ref{fig:Venusaur_Phot_Evolution}) has shown no evidence for outbursts or other evolution. Additionally, the MIR behaviour and optical emission-line properties (e.g., the very prominent \fspectralline{O}{iii}{5007} emission) of Venusaur strongly indicate the dominance of AGN activity, with a lack of detection in either VLA-FIRST or VLASS suggesting it is radio-quiet.

\subsubsection{DESI 39633255741260888 : Charizard}
\label{sec:Charizard}

\begin{figure*}
    \centering
    \includegraphics[width=\textwidth]{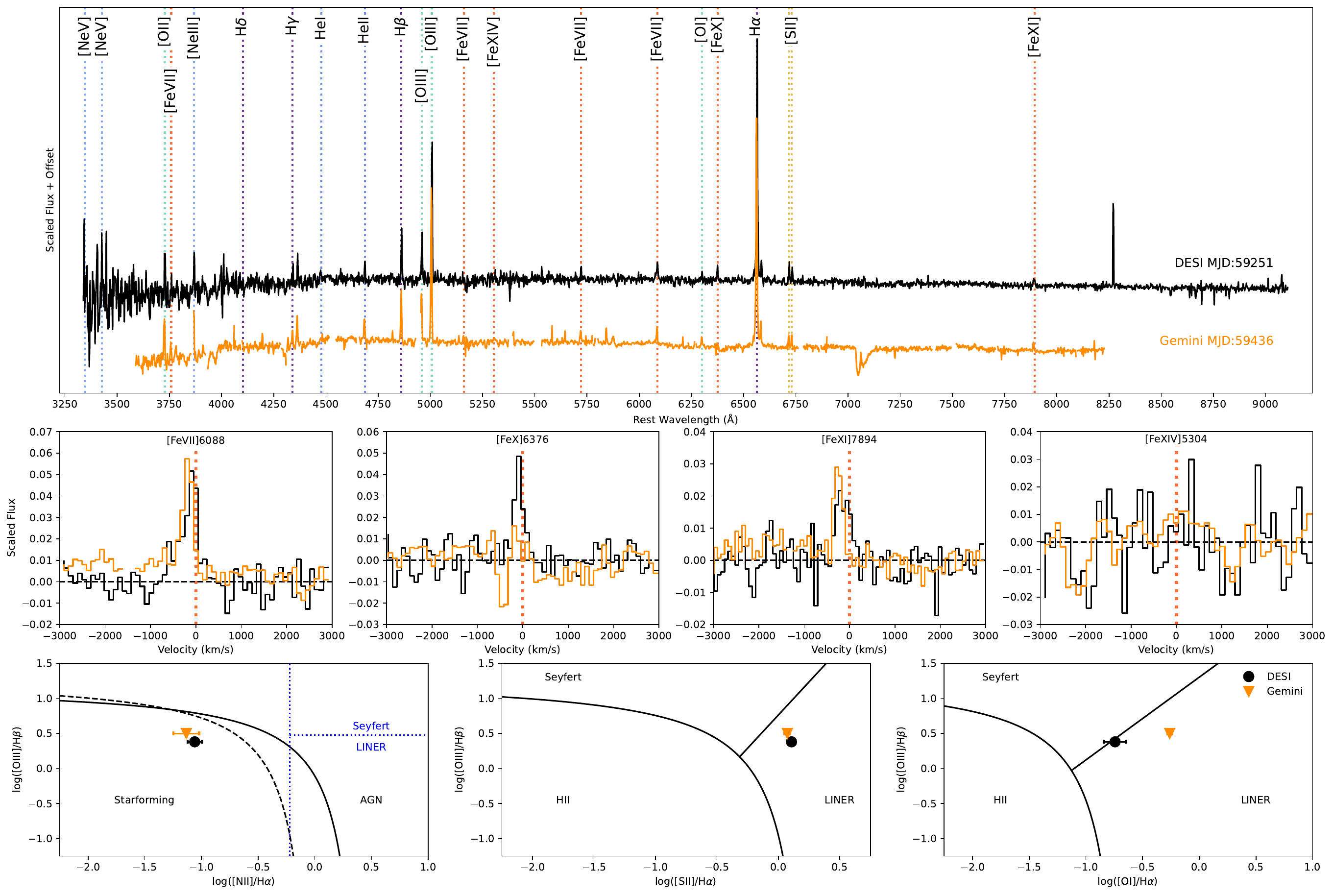}
    \caption{Spectra and BPT diagrams of Charizard. All spectra have been rebinned to 2~\AA\ bin$^{-1}$ but have not been smoothed. 
    \textit{Top row:} Spectral observations of Charizard from both DESI and Gemini+GMOS. Prominent emission lines are labelled (coloured by element). Spectra have been normalised relative to the flux of the strongest feature and then offset from each other for clarity. \textit{Middle row:} Comparison between the strongest Fe CrLs in velocity space, following a calibration matching each spectrum's mean flux at 2000--3000~\kms. Fe emission is unchanged between the two spectra, though this is not possible to confirm for \Fex\ which suffered from telluric contamination in the Gemini+GMOS spectrum. \textit{Bottom row:} BPT diagnostic diagrams for the DESI and Gemini+GMOS spectra showing broadly similar results. The difference in the \Oi\ / H$\alpha$ ratio is driven by a somewhat broader \Oi\ line observed in the Gemini spectrum.}
    \label{fig:Charizard_Spec_Evolution}
\end{figure*}

\begin{figure*}
    \centering
    \includegraphics[width=\textwidth]{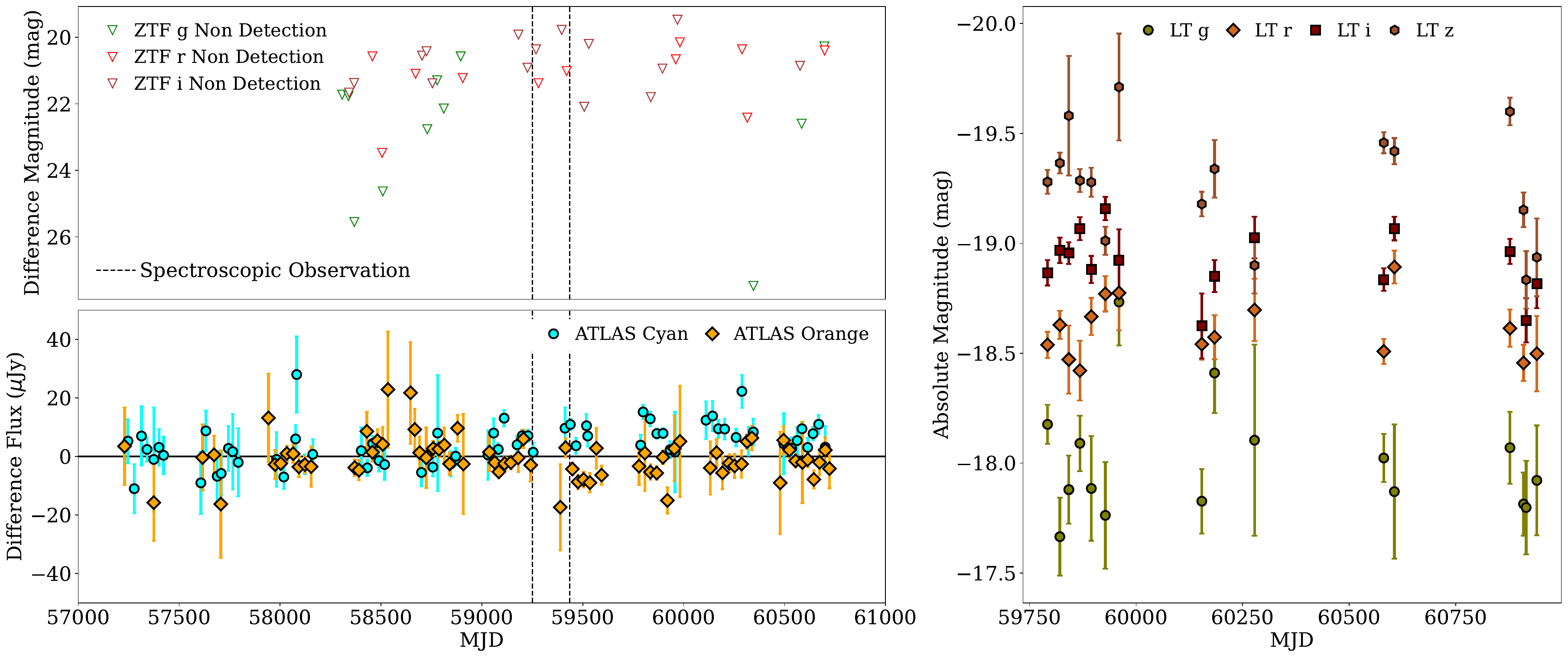}
    \caption{Photometric evolution of Charizard. \textit{Top left:} ZTF forced photometry difference magnitude light curve. \textit{Bottom left:} ATLAS forced photometry light curve presented in flux space. \textit{Right:} LT absolute magnitude light curve.}
    \label{fig:Charizard_Phot_Evolution}
\end{figure*}

Charizard was identified early in our analysis of the EDR sample by the presence of multiple clear CrLs (though lacking \Fexiv\ emission). We triggered targeted follow-up spectroscopy with Gemini+GMOS and optical photometry with LT+IO:O. The Gemini spectra revealed no change in the CrL-emission lines (see Fig.~\ref{fig:Charizard_Spec_Evolution}), but note the contamination affecting the GMOS observation of \Fex\ which makes its nondetection unreliable. Optical photometry also exhibits no signs of outburst or significant fading activity (Fig.~\ref{fig:Charizard_Phot_Evolution}). The LT observations do show some indications of repeated fading and rebrightening activity in the \textit{riz} bands, but the large uncertainties make such trends difficult to confirm. Additionally, such behaviour is not observed in ATLAS or ZTF observations.

In the MIR (Fig.~\ref{fig:Merged_MIR}), Charizard displays repeated cycles of brightening and fading by a few 0.1~mag in both the \textit{W1} and \textit{W2} bands, consistent with AGN activity and further supported by a \WOneminusWTwo\ colour above the \citet{stern_2012_MIDINFRAREDSELECTIONACTIVE} AGN activity threshold. As with Venusuar, the lack of detections in either VLA-FIRST or VLASS suggest that this activity is radio quiet.

\subsubsection{DESI 39633255741260888 : Fearow}
\label{sec:Fearow}

\begin{figure*}
    \centering
    \includegraphics[width=\textwidth]{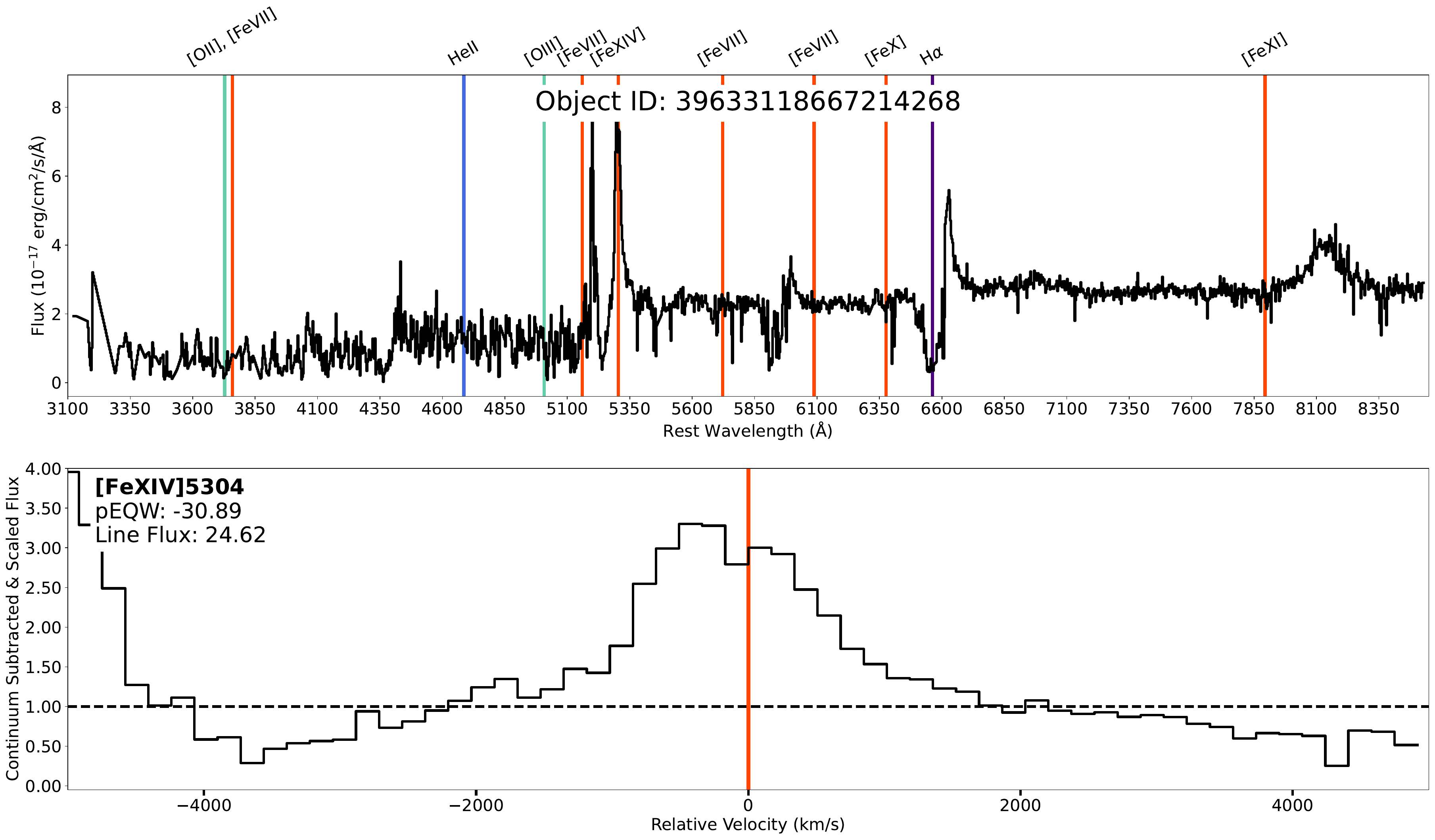}
    \caption{\textit{Top:} DESI spectrum of Fearow. At the erroneous redshift determined by \redrock\ ($z = 0.152$, compared to the correct $z = 3.928$) a strong \Nv\ feature appears as \Fexiv. Emission lines displayed are shown if the \redrock-determined redshift was accurate. In reality, these features are (in order of increasing wavelength) Ly$\alpha$, $\Nv$, $\Civ$, and a broad blueshifted \Ciii\ feature. \textit{Bottom:} Region local to the ``\Fexiv'' feature.}
    \label{fig:Fearow_Spec}
\end{figure*}

\begin{figure*}
    \centering
    \includegraphics[width=\textwidth]{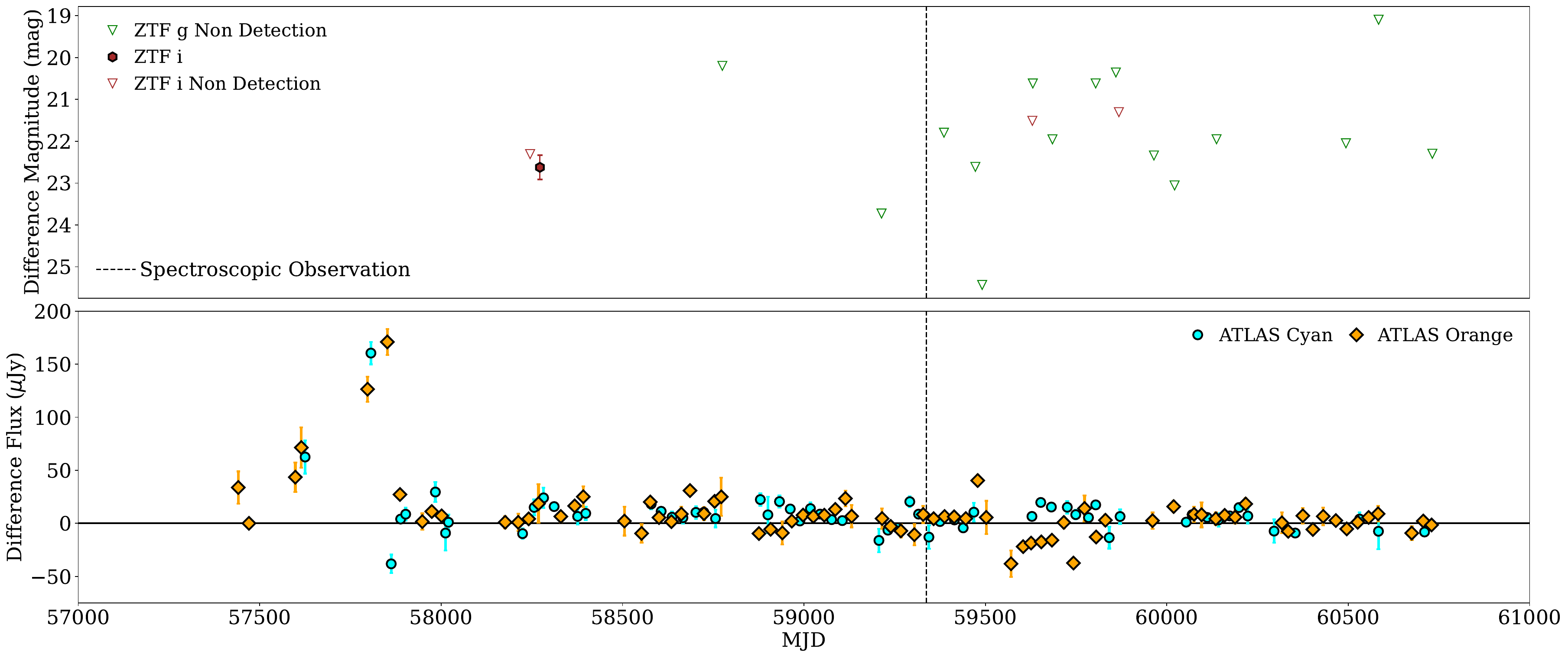}
    \caption{Photometric evolution of Fearow. \textit{Top:} ZTF forced photometry difference magnitude light curve. \textit{Bottom:} ATLAS forced photometry light curve presented in flux space.}
    \label{fig:Fearow_Phot_Evolution}
\end{figure*}

Fearow was identified by \sleipnir\ as an ECLE candidate based on the detection of an extremely strong \Fexiv\ feature, with a pEQW of $-30.89$~\AA, by far the strongest CrL feature detected in our search.

Visual inspection revealed an unusual spectrum with multiple strong emission features. Fearow lacks both 2MASS NIR and \textit{WISE} MIR detections in any band. Additional optical photometry from ZTF shows no sign of outburst behaviour over the course of the survey. ATLAS observations display a large outburst, ($\sim$ MJD~58000). However, following an inspection of the difference imaging, we attribute this to issues with the image subtraction at these epochs rather than true transient activity.

Following a reexamination of the spectrum, we conclude that Fearow is a broad-line QSO that experienced a redshift-measurement failure during processing by \textsc{REDROCK}. We estimate that the true redshift of the galaxy is $z = 3.928$, with the detected ``\Fexiv'' feature in reality produced by \Nv. The strong emission features displayed in the spectrum are thus (in order of increasing wavelength) Ly$\alpha$, \Nv, \Civ, and a broad and blueshifted \Ciii\ feature. As the detection of \Fexiv\ has been determined to be a false positive, Fearow is not included in the EDR CrL Object Sample.

\subsubsection{DESI 39633066745924399 : Arbok}
\label{sec:Arbok}

\begin{figure*}
    \centering
    \includegraphics[width=\textwidth]{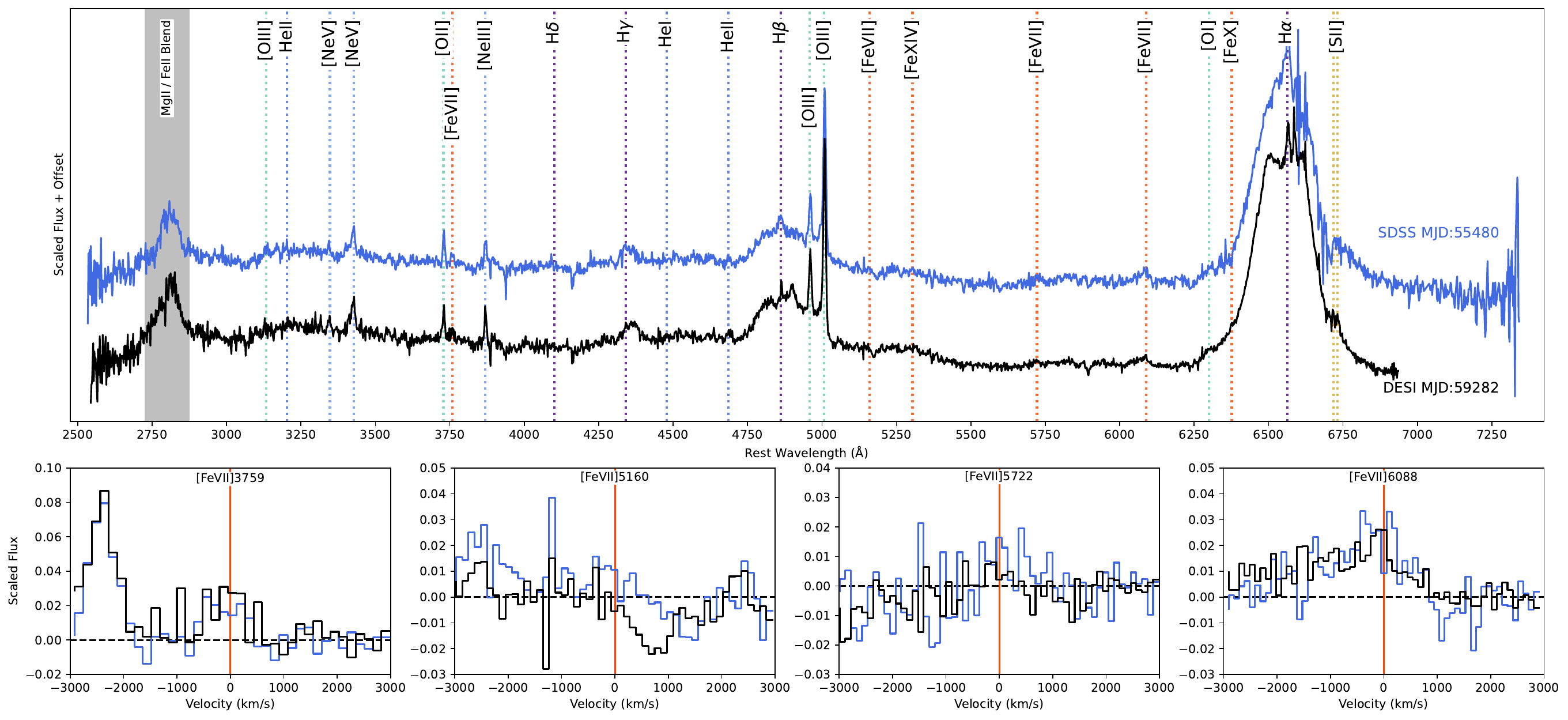}
    \caption{Spectra and BPT diagrams of Arbok. All spectra have been rebinned to 2~\AA\ bin$^{-1}$ but have not been smoothed. 
    \textit{Top row:} Spectra of Arbok from both SDSS-BOSS and DESI. Prominent emission lines are labelled (coloured by element). Spectra have been normalised relative to the flux of the strongest feature and then offset from each other for clarity. \textit{Bottom row:} Comparison between the spectral lines of most interest in velocity space, following a calibration matching each spectrum's mean flux at 2000--3000~\kms. \Fevii\ emission, whilst weak, is present largely unchanged in both spectra. Owing to the significant broad H features present, we do not attempt line deblending to produce BPT diagrams for Arbok.}
    \label{fig:Arbok_Spec_Comp}
\end{figure*}

\begin{figure*}
    \centering
    \includegraphics[width=\textwidth]{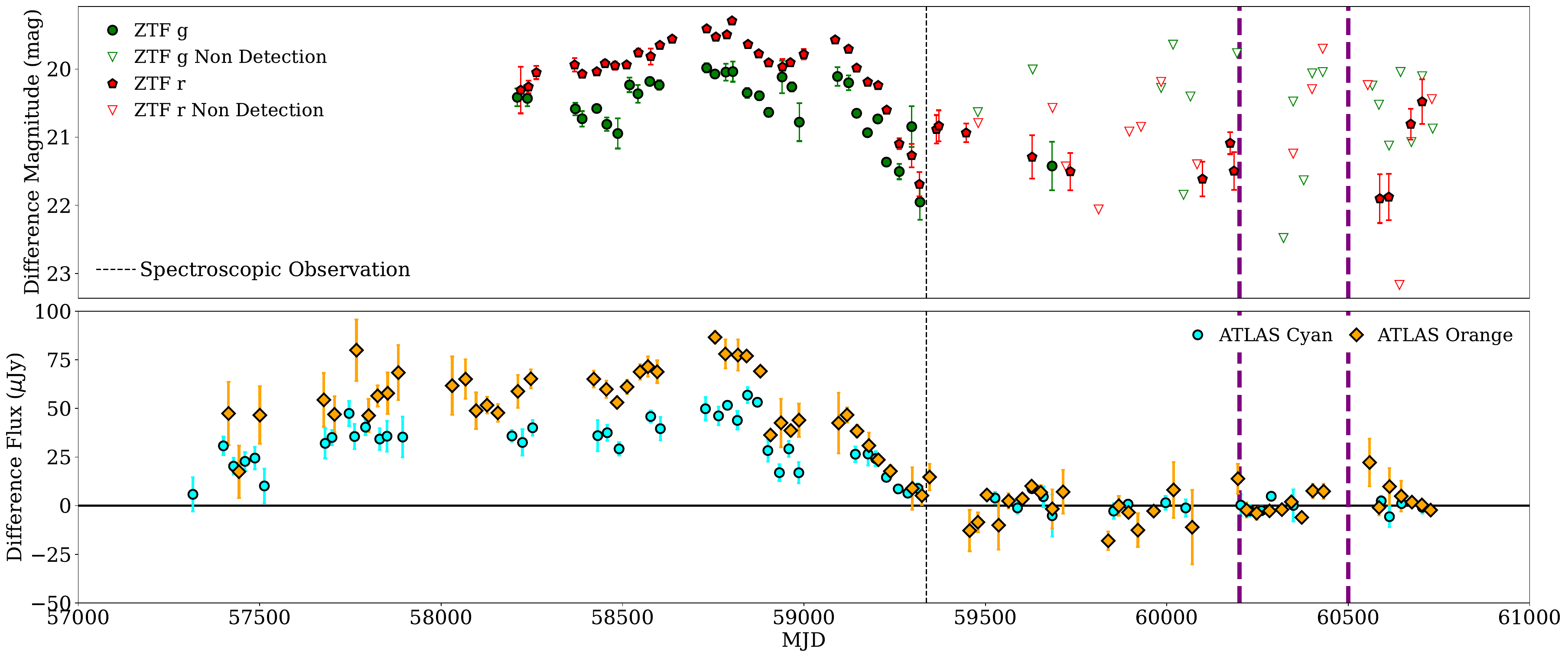}
    \caption{Photometric evolution of Arbok. Observational time frame used to rereference the forced photometry is shown within the thick dashed purple lines; selected to use the object's ``low'' state as the relative baseline. 
    \textit{Top:} ZTF forced photometry difference magnitude light curve. \textit{Bottom:} ATLAS forced photometry light curve presented in flux space.}
    \label{fig:Arbok_Phot_Evolution}
\end{figure*}

Arbok was identified by \sleipnir\ as a potential ECLE based on the presence of weak but consistent \Fevii\ emission lines. A visual inspection of the spectrum confirmed this automated assessment, whilst also revealing the presence of very broad H features.

Arbok was also spectroscopically observed by the SDSS BOSS (MJD~55480 i.e., $\sim$~10~yr prior to the DESI spectrum). The SDSS automatic analysis pipeline incorrectly identified the broad H$\alpha$ feature as Ly$\alpha$ emission, resulting in an erroneous redshift determination of $z=6.6$, which excluded the spectrum from our previous BOSS ECLE search \citep{callow_2025_rateextremecoronal}. A comparison between the two spectra, including detailed comparisons of the \Fevii\ emission lines, is shown in Fig.~\ref{fig:Arbok_Spec_Comp}. 
We note that the SDSS Legacy and DESI spectra were obtained with fibers of differing diameter ($2''$ and $1.5''$, respectively). As such, the two spectra do not explore identical regions of the galaxy despite both being centred on the nucleus. We have previously explored the potential effect this aperture-size difference may have on the interpretation of ECLEs through our analysis of synthetic aperture spectra derived from a MaNGA IFU spectral dataset of the host of the CrL-TDE AT~2018dyk and found it to be negligible \citep{clark_2025_2018dyktidaldisruption}. While we do not have corresponding MaNGA data of Arbok for a similar level of comparison, we find it unlikely that the aperture-size difference would significantly influence the overall spectral properties observed. We do, however, note that the redshift of Arbok is $\sim$~0.4 compared to the $\sim$~0.04 redshift of AT~2018dyk, which means the physical size differences between the apertures are larger in this case.

The spectral comparison reveals the optical spectrum of Arbok to be largely unchanged over the intervening 10.5~yr. The weak \Fevii\ coronal line emission was already present at the time of the SDSS spectrum, in addition to prominent \Neiii\ and \Nev\ lines. Variability in the broad H$\alpha$ and H$\beta$ complexes is observed with changes in the line profiles, with the features being relatively stronger in the DESI spectrum. Despite this, their presence in the SDSS spectrum confirms that they are not related to the occurrence of a transient outburst.

The optical photometry from ATLAS and ZTF shows two clear states in the object's luminosity: a ``high'' state from the start of optical observations until $\sim$~MJD~59000, followed by a decline across all optical bands until a ``low'' state is reached at $\sim$~MJD~59400. During this state change, a colour change is also observed, with the colour in the high state (maximal \textit{g}$-$\textit{r} $\approx 1$) being significantly redder than in the low state (\textit{g}$-$\textit{r} $\approx 0$). The DESI spectrum obtained on MJD~59282 lies close to the end of this state change. No associated radio sources are observed in either VLA-FIRST or VLASS. The lack of long-term spectral variability of the coronal lines, optical emission-line diagnostics, and both NIR and MIR photometric diagnostics all indicate that Arbok is driven by AGN activity.

\subsubsection{DESI 39633255741260888 : Sandslash}
\label{sec:Sandslash}

\begin{figure*}
    \centering
    \includegraphics[width=\textwidth]{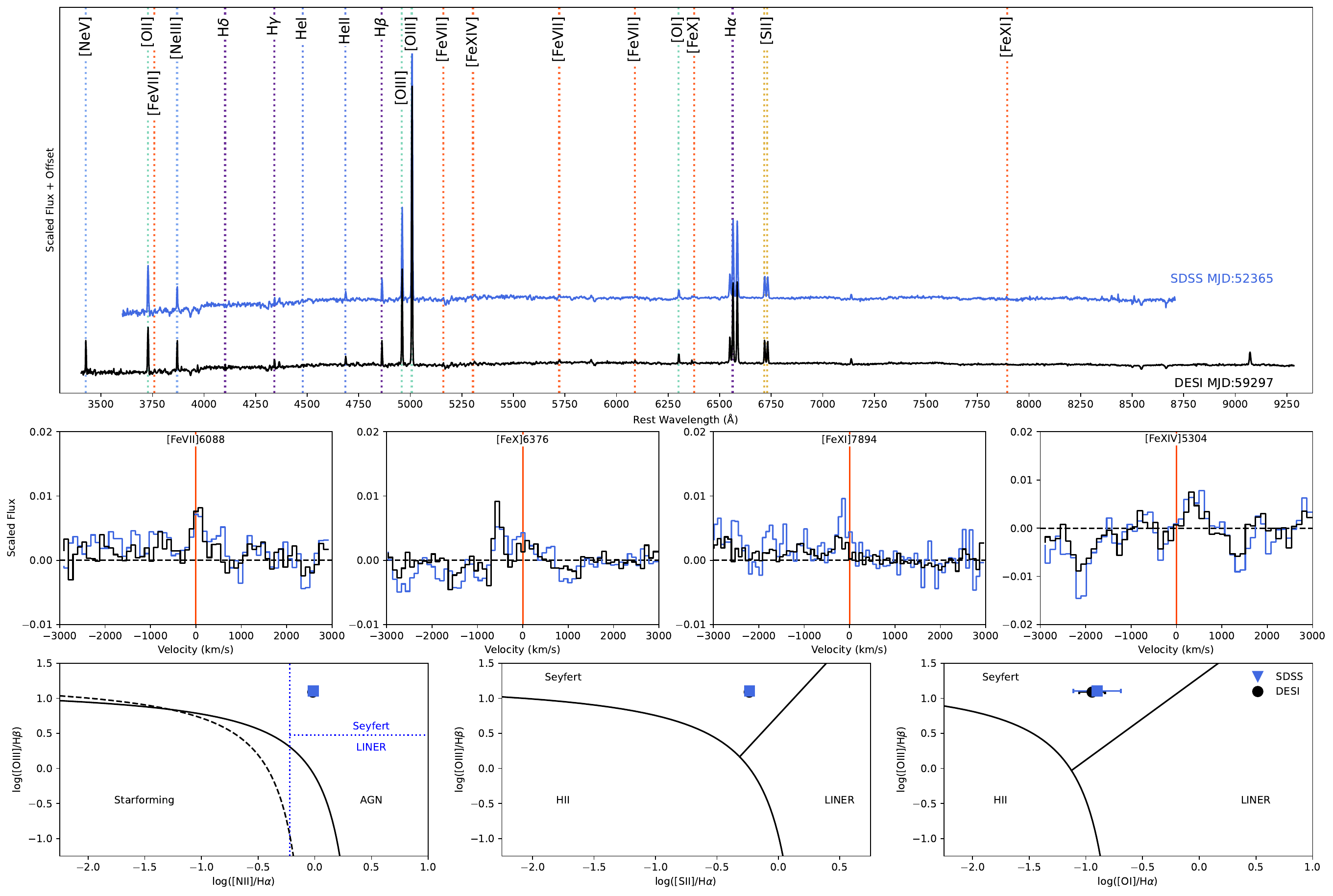}
    \caption{Spectra and BPT diagrams of Sandslash. All spectra have been rebinned to 2~\AA\ bin$^{-1}$ but have not been smoothed.
    \textit{Top row:} Spectra of Sandslash from both SDSS-Legacy and DESI. Prominent emission lines are labelled (coloured by element). Spectra have been normalised relative to the flux of the strongest feature and then offset from each other for clarity. \textit{Middle row:} Comparison between the spectra lines of most interest in velocity space, following a calibration matching each spectrum's mean flux at 2000--3000~\kms. \Fevii\ and \Fex\ emission are close matches despite the $\sim$~19~yr between the two spectra. Whilst still present in the DESI spectrum, \Fexi\ is somewhat reduced, indicating that some variability in the Fe CrLs displayed by AGNs may be expected over long timescales. \textit{Bottom row:} BPT diagnostic diagrams for the SDSS and DESI spectra.}
    \label{fig:Sandslash_Spec_Comparison}
\end{figure*}

\begin{figure*}
    \centering
    \includegraphics[width=\textwidth]{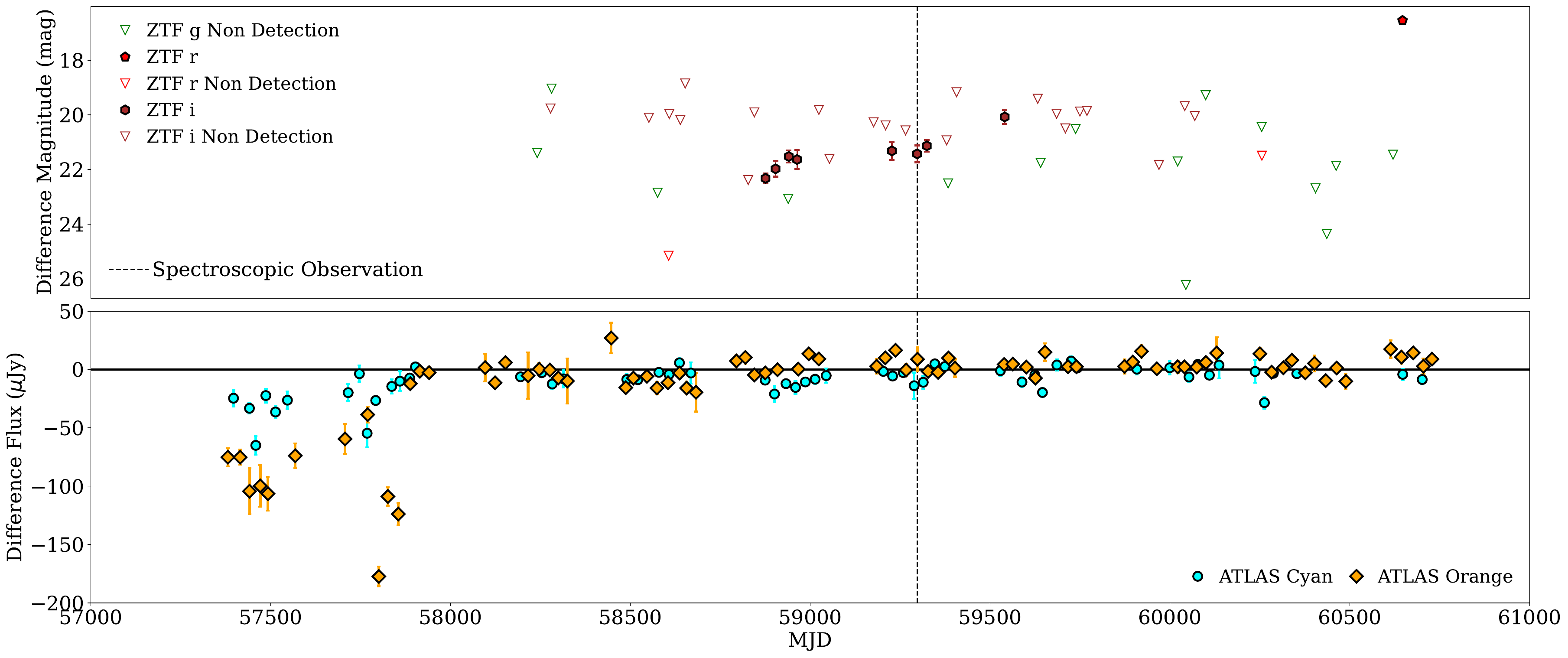}
    \caption{Photometric evolution of Sandslash. \textit{Top:} ZTF forced photometry difference magnitude light curve. \textit{Bottom:} ATLAS forced photometry light curve presented in flux space.}
    \label{fig:Sandslash_Phot_Evolution}
\end{figure*}

Like Arbok, Sandslash was identified by \sleipnir\ as a potential ECLE through the presence of weak but clearly detected \Fevii\ emission lines. Visual inspection confirmed these \Fevii\ lines, along with additional weak (below the pEQW detection thresholds of \sleipnir) \Fex, \Fexi, and \Fexiv\ emission.

Also clearly visible from a cursory inspection of the spectrum are strong emission signatures associated with AGN activity, such as \Oiii\ and \Sii. All three of the usual BPT diagnostics are well within the parameter space occupied by Seyfert AGNs.

Sandslash also has an archival SDSS-Legacy spectrum obtained on MJD~52365 (i.e., $\sim$~19~yr prior to the DESI spectrum; see Fig.~\ref{fig:Sandslash_Spec_Comparison}). The spectral comparison reveals the observed CrLs were also present at the time of the SDSS spectrum, with the SDSS and DESI spectra very similar overall, though with the DESI spectrum having a slightly bluer continuum. The weakness of the overall features is the primary reason why the object was not identified in our previous SDSS-based search. We note that the \Fexi\ emission is stronger in the SDSS spectrum, indicating that some variability in the Fe CrLs displayed by AGNs may be expected when explored over sufficiently long timescales.

Despite this spectral similarity, Sandslash was selected for further analysis as its MIR behaviour and colour are more typical of a quiescent galaxy rather than an AGN. AllWISE \WOneminusWTwo\ vs. \WTwominusWThree\ colour information provides an unclear classification, with Sandslash occupying a parameter region consistent with a range of galaxy types (spirals, LIRGs, ULIRGs, LINERs, and starburst galaxies). More recent NEOWISE \WOneminusWTwo\ photometry suggests a lack of strong AGN activity, with a stable value of $\sim$~0.24~mag, significantly below the AGN identification threshold of 0.8~mag developed by \citet{stern_2012_MIDINFRAREDSELECTIONACTIVE}. We remind the reader that this threshold is not a fully complete demarcation between AGNs and non-AGNs, with \citet{stern_2012_MIDINFRAREDSELECTIONACTIVE} indicating an AGN completeness of 73~per cent for objects above their threshold. Sandslash additionally lies outside the \citet{mateos_2012_UsingBrightUltrahard} AGN wedge (unsurprising, given the similar \WOneminusWTwo\ threshold), which becomes less sensitive at lower AGN luminosities. Sandslash's host galaxy also has a somewhat disturbed morphology (Fig.~\ref{fig:Legacy_Image_Grid}) that could indicate possible interaction or mergers that may affect the MIR AGN diagnostic. Furthermore, Sandslash does not display either an \WOneminusWTwo\ outburst or long-term decline expected of a TDE-ECLE. Sandslash is not associated with a radio source in either VLA-FIRST or VLASS, but is coincident with an isolated and clearly detected radio source in LoTSS-DR2 (ILTJ113040.22+503704.7), further supporting its classification as being AGN-related.

Based on the lack of large spectroscopic or photometric evolution and radio detection, we classify Sandslash as the result of CrL-AGN activity (albeit one with unusual MIR colours) rather than being TDE-related. 

\subsubsection{DESI 39633290029695741 : Nidoqueen}
\label{sec:Nidoqueen}

\begin{figure*}
    \centering
    \includegraphics[width=\textwidth]{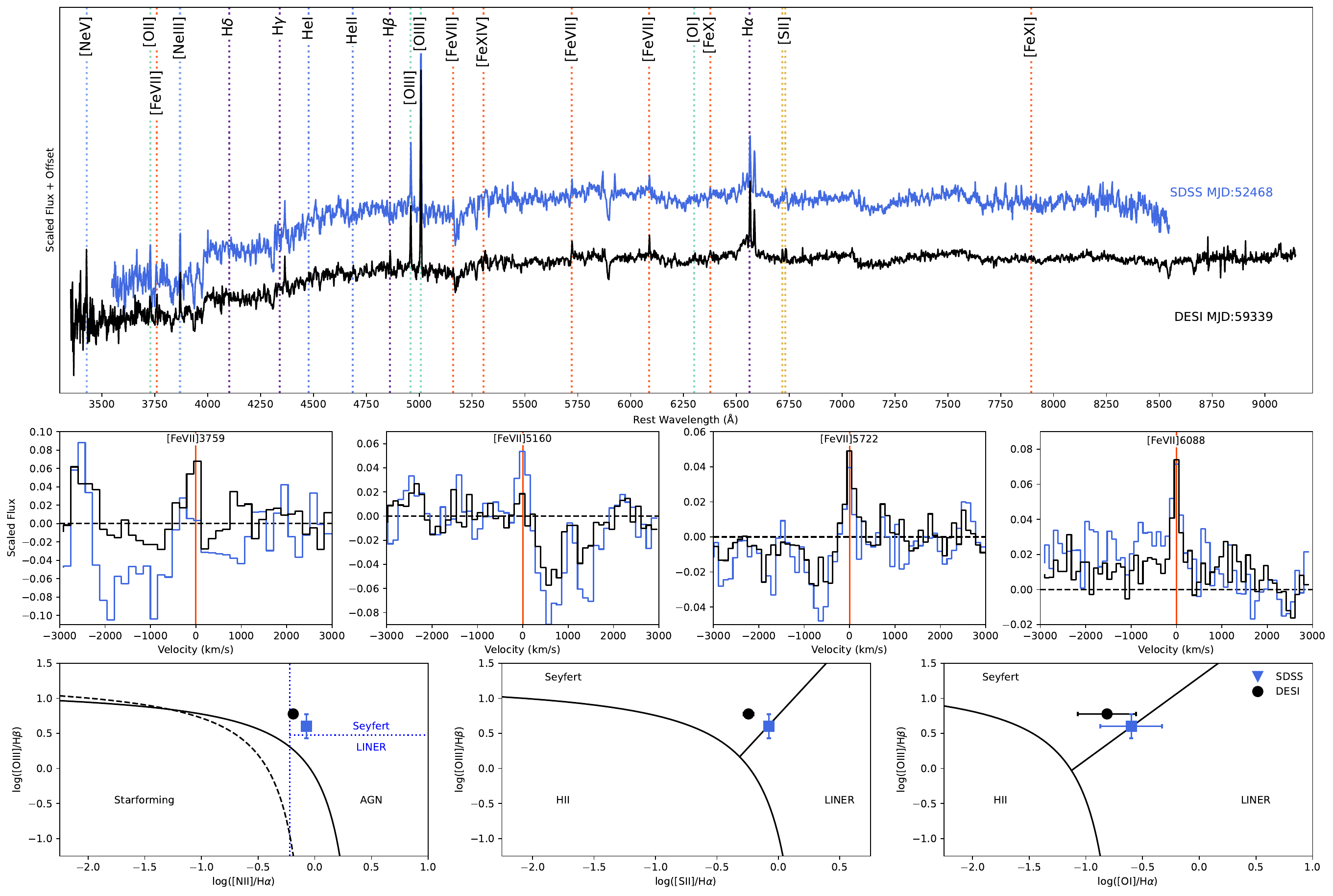}
    \caption{Spectra and BPT diagrams of Nidoqueen. All spectra have been rebinned to 2~\AA\ bin$^{-1}$ but have not been smoothed.
    \textit{Top row:} Spectra of Nidoqueen from both SDSS-Legacy and DESI. Prominent emission lines are labelled (coloured by element). Spectra have been normalised relative to the flux of the strongest feature and then offset from each other for clarity. 
    \textit{Middle row:} Comparison between the spectral lines of most interest in velocity space, following a calibration matching each spectrum's mean flux at 2000--3000~\kms. \Fevii\ emission is present in both spectra showing little evolution despite the $\sim$~19~yr between the two spectra. \textit{Bottom row:} BPT diagnostic diagrams for the SDSS and DESI spectra.}
    \label{fig:Nidoqueen_Spec_Comparison}
\end{figure*}

\begin{figure*}
    \centering
    \includegraphics[width=\textwidth]{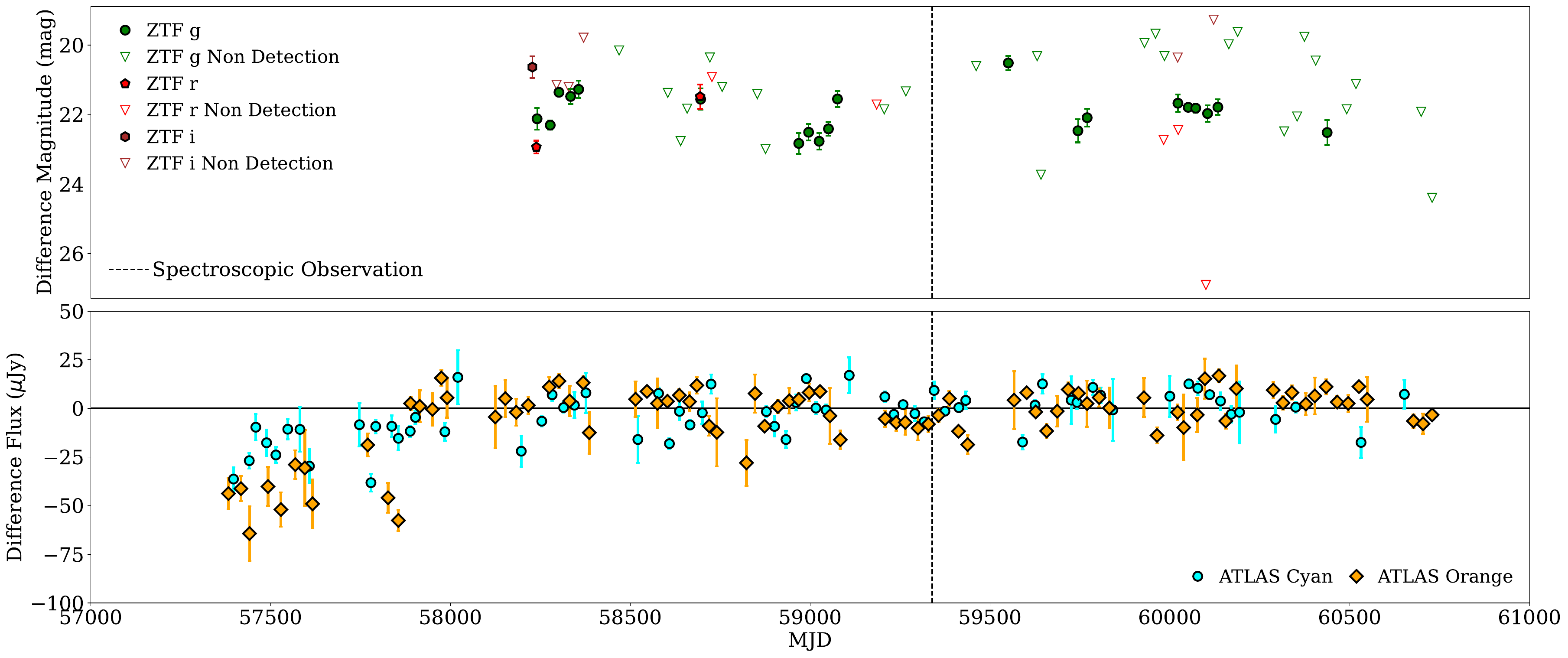}
    \caption{Photometric evolution of Nidoqueen. \textit{Top:} ZTF forced photometry difference magnitude light curve. \textit{Bottom:} ATLAS forced photometry light curve presented in flux space.}
    \label{fig:Nidoqueen_Phot_Evolution}
\end{figure*}

Nidoqueen was again selected by \sleipnir\ for the presence of weak but clearly detected \Fevii\ emission lines, which were confirmed by visual inspection.

As with Arbok and Sandslash, Nidoqueen was spectroscopically observed by the SDSS-Legacy on MJD~52468 (19~yr prior to the DESI observation) and the \Fevii\ emission is seen to be persistent, accompanied by little overall spectral variability other than a somewhat bluer continuum in the DESI spectrum. Additionally, the wider wavelength coverage of the DESI spectrum allows for the detection of \Nev. Emission-line diagnostics indicate the presence of a narrow-line AGN, given the high strength of the \Oiii\ emission. ATLAS and ZTF photometry shows no signs of significant variability across the time period of observations. 

As with Sandslash, Nidoqueen was further analysed owing to its MIR colour being atypical of an AGN, showing a \WOneminusWTwo\ colour of $\sim$~0.3~mag at all epochs. Additionally, its NIR colours were not suggestive of the presence of an AGN (see Fig.~\ref{fig:NIR_Phot_Diagnostic}). Similar to Sandslash, Nidoqueen's host galaxy has a somewhat disturbed morphology in the DESI Legacy survey imaging (Fig.~\ref{fig:Legacy_Image_Grid}), which may be at least partly responsible for its unusual NIR and MIR colours through mergers or interaction. Whilst Nidoqueen is not within the sky-coverage region of LoTSS-DR2, it is not associated with a radio source in either VLA-FIRST or VLASS, suggesting its AGN activity is radio-quiet.

As with Sandslash, based on the lack of spectroscopic or photometric evolution, we classify Nidoqueen as the result of CrL-AGN activity (albeit from one with unusual MIR colours) rather than being linked to a TDE. 

\section{Analysis}
\label{sec:Analysis}

\subsection{MIR behaviour comparison}
\label{sec:MIR_Comp}

\begin{figure*}
    \centering
    \includegraphics[width=0.95\textwidth]{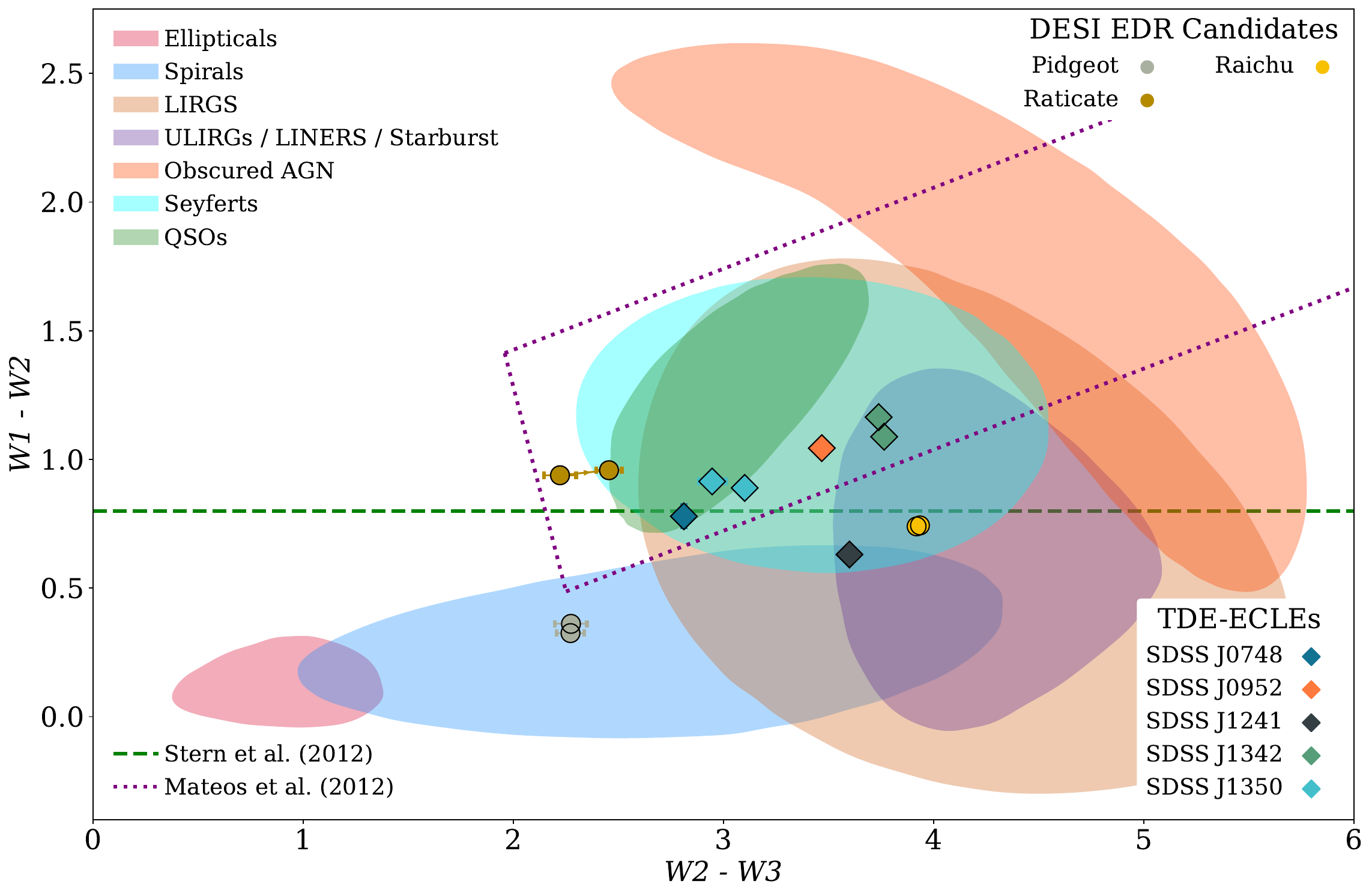}
    \caption{AllWISE colour-colour plot showing the DESI EDR TDE-ECLE candidates (circles) in comparison with the \citet{wang_2012_EXTREMECORONALLINE} sample of TDE-ECLEs (diamonds). Regions have been sourced from \protect\citet{wright_2010_WIDEFIELDINFRAREDSURVEY}. The AGN identification cuts from \citet{stern_2012_MIDINFRAREDSELECTIONACTIVE} and \citet{mateos_2012_UsingBrightUltrahard} are included as a green dashed line and purple dotted lines, respectively. Uncertainties in both axes are included but are generally smaller than the points.}
    \label{fig:AllWISE_Colours}
\end{figure*}

We compare the MIR outburst behaviour of the three candidates we identify as being most likely TDE-related to the behaviour of the CrL object sample described by \citet{clark_2025_2018dyktidaldisruption}. First, we compare the AllWISE \WOneminusWTwo\ vs. \WTwominusWThree\ colours of the candidates to those of the \citet{wang_2012_EXTREMECORONALLINE} sample, as shown in Fig.~\ref{fig:AllWISE_Colours}.

During the time of the AllWISE observations, Pidgeot was in a quiescent state and displayed the MIR colours of a star-forming galaxy without any significant AGN activity (noting the contamination from the star in close observational proximity, as previously described). Raichu occupied a similar location in this parameter space as SDSS~J1241+4426, which also showed a slow consistent decline in both the \textit{W1} and \textit{W2} bands and an increasingly blue colour. Raticate was in an active/outburst state during the AllWISE observations, displaying MIR colours more typical of AGNs, and is found in both the \citet{stern_2012_MIDINFRAREDSELECTIONACTIVE} and \citet{mateos_2012_UsingBrightUltrahard} AGN regions, similar to the TDE-ECLEs in the  \citet{wang_2012_EXTREMECORONALLINE} sample.

As in \citet{clark_2025_2018dyktidaldisruption}, we measure and compare the maximum difference between outburst peak and pre-outburst quiescence ($\Delta$ values) for both MIR bands and MIR colour for the DESI EDR TDE-ECLEs (Table~\ref{tab:Delta_Values} and Fig.~\ref{fig:MIR_Delta_Fig}).

\begin{table}
\centering
\caption{Peak changes in absolute magnitude and colour of the MIR outbursts displayed by the DESI EDR TDE-ECLEs.}
\label{tab:Delta_Values}
\begin{tabular}{lccc}
\hline
\textbf{Object} & \textbf{$\Delta$\textit{W1}} & \textbf{$\Delta$\textit{W2}} & \textbf{$\Delta$(\textit{W1$-$W2})} \\ \hline
Pidgeot $^{1}$&  $-0.66$~$\pm$~0.01  &  $-0.88$~$\pm$~0.01 &  0.30~$\pm$~0.01 \\
Raticate $^{2}$&  $-1.32$~$\pm$~0.02  &  $-1.66$~$\pm$~0.02 &  0.49~$\pm$~0.04 \\
Raichu&  <~-0.12  &  $< -0.25$ & $< 0.15$ \\
\hline
\end{tabular}
\begin{flushleft}
\textit{Notes:} In all cases, values for each band and the overall observed peak colour change are determined independently and do not necessarily occur at the same phase.\\
$^{1}$ Pidgeot displays a long-term decline in both \textit{W1} and \textit{W2} bands apparently independent of its outbursts. These measured $\Delta$ values are given following the removal of this long-term decline using a linear fit.\\
$^{2}$ Raticate displays two outbursts in \textit{WISE} data. Measured $\Delta$ values here reflect the second outburst for which the rise, peak, and decline have all been observed. The quiescent period used to set the baseline for these measurements is MJD 57048--58304. \\
\end{flushleft}
\end{table}
\raggedbottom

The $\Delta$ values for Raichu were measured in the same manner as the TDE-ECLEs explored by \citet{clark_2025_2018dyktidaldisruption}, with the assumption that the original outburst occurred well before the start of MIR observations, so the measured values reflect lower limits only, as only the late-time decline phase rather than the outburst peak itself is observed. It is thus not included in the determination of the overall sample's behavioural relations.

As previously described, the MIR behaviour of Raichu is most similar to that of SDSS~J1241+4426 from the original \citet{wang_2012_EXTREMECORONALLINE} ECLE sample, though the long-term per-band declining light curves and blueward colour trends displayed by Raichu are smaller overall across the same time frame ($\Delta$\textit{W1} $< -0.12$, $\Delta$\textit{W2} $< -0.25$, and $\Delta$(\WOneminusWTwo) $> 0.15$~mag). SDSS~J1241+4426 already displayed the smallest changes in both bands and colour evolution ($\Delta$\textit{W1} $< -0.30$, $\Delta$\textit{W2} $< -0.61$, and $\Delta$(\WOneminusWTwo) $> 0.35$~mag) compared to the means observed in the overall TDE-ECLE sample ($\Delta$\textit{W1} $< -0.65$, $\Delta$\textit{W2} $< -1.38$, and $\Delta$(\WOneminusWTwo) $> 0.77$~mag; \citealt{clark_2025_2018dyktidaldisruption}). 

As such, Raichu sits somewhat below the expected $\Delta$\textit{W2} vs. $\Delta$\textit{W1} relation, though it remains above the one-to-one ratio (Fig.~\ref{fig:MIR_Delta_Fig}). Given the lack of data from the start of the outburst, we cannot determine if this represents a deviation from the expected behaviour at low outburst intensities or if the late-time behaviour observed does not represent the evolution as a whole. Given their unusual behaviour compared to the rest of the sample, further observations of both Raichu and SDSS~J1241+4426 would be desired to better determine if they are outliers purely in duration, or if additional physical differences are involved.

The remaining two objects required additional analysis prior to the determination of their $\Delta$ values. As described in Section~\ref{sec:Pidgeot}, Pidgeot displays a long-term decline and blueward MIR colour trend in addition to the expected outburst from the occurrence of a TDE. To isolate this long-term decline from the outburst, we first remove the observations from the start of the outburst (MJD~57215) to its apparent end, as determined by the last point to be on a consistent decline (MJD~58501). We perform a linear fit to the remaining data points and use these fits to normalise the light curves without the long-term behaviour. Following this step, we return to the method used by \citet{clark_2025_2018dyktidaldisruption}, with the first observation post-outburst in both bands now identified as occurring at MJD~58298.

From these long-term linear fits, we find a \textit{W1} decline of 3.9$\times$10$^{-5}$~mag~d$^{-1}$ and a \textit{W2} decline of 6.8$\times$10$^{-5}$~mag~d$^{-1}$, though we note that the increasing rate of decline in \textit{W2} following MJD~60000 is not well described by this general fit. 
The $\Delta$ values measured for Pidgeot following this correction are within the expected parameter range of CrL-TDEs, further supporting its classification as being TDE-related (see Fig.~\ref{fig:MIR_Delta_Fig}).

As noted in Section~\ref{sec:Raticate}, Raticate displays two distinct MIR outbursts, with the first beginning, and likely peaking, prior to the start of MIR observations. We base our $\Delta$ analysis on the second outburst, for which the start, peak, and decline are all observed.

We set the quiescent baseline as the mean observed magnitudes in the MJD range 57048--58304, with the first observation in the second outburst occurring at 58507. As with Pidgeot, we find the measured $\Delta$ values to be within the expected parameter ranges for CrL-TDE outbursts, supporting Raticate's classification as being at least partially driven by a TDE (see Fig.~\ref{fig:MIR_Delta_Fig}).

\afterpage{
\clearpage
\begin{landscape}
\centering

\begin{figure}
    \centering
    \includegraphics[width=\columnwidth]{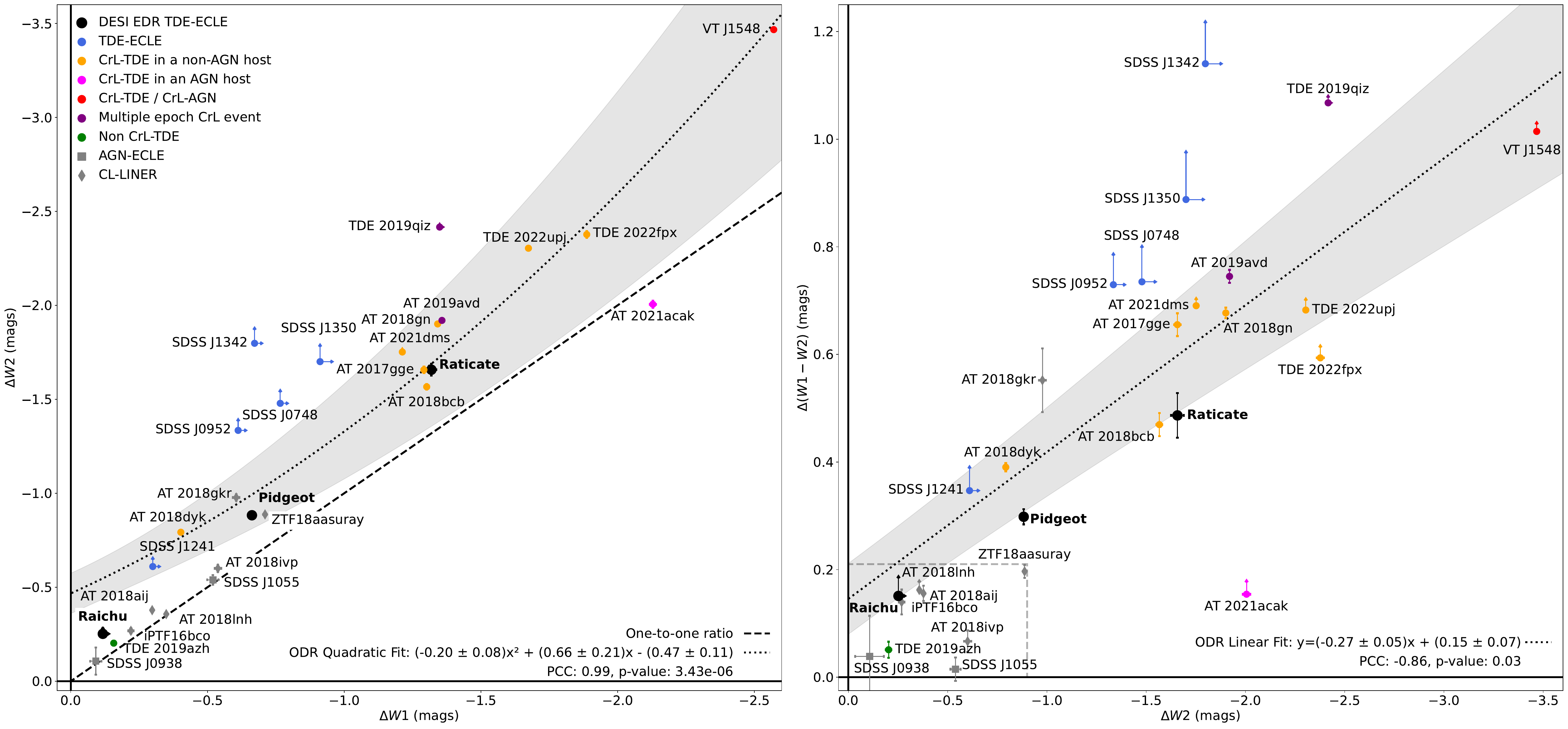}
    \caption{\textit{Left:} Comparison between the maximum change in \textit{W1} and \textit{W2} between the DESI EDR TDE-ECLEs (black circles) and a range of comparison objects. Dashed black line shows a 1-to-1 relation. Dotted line shows the orthogonal distance regression (ODR) best fitting quadratic for the CL-TDEs excluding objects with only upper limits, with the shaded region displaying the 1~$\sigma$ fitting uncertainty region. A quadratic fit to the data was preferred at a >~5~$\sigma$ level when tested against a constant and a linear fit using a likelihood ratio test, supported through AIC comparisons. A strong positive correlation is also supported by the Pearson's correlation coefficient with a value of 0.99 and corresponding p-value of 3.4~$\times$~10$^{-6}$.
    {\it Right:} Comparison between the maximum change in \textit{W1} - \textit{W2} colour versus maximum change in \textit{W2} brightness. A trend is visible in the behaviour of the CrL-TDEs and TDE-ECLEs with those objects with brighter MIR outbursts (larger $\Delta$\textit{W2} values) also showing larger shifts to MIR redder colours. Dotted line shows the ODR linear fit for this trend after excluding objects that only have upper limits on either variable, with the shaded region displaying the 1~$\sigma$ fitting uncertainty region. Restricting the fit to objects with measurable values for both parameters (6), excluding those constrained only by upper or lower limits supports this assessment with the linear relation having a Pearson's correlation coefficient with a value of -0.86 and corresponding p-value of 0.03. We note that formally the statistical tests prefer a quadratic fit to the data, but given the lack of constraining points at low and high values of $\Delta$\textit{W2}, this fit is not supported by the overall dataset when those objects measured as limits are considered. The region of parameter space occupied by the CL-LINERs from \citet{frederick_2019_NewClassChanginglook} and AGN-ECLEs (excluding AT~2018gkr) is demarcated by the dashed grey box. }
    \label{fig:MIR_Delta_Fig}
\end{figure}

\end{landscape}
    \clearpage
}

We fit a quadratic polynomial (preferred over a fixed constant or linear fit by a maximum likelihood and AIC comparison; see Table~\ref{tab:Appendix_Fit_Stats_Report}) to our measurements, alongside those from \citet{clark_2025_2018dyktidaldisruption}. The best-fitting curve continues to reflect the increased magnitude of the outbursts in the \textit{W2} band compared to \textit{W1}, 

\begin{equation}
\label{eqn:delta_w1_vs_w2}
 \Delta \mathit{W2} = (-0.2\pm0.1)\Delta \mathit{W1}^{2} + (0.7\pm0.2)\Delta \mathit{W1} - (0.5\pm0.1)\, .
\end{equation}
\noindent
The updated relation linking the increasing redness of a MIR outburst to an increasing \textit{W2} luminosity is 

\begin{equation}
\label{eqn:delta_w1_minus_w2_vs_w2}
\Delta(\mathit{W1}-\mathit{W2}) = (-0.27\pm0.05)\Delta \mathit{W2} + (0.15\pm0.07)\, .
\end{equation}

The addition of the TDE-ECLEs studied in this work to the sample from \citet{clark_2025_2018dyktidaldisruption} raises the statistical significance of the fit (when restricted to those objects with measurable values for both parameters), which now results in a Pearson's correlation coefficient of $-0.86$ and a $p$-value of 0.03. We note that formal statistical tests indicate that this relation is best fitted by a quadratic rather than a linear form, but this preference is due to the lack of constraining points at low and high values of $\Delta$\textit{W2}. Such a fit is not supported by the overall dataset when those objects measured as limits are considered. A full statistical report of these fits is provided in Table~\ref{tab:Appendix_Fit_Stats_Report}. These relations are similar to the trend identified for the MIRONG sample by \citet{jiang_2021_InfraredEchoesOptical}, where again those objects showing larger outbursts in \textit{W2} (i.e., more negative $\Delta$\textit{W2} values) also displayed larger (i.e., redder) $\Delta$(\WOneminusWTwo) values. A more in-depth exploration of the MIR behaviour of these and other similar transients is planned for a future work.

\subsection{Rate estimation}
\label{sec:Rates}

The rate of TDE-ECLEs was initially estimated by \citet{wang_2012_EXTREMECORONALLINE}, assuming all seven of their ECLEs were linked to TDEs. \citet{callow_2024_rateextremecoronal} examined the SDSS-Legacy dataset first searched by \citet{wang_2012_EXTREMECORONALLINE} and conducted a complete rates analysis. Their ECLE rate, at a median redshift of $\sim$~0.1, was consistent with the low end of TDE rates and implied that 10--40 per cent of TDEs create ECLEs. This result was corroborated by \citet{callow_2025_rateextremecoronal}, who measured a similar ECLE rate based on a single variable ECLE discovered in the BOSS LOWZ survey at $z \approx 0.3$. Here, we use the three TDE-ECLEs described above and the detection efficiency measured in Section~\ref{sec:Detection_Efficiency} to determine the ECLE rate at the DESI EDR median redshift of $\sim$~0.2.

\subsubsection{Visibility time}
\label{sec:VisibilityTime}

In order to estimate the rate of ECLEs using our DESI EDR sample, we must first calculate the visibility time for each galaxy in the sample, which is the length of time an ECLE would be detectable at that galaxy's redshift. We do this using the method developed by \citet{callow_2024_rateextremecoronal}. This process starts with sampling a TDE peak X-ray luminosity from the \citet{sazonov_2021_Firsttidaldisruption} luminosity function (LF), which is parameterized as
\begin{equation}
	\frac{\mathrm{d}\dot{N}}{\mathrm{d}\log_{10}L}=\dot{N}_0(L/L_0)^a\, ,
	\label{eq:vv_lf}
\end{equation}
where $\dot{N}$ is the volumetric rate and $L$ is the peak luminosity of the TDE.
The constants are $L_0=10^{43}\ \mathrm{erg\ s^{-1}}$, $\dot{N}_0=(1.4\pm0.8)\times10^{-7}\ \mathrm{Mpc^{-3}~yr^{-1}}$, and $a=-0.6\pm0.2$.
As the LF gives the rate of TDEs of a particular peak luminosity, we normalise it over the range of luminosities used to construct the LF, $10^{42.7}-10^{44.9}\ \mathrm{erg\ s^{-1}}$, to create a probability distribution.

We convert the peak luminosity to a peak coronal line strength using spectroscopic observations of the coronal line TDE, AT~2017gge \citep{onori_2022_NuclearTransient2017gge,callow_2024_rateextremecoronal}. The ratio between the peak coronal line strength, $S_\mathrm{max}$, and AT~2017gge's peak coronal line strength at 218~d post-discovery, $S_\mathrm{gge}$, is required to be equal to the ratio between the sampled peak luminosity and AT~2017gge's peak X-ray luminosity, $L_\mathrm{gge}$:
\begin{equation}
	\frac{S_\mathrm{max}}{S_\mathrm{gge}}=\frac{L_\mathrm{max}}{L_\mathrm{gge}}\, .
	\label{eq:peaks}
\end{equation}

The peak coronal line strength is then evolved over a period of 10~yr according to a power law. This time period was selected as it was the length of time over which the coronal lines of the \citet{wang_2012_EXTREMECORONALLINE} ECLEs faded, and a power law was used because it is the predicted shape of a TDE light-curve decline. The power-law index was sampled from the range of theoretical light-curve indices, $-5/12$ to $-5/3$.
The strength curve is then redshifted according to the galaxy's redshift, which, due to cosmological time dilation, increases the time over which the coronal lines are visible.

The visibility time, $t_\mathrm{v}$, for each galaxy is then
\begin{equation}
	t_\mathrm{v}=\int\epsilon[S(t)]\mathrm{d}t\, ,
	\label{eq:vis_time}
\end{equation}
where $\epsilon(S)$ is the detection efficiency as a function of coronal line strength measured in Section \ref{sec:Detection_Efficiency} and the integral runs over the full time over which the strength evolution is modelled.

\subsubsection{Galaxy-normalised rate}
\label{sec:GalaxyRate}

The galaxy-normalised ECLE rate is
\begin{equation}
	R_\mathrm{G}=\frac{N_{\mathrm{ECLE}}}{\sum_{i=1}^{N_\mathrm{g}}t_{\mathrm{v},i}}\, ,
	\label{eq:gal_rate}
\end{equation}
where $N_{\mathrm{ECLE}}$ is the number of ECLEs detected, $N_\mathrm{g}$ is the number of galaxies searched, and $t_{\mathrm{v},i}$ is the visibility time of the \textit{i}-th galaxy.

We detected three ECLEs in our EDR sample, which means that the errors on the resultant rate are dominated by, but do not entirely match, the Poisson uncertainty on three detections, $N_{\mathrm{ECLE}}=3.0^{+2.9}_{-1.6}$.
In order to account for the statistical uncertainties on the parameters used in the total visibility-time calculation, we conducted a Monte Carlo simulation by repeating the visibility-time calculation $500$ times and taking the peak of the resulting distribution of visibility times as the value used in the rate calculation. The $1\sigma$ standard deviation of the distribution is then used as the error on the total visibility time. This results in a galaxy-normalised rate of \galrate.

\begin{figure}
    \centering
    \includegraphics[width=\linewidth]{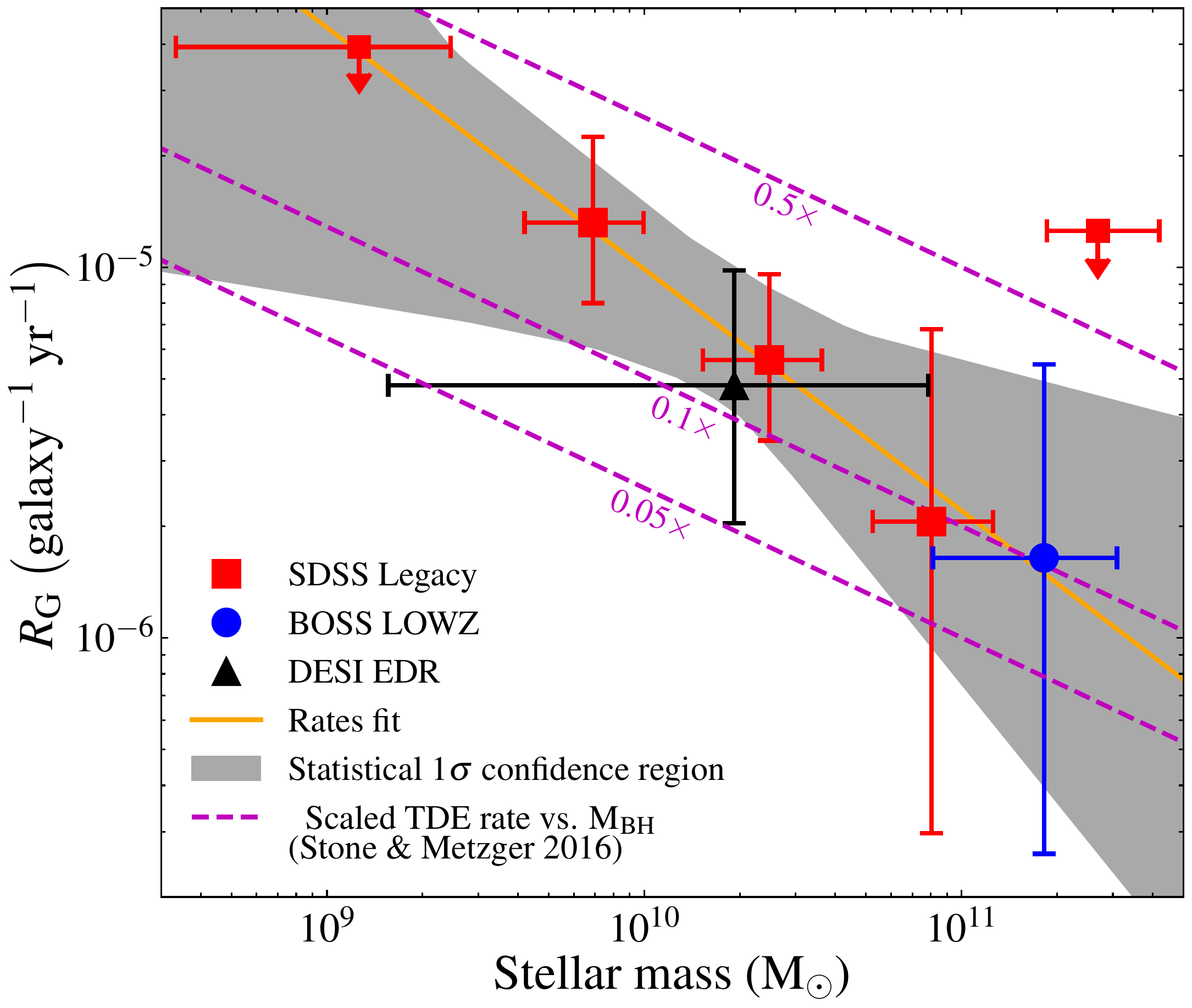}
    \caption{Galaxy-normalised ECLE rates as a function of galaxy stellar mass for SDSS Legacy (red squares), BOSS LOWZ (blue square), and DESI EDR (black square). Vertical error bars show the statistical errors on the rates derived using the Monte Carlo simulations (see text), and the horizontal error bars denote the range within each mass bin that $68$ per cent of the galaxies fall. The points marked with downward arrows are $2\sigma$ upper bounds on the rates calculated using the upper Poisson error on zero detections. The solid orange line shows the power-law fit to the Legacy, LOWZ, and EDR rates, and the shaded area is the $1\sigma$ confidence region. The dashed purple line shows the TDE rate vs. black-hole mass relation calculated by \citet{stone_2016_EnhancedRateTidal}, scaled by $0.05$, $0.1$, and $0.5$.}
    \label{fig:gal_rate_mass_rel}
\end{figure}

We add this rate to the galaxy-normalised rate vs. mass relation developed by \citet{callow_2024_rateextremecoronal, callow_2025_rateextremecoronal} at the mean stellar mass of EDR in Figure~\ref{fig:gal_rate_mass_rel}.
This rate is consistent with the relation calculated using the rates from SDSS Legacy and BOSS LOWZ.
The values of the parameters of the fit change to $a=-0.7\pm0.3$ and $b=1.5^{+3.1}_{-3.5}$, with a reduced $\chi^2=0.14$.

We caution that the $\chi^2$ distribution used to determine the power-law fit parameters and confidence region is a biased estimator owing to the inclusion of Poisson uncertainties stemming from the small number of ECLEs per bin. Hence, the errors on the parameters and the confidence region are likely underestimated. However, we note that the shape of the power-law fit and the theoretical TDE rate vs. mass relation are still qualitatively consistent.

\subsubsection{Mass-normalised rate}

\begin{figure}
	\centering
	\includegraphics[width=\columnwidth]{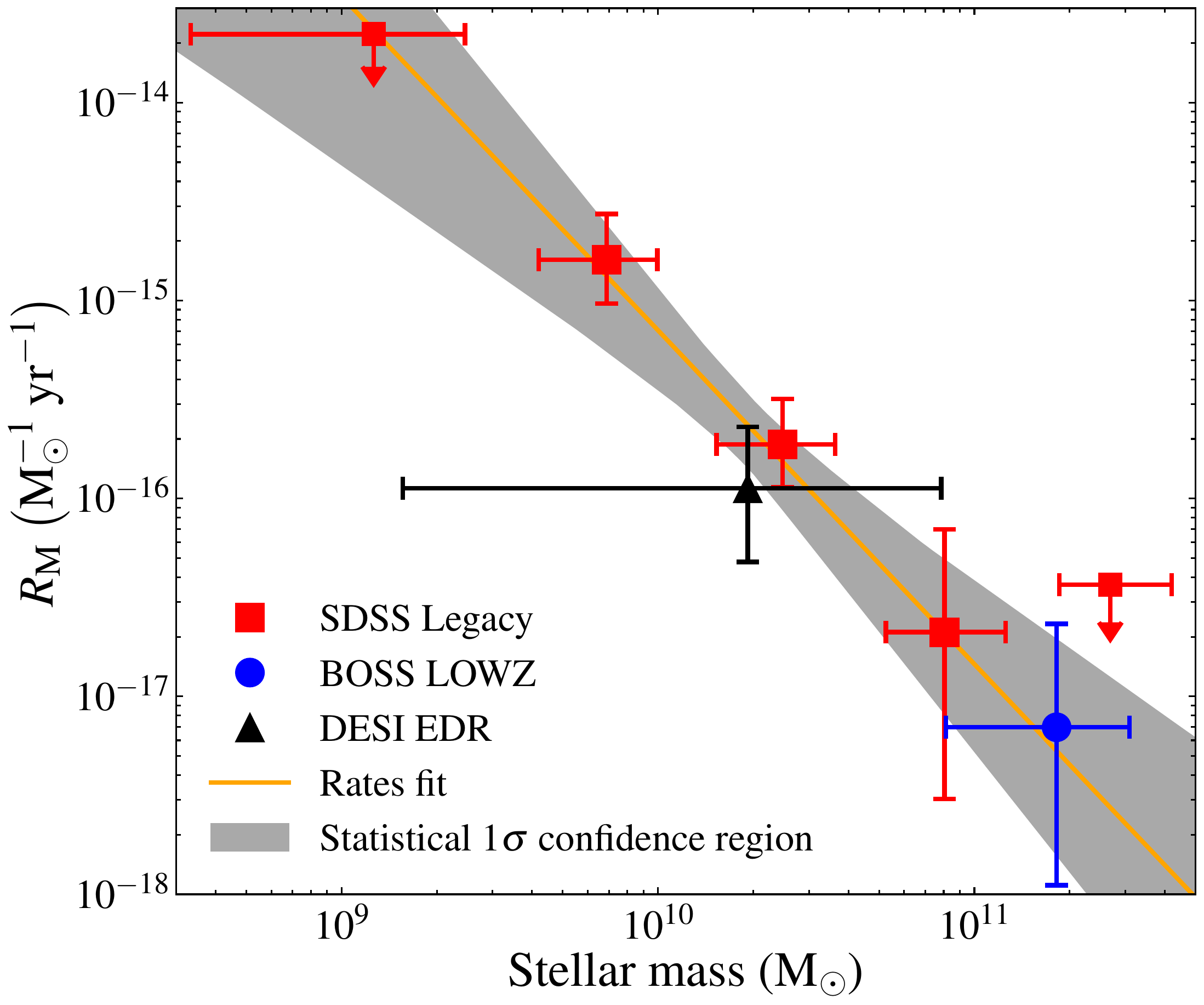}
	\caption{Mass-normalised ECLE rates as a function of galaxy stellar mass for SDSS Legacy (red squares), BOSS LOWZ (blue square) and DESI EDR (black square).
	Vertical error bars show the statistical errors on the rates derived using the Monte Carlo simulations (see text), and the horizontal error bars denote the range within each mass bin that $68$ per cent of the galaxies fall.
	The points marked with downward arrows are $2\sigma$ upper bounds on the rates calculated using the upper Poisson error on zero detections.
	The solid orange line shows the power-law fit to the Legacy, LOWZ, and EDR rates, and the shaded area is the $1\sigma$ confidence region.
	}
	\label{fig:mass_rate_mass_rel}
\end{figure}

The mass-normalised rate is calculated by weighting each galaxy's visibility time by its stellar mass taken from the \fsf\ VAC.
We perform the same Monte Carlo process of calculating the total visibility time $500$ times and taking the peak of the distribution as the true value.
This gives a mass-normalised rate of \massrate.
We add this rate to the mass-normalised rate vs. stellar mass relation from \citet{callow_2024_rateextremecoronal, callow_2025_rateextremecoronal} in Figure~\ref{fig:mass_rate_mass_rel}.
The power-law fit produces values of $a=-1.7^{+0.4}_{-0.3}$ and $b=2^{+3}_{-4}$, with a reduced $\chi^2=1.5$.
The resultant relation is consistent with the one measured using the SDSS Legacy and BOSS LOWZ samples.

\begin{figure*}
	\centering
	\includegraphics[width=0.95\textwidth]{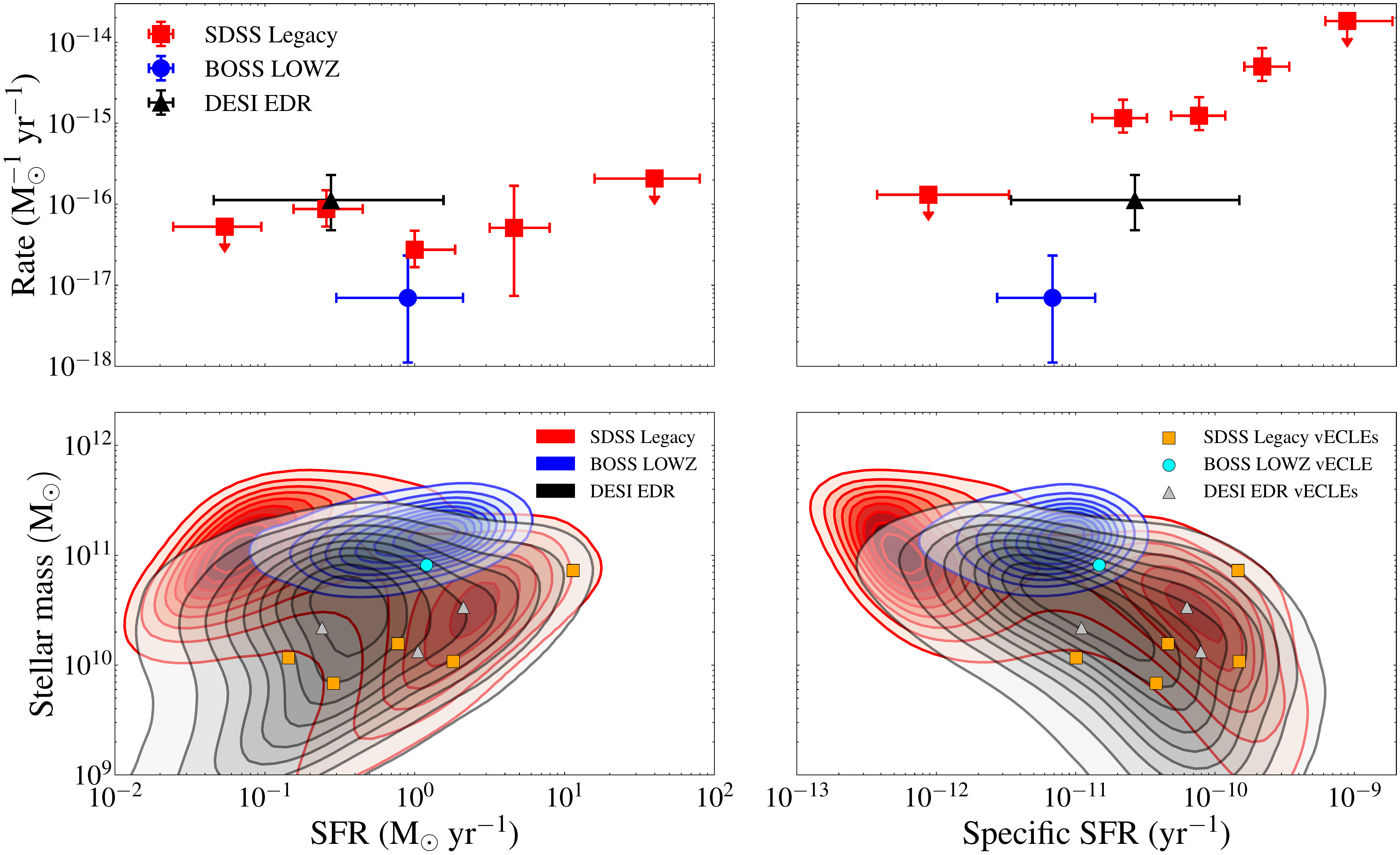}
	\caption[Relation between mass-normalised vECLE rate and SFR and sSFR, with the DESI EDR rate added]{\textit{Top}: Mass-normalised ECLE rates as a function of galaxy SFR (left) and sSFR (right) for SDSS Legacy (red squares), BOSS LOWZ (blue circle), and DESI EDR (black triangle).
	Vertical error bars show the statistical errors on the rates, while the horizontal error bars denote the SFR and sSFR ranges comprising $68$ per cent of the galaxies in each bin.
	The points marked with downward arrows are $2\upsigma$ upper limits on the rates calculated using the upper Poisson error on zero detections.
	\textit{Bottom}: Contours showing the density of galaxies in the SDSS Legacy (red), BOSS LOWZ (blue), and DESI EDR (black) samples in decrements of $10$ per cent.
	The ECLEs detected in SDSS Legacy by \citet{wang_2012_EXTREMECORONALLINE} are shown as orange squares, the ECLE detected in BOSS LOWZ as a cyan circle, and the ECLEs detected in DESI EDR as grey triangles.}
	\label{fig:rate_sfr_relation}
\end{figure*}

We also investigate the relation between the mass-normalised ECLE rate and star-formation rates (SFRs) and specific SFRs (sSFRs) of the galaxies in the sample to test whether they strengthen the tentative positive correlation between rate and sSFR found by \citet{callow_2025_rateextremecoronal}.
This is shown in Figure~\ref{fig:rate_sfr_relation}.
The mass-normalised rate continues to show no dependence on SFR, whereas the DESI EDR rate is consistent with the positive correlation with sSFR found previously.
A likelihood ratio test continues to prefer a first-order polynomial over a zeroth-order polynomial at a $>2\sigma$ confidence level.
We also measure a Pearson $r$ coefficient of $0.88$ with a $p$-value of $0.047$, indicating a hint of a correlation.
More measurements at a higher precision are needed to claim statistical significance of this relation.
The DESI ECLEs also lie in similar regions of the SFR/sSFR vs. stellar mass phase space as the previously discovered ECLEs.
These results of no dependence on SFR and a possible correlation with sSFR are very similar to the relations found for Type Ia supernovae by \citet{graur_2015_unifiedexplanationsupernova}.
Those correlations were explained as a consequence of the galaxy age-mass relation, in combination with the additional correlation between galaxy stellar mass and either SFR or sSFR, and the delay-time distribution of Type Ia supernovae.
Given the similarities between the relations in this work and \citet{graur_2015_unifiedexplanationsupernova}, future studies may find a similar explanation is suitable for ECLEs and, by extension, TDEs.

\subsubsection{Volumetric rate}

We calculate the volumetric ECLE rate by multiplying the mass-normalised rate by the cosmic stellar mass density.
This is calculated by integrating the galactic stellar mass function (GSMF), accounting for the mass distribution of EDR and the ECLE rate vs. stellar mass relation by including both of them in the calculation.
As the \citet{baldry_2012_GalaxyMassAssembly} GSMF has been shown to be valid out to $z\approx0.5$, which covers the full redshift range of our EDR sample, we continue to use it for this calculation.
The volumetric rate is then calculated to be \volrate.

We also investigate the contribution of each source of uncertainty to the overall values by repeating the Monte Carlo simulations and varying each source of uncertainty individually.
The resultant error budget is presented in Table~\ref{tab:uncertainties}.
The total uncertainty percentages are the linear sum of the total statistical and systematic uncertainties divided by the corresponding rate value.

\begin{table}
	\centering
    \caption{ECLE rate uncertainty percentages.}
	\begin{tabular}{lc}
		\hline
		Uncertainty 				& Percentage of rate \\
		\hline
		\multicolumn{2}{c}{\textbf{Galaxy-normalised rate}} \\
		Total						& $+104/-58$ \\
		\multicolumn{2}{c}{\textit{Statistical}} \\
		Poisson						& $+97/-54$ \\
		AT~$2017$gge peak CrL strength	& $+3/-4$ \\
		AT~$2017$gge peak luminosity	& $+31/-13$ \\
		Range of power-law indices	& $+24/-14$ \\
		\multicolumn{2}{c}{\textbf{Mass-normalised rate}} \\
		Total						& $+104/-58$ \\
		Galaxy stellar masses 		& $\pm1$ \\
		\multicolumn{2}{c}{\textbf{Volumetric rate}} \\
		Total						& $+593/-70$ \\
		GSMF parameters 			& $+174/-65$ \\
		Rate mass trend fit			& $+490/-1$ \\
		\hline
	\end{tabular}
	\label{tab:uncertainties}
\end{table}

The positive uncertainty on the volumetric rate is proportionally much larger than the corresponding uncertainties on the galaxy- and mass-normalised rates.
The uncertainty budget shows that the main contributing factor to this large uncertainty is the fit to the mass-normalised rate vs. galactic stellar mass trend.
This is because the stellar masses of the SDSS Legacy ECLE host galaxies and the mean stellar masses of the BOSS LOWZ and DESI EDR samples are all within the range $\sim10^9-10^{11}~\msol$.
Therefore, the fit to the trend is best constrained in this range and the confidence region of the fit grows outside the range.
DESI EDR contains galaxies with stellar masses down to $\sim10^7~\msol$, well outside the well-constrained region.
Calculating the volumetric rate requires integrating the rate-mass relation over the full mass range of the galaxy sample.
Therefore, the large uncertainties in the rate-mass fit at the low stellar masses that DESI EDR contains propagate through to the volumetric rate uncertainties.
A similar effect is seen in the SDSS Legacy volumetric rate uncertainties, although not to as great an extent, as SDSS Legacy did not contain as many lower-mass galaxies as DESI EDR.

\subsubsection{Comparison with previous rates}

\begin{figure}
	\centering
	\includegraphics[width=\columnwidth]{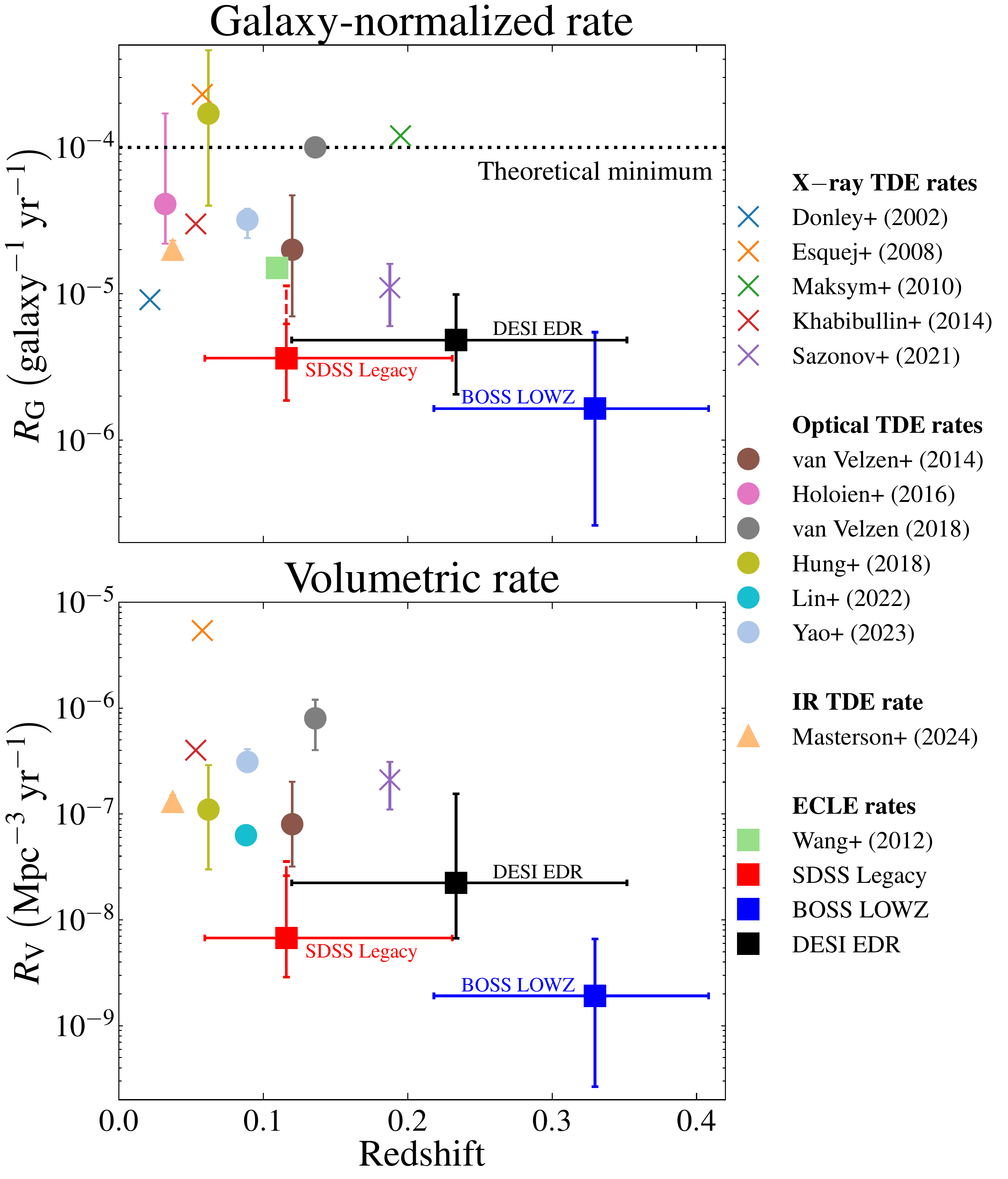}
	\caption{Comparisons of our galaxy-normalised (left) and volumetric (right) ECLE rates with TDE and ECLE rates from the literature.
	TDE rates derived from X-ray surveys are shown as crosses \citep{donley_2002_LargeAmplitudeXRayOutbursts,esquej_2008_Evolutiontidaldisruption,maksym_2010_TidalDisruptionFlare,khabibullin_2014_Stellartidaldisruption,sazonov_2021_Firsttidaldisruption}, those from optical/UV surveys as circles \citep{vanvelzen_2014_MeasurementRateStellar,holoien_2016_ASASSN15oirapidlyevolving,vanvelzen_2018_MassLuminosityFunctions,hung_2018_SiftingSapphiresSystematic,lin_2022_LuminosityFunctionTidal,yao_2023_TidalDisruptionEvent}, and IR surveys as triangles \citep{masterson_2024_NewPopulationMidinfraredselected}. Our previous SDSS Legacy \citep{callow_2024_rateextremecoronal} and BOSS LOWZ \citep{callow_2025_rateextremecoronal} ECLE rates are indicated as red and blue squares, respectively, while the Wang et al. (2012) overestimated galaxy-normalised ECLE rate is shown in green.
	Error bars are displayed if available.
	The statistical errors of our ECLE rates are denoted by the solid error bars and the systematic errors by the dashed error bars.
	The horizontal error bars show the range of redshifts spanned by $68$ per cent of the galaxies in the spectroscopic samples used.
	The dotted horizontal line marks the theoretical minimum TDE rate calculated by \citet{wang_2004_RevisedRatesStellar}.}
	\label{fig:rate_comp}
\end{figure}

In Figure~\ref{fig:rate_comp}, we compare our galaxy-normalised and volumetric ECLE rates from SDSS Legacy, BOSS LOWZ, and DESI EDR to TDE rates from the literature \citep{donley_2002_LargeAmplitudeXRayOutbursts,esquej_2008_Evolutiontidaldisruption,maksym_2010_TidalDisruptionFlare,khabibullin_2014_Stellartidaldisruption,vanvelzen_2014_MeasurementRateStellar,holoien_2016_ASASSN15oirapidlyevolving,hung_2018_SiftingSapphiresSystematic,vanvelzen_2018_MassLuminosityFunctions,sazonov_2021_Firsttidaldisruption,lin_2022_LuminosityFunctionTidal,yao_2023_TidalDisruptionEvent,masterson_2024_NewPopulationMidinfraredselected} and the ECLE rate estimated by \citet{wang_2012_EXTREMECORONALLINE}.
The EDR ECLE galaxy-normalised rate is consistent with both the SDSS and BOSS rates, but is still roughly half an order of magnitude lower than the \citet{wang_2012_EXTREMECORONALLINE} estimate (which counted two nonvariable ECLEs among its seven objects).
The volumetric rate from EDR is also consistent with the SDSS Legacy rate, but is higher than the BOSS LOWZ rate.
Given that we control for the different surveys selecting galaxies in different mass regimes, this could be due to ECLE rates being lower at higher redshifts, aligning with theoretical work on the evolution of the TDE rate with redshift \citep{kochanek_2016_Abundanceanomaliestidal}.
However, more precise rates at higher redshifts are required to check whether this is truly the case.

The galaxy-normalised rate is over an order of magnitude lower than the theoretical minimum TDE rate, and all three ECLE rates are lower than all the observational TDE rates, strengthening our proposition that only a subset of TDEs produce CrLs.
By comparing the EDR rates to the theoretical minimum TDE rate and the lowest observational galaxy-normalised and volumetric TDE rates, we can  constrain the percentage of TDEs that create strong CrLs.
The \citet{wang_2004_RevisedRatesStellar} theoretical minimum gives $5^{+5}_{-3}$ per cent, whereas the galaxy-normalised rate from \citet{donley_2002_LargeAmplitudeXRayOutbursts} yields $50^{+50}_{-30}$ per cent, and the volumetric rate from \citet{lin_2022_LuminosityFunctionTidal} is $36^{+64}_{-25}$ per cent.
All of these ratios are consistent with the values calculated using the SDSS and BOSS rates with the conclusion that roughly 5--50 per cent of TDEs result in the creation of ECLEs.
However, these ratios are calculated under the assumption that all ECLEs are the result of TDEs.
If other transients are able to create strong, variable CrLs, then the ratio of TDEs that produce CrLs would be lower.
\citet{komossa_2024_ExtremesContinuumEmissionLine} investigated extreme AGN flares, a subset of which can appear similar to TDEs with a large increase in X-ray luminosity that fades over the following months to years.
They discuss two such flares that also developed CrLs, which then continued to vary or fade.
However, it is unclear whether these flares were the result of variability of the AGN itself or due to a TDE occurring in the AGN.
Another possible example of this is the transient AT~$2019$aalc, which showed strong transient CrLs similar to those in this work and CrL-TDEs \citep{milanveres_2024_BackdeadAT2019aalc}.
It has been classified as a Bowen fluorescence flare (a type of AGN flare) owing to Bowen fluorescence lines in the transient spectrum and an AGN-like pre-transient spectrum.
Again, the progenitor of this flare is not understood, so it is possible that AT~$2019$aalc is a TDE occurring in an AGN.
Given these examples, it is possible that not all ECLEs are created by TDEs, and so the ratios calculated in this work can be considered as upper limits on the true values.

\section{Conclusions}
\label{sec:Conclusions}

In this work we conducted a search for new ECLEs within DESI's EDR, with a focus on identifying those linked to the occurrence of TDEs. We have described the details of the search, including the custom Python code (\sleipnir) used to perform the search, highlighting the primary sources of potential contamination in similar analyses and how these can be mitigated.

Our search of the DESI EDR has identified a total of 208 galaxies displaying Fe CrLs. Using archival and follow-up imaging and spectroscopy in the optical, NIR, and MIR, we have classified three of these ECLEs as being linked to TDEs (i.e., TDE-ECLEs). The other 205 were found to be consistent with AGN activity.

Based on the identification of these objects, we have been able to further strengthen the claim made by \citet{clark_2025_2018dyktidaldisruption} that TDEs displaying MIR outbursts show a colour-luminosity relationship in their MIR evolution. Specifically, objects with brighter MIR flares exhibit more significant reddening at outburst. We also calculate the rates of TDE-linked ECLEs at the DESI median redshift of 0.2, finding a galaxy-normalised rate of \galrate, a mass-normalised rate of \massrate, and a volumetric rate of \volrate. These rates are consistent with those measured for other surveys. More precise rates, such as those that could be measured from an ECLE survey of the complete DESI sample, are required to ascertain whether TDE-ECLE rates decline with redshift, as predicted by \citet{callow_2025_rateextremecoronal} under the assumption that TDEs are the dominant source of variable extragalactic coronal lines and following previous studies that TDE rates generally decrease with increasing redshift. Based on the number of TDE-ECLEs discovered in DESI EDR and the rates measured here, we expect the full DESI survey to yield $\sim 35$--$50$ new TDE-ECLEs.

The search has also revealed more than 200 AGNs that display CrLs; given their nature as the primary astrophysical contaminant in ECLE searches, they will serve as an important reference for future work in screening them from TDE samples. This work also highlights the necessity for long-term, multiwavelength follow-up observations of nuclear transients on years-long timescales.

\section*{Acknowledgements}
\label{sec:Acknowledgements}

We thank the anonymous referee for their constructive comments and feedback during the review process. We are also grateful to Yi Yang for his role in obtaining the follow-up Keck spectra used here.

P.C. was supported by the Science \& Technology Facilities Council grant ST/Y001850/1. O.G. and P.C. were also supported by the Science \& Technology Facilities Council [grants ST/S000550/1 and ST/W001225/1].

A.V.F.’s research group at U.C. Berkeley was supported by the Christopher R. Redlich Fund, 
Gary and Cynthia Bengier, Clark and Sharon Winslow, Alan Eustace and Kathy Kwan (W.Z. is a Bengier-Winslow-Eustace Specialist in Astronomy), Timothy and Melissa Draper, Briggs and Kathleen Wood, Ellen and Alan Seelenfreund (T.G.B. is Draper-Wood-Seelenfreund Specialist in Astronomy), and numerous other donors.
Some of the data presented herein were obtained at the W. M. Keck Observatory, which is operated as a scientific partnership among the California Institute of Technology, the University of California, and NASA; the observatory was made possible by the generous financial
support of the W. M. Keck Foundation. The authors wish to recognize and acknowledge the very significant cultural role and reverence that the summit of Maunakea has always had within the indigenous Hawaiian community; we are most fortunate to have the opportunity to conduct observations from this mountain. We thank the Keck staff for their assistance.

This material is based upon work supported by the U.S. Department of Energy (DOE), Office of Science, Office of High Energy Physics, under Contract No. DE-AC02-05CH11231, and by the National Energy Research Scientific Computing Center, a DOE Office of Science User Facility under the same contract. Additional support for DESI was provided by the U.S. National Science Foundation (NSF), Division of Astronomical Sciences under Contract No. AST-0950945 to the NSF’s National Optical Infrared Astronomy Research Laboratory; the Science and Technology Facilities Council of the United Kingdom; the Gordon and Betty Moore Foundation; the Heising-Simons Foundation; the French Alternative Energies and Atomic Energy Commission (CEA); the National Council of Humanities, Science and Technology of Mexico (CONAHCYT); the Ministry of Science, Innovation and Universities of Spain (MICIU/AEI/10.13039/501100011033), and by the DESI Member Institutions: \url{https://www.desi.lbl.gov/} collaborating-institutions. Any opinions, findings, and conclusions or recommendations expressed in this material are those of the author(s) and do not necessarily reflect the views of the U. S. National Science Foundation, the U. S. Department of Energy, or any of the listed funding agencies. The authors are honored to be permitted to conduct scientific research on I'oligam Du'ag (Kitt Peak), a mountain with particular significance to the Tohono O’odham Nation.

The Liverpool Telescope is operated on the island of La Palma by Liverpool John Moores University in the Spanish Observatorio del Roque de los Muchachos of the Instituto de Astrofisica de Canarias with financial support from the UK STFC (Project IDs PL22A04, PL22B03, PL23A03, PL23B07, PL24A13, PL24B03, and PL25A01; PI P. Clark).
Several spectra presented in this work were obtained with ALFOSC, which is provided by the Instituto de Astrofisica de Andalucia (IAA) under a joint agreement with the University of Copenhagen and NOT (Proposal ID 70-108, PI P. Clark). 

The DESI Legacy Imaging Surveys consist of three individual and complementary projects: the Dark Energy Camera Legacy Survey (DECaLS), the Beijing-Arizona Sky Survey (BASS), and the Mayall z-band Legacy Survey (MzLS). DECaLS, BASS and MzLS together include data obtained, respectively, at the Blanco telescope, Cerro Tololo Inter-American Observatory, NSF’s NOIRLab; the Bok telescope, Steward Observatory, University of Arizona; and the Mayall telescope, Kitt Peak National Observatory, NOIRLab. NOIRLab is operated by the Association of Universities for Research in Astronomy (AURA) under a cooperative agreement with the U.S. National Science Foundation. Pipeline processing and analyses of the data were supported by NOIRLab and the Lawrence Berkeley National Laboratory (LBNL). Legacy Surveys also uses data products from the Near-Earth Object Wide-field Infrared Survey Explorer (NEOWISE), a project of the Jet Propulsion Laboratory/California Institute of Technology, funded by the National Aeronautics and Space Administration. Legacy Surveys was supported by: the Director, Office of Science, Office of High Energy Physics of the U.S. Department of Energy; the National Energy Research Scientific Computing Center, a DOE Office of Science User Facility; the U.S. National Science Foundation, Division of Astronomical Sciences; the National Astronomical Observatories of China, the Chinese Academy of Sciences and the Chinese National Natural Science Foundation. LBNL is managed by the Regents of the University of California under contract to the U.S. Department of Energy. The complete acknowledgments can be found at \url{https://www.legacysurvey.org/acknowledgment/}.

This research has made use of the VizieR catalogue access tool, CDS, Strasbourg Astronomical Observatory, France (DOI: \url{10.26093/cds/vizier}).
NOIRLab IRAF is distributed by the Community Science and Data Center at NSF NOIRLab, which is managed by the Association of Universities for Research in Astronomy (AURA) under a cooperative agreement with the U.S. National Science Foundation.


\section*{Data Availability}
The data underlying this work are available in the article and in its online supplementary material available through Zenodo \citep{clark_2026_SupplementarydataEarly}.


Additionally, all DESI spectra utilised here are part of the public ``Early Data Release'', details of which (and how to access the data) can be found within \citet{desicollaboration_2024_EarlyDataRelease} and through the DESI data \href{https://data.desi.lbl.gov/doc/releases/edr/}{website}.
The public version of the \sleipnir\ code will be released through \href{https://github.com/Lightbulb500/SLEIPNIR}{GitHub} following this work's publication.



\bibliographystyle{mnras}
\bibliography{MyLibrary.bib} 



\clearpage
\appendix

\section{\fsf\ value-added catalogue information}
\label{subsec:Appendix_FSF}

In this Appendix we outline the specific parameters retrieved from the \fsf\ VAC used in this work for our comparative host-galaxy analysis.

\begin{table*}
\centering
\caption{Parameters retrieved from the \fsf\ EDR VAC. Parameter descriptions from the \fsf\ documentation.${^1}$ Version used: v3.2}
\label{tab:fastspecfit_info}
\begin{tabular}{cc}
\hline
\textbf{Parameter} & \textbf{Description} \\ \hline
\texttt{TARGETID} &   DESI Target ID\\
\texttt{Z} & Recorded redshift ${^2}$\\
\texttt{ZWARN} & \redrock\ redshift warning bit\\
\texttt{SFR} & Instantaneous star-formation rate ($h=1.0$, \citet{chabrier_2003_GalacticStellarSubstellar} initial mass function) \\
\texttt{LOGMSTAR} & Logarithmic stellar mass ($h=1.0$, \citet{chabrier_2003_GalacticStellarSubstellar} initial mass function).\\
\texttt{LOGMSTAR$\_$IVAR} & Inverse variance of \texttt{LOGMSTAR}. \\
\hline
\end{tabular}
\begin{flushleft}
\textit{Notes:} ${^1}$ Documentation available here \url{https://fastspecfit.readthedocs.io/en/latest/fastspec.html}\\
${^2}$ Obtained from \redrock, or from \textsc{QuasarNET} \citep{green_2025_UsingActiveLearning} for QSO targets
\end{flushleft}
\end{table*}

\section{Detection Efficiency Comparison}
\label{subsec:Appendix_DetEff}

Here we show the comparative performance of \sleipnir\ when deployed on the DESI EDR sample compared with the SDSS Legacy and BOSS LOWZ samples used by \cite{callow_2024_rateextremecoronal} and \cite{callow_2025_rateextremecoronal}, respectively.

\begin{figure}
    \centering
    \includegraphics[width=0.95\columnwidth]{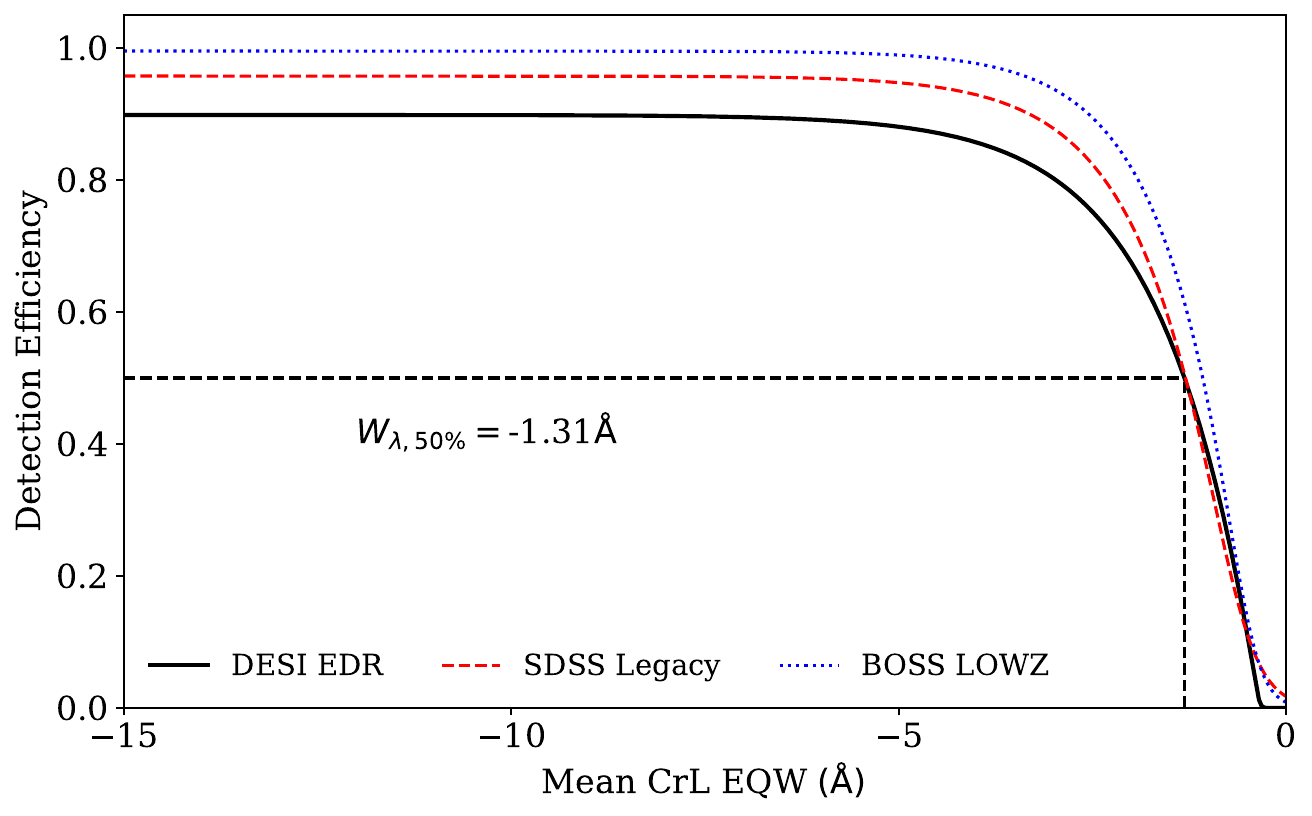}
    \caption{\sleipnir\ detection efficiency as a function of the average equivalent width of the coronal lines for DESI EDR and the comparison SDSS Legacy and BOSS LOWZ galaxy samples \citep{callow_2024_rateextremecoronal,callow_2025_rateextremecoronal}. Curves are generalised sigmoid fits to the underlying distributions.}
    \label{fig:DetEff}
\end{figure}

\section{Detailed Crossmatching Results}
\label{sec:Crossmatch_Results}

Here we provide the full results of the sample crossmatching as described in Section~\ref{sec:Crossmatch_Results}.

\subsection{TNS crossmatch}
\label{sec:Appendix_TNS_Crossmatch}

We first provide the full results of the crossmatch between the \sleipnir\ Input Galaxy Sample and the TNS database in Table~\ref{tab:Appendix_Full_TNS_Crossmatch}. 1,219 crossmatches were obtained, of which six are within our final EDR CrL Object Sample classified as CrL-AGN. This crossmatch was completed against the full public TNS database as of \TNSMatchDate.

\label{sec:Appendix_TNS_Crossmatch_Results}
\begin{table*}
\caption{The results of the TNS crossmatch to the \sleipnir\ Input Galaxy Sample. Crossmatch performed on the full TNS database as of \TNSMatchDate. A small portion of the table is included here for guidance of form. The full machine-readable version is included in the supplementary material.}
\label{tab:Appendix_Full_TNS_Crossmatch}
\begin{tabular}{lccc}
\hline
\textbf{DESI ID} & 
\textbf{CrL Classification}&
\textbf{TNS Name} & \textbf{TNS Classification} \\ \hline
39633453855017885 & CrL-AGN & AT 2021vje & None \\
39633304189666247 & CrL-AGN & AT 2019jw & None \\
39633300653869008 & CrL-AGN & AT 2021lar & None \\
39628506220463012 & CrL-AGN & AT 2018mer & None \\
39627782338120416 & CrL-AGN & AT 2021mks & None \\
... & ... & ... \\ \hline
\end{tabular}
\end{table*}

\subsection{Literature nuclear transient crossmatch}
\label{sec:Appendix_Clark_Crossmatch_Results}

Next, Table~\ref{tab:Clark_Crossmatch_Results} provides the full results of the crossmatch between the \sleipnir\ Input Galaxy Sample and our expanded database of nuclear transients. 11 crossmatches were obtained. None of these nuclear transients was seen to display CrLs in their DESI EDR spectra.

\begin{table*}
\caption{Crossmatch results between the full \sleipnir\ Input Galaxy Sample and a literature-sourced database of TDEs and other nuclear transients. No crossmatched object displayed coronal lines in its DESI EDR spectrum/spectra.}
\label{tab:Clark_Crossmatch_Results}
\begin{tabular}{lccc}
\hline
\textbf{DESI ID} & \textbf{Object Name} & \textbf{Classification} & \textbf{Classification Source}\\ \hline
39633451355212798 & AT 2019baf & TDE & \citet{yao_2023_TidalDisruptionEvent}\\
39633342957617640 & AT 2019mha & TDE & \citet{hammerstein_2023_FinalSeasonReimagined} \\
39633290046475104 & GALEX D3-13 & TDE & \citet{gezari_2006_UltravioletDetectionTidal}\\
39627793834706352 & J121116 & TDE & \citet{qin_2022_LinkingExtragalacticTransients}\\
39633390168704930 & J123715 & TDE & \citet{qin_2022_LinkingExtragalacticTransients} \\
39633300964248835 & RX J1420.4+5334 & TDE & \citet{greiner_2000_RXJ14204+5334another}\\
39628433277325770 & RX J1301.9+2747 & QPE & \citet{giustini_2020_Xrayquasiperiodiceruptions}\\
39627835576422677 & SDSS J100350.97+020227.6 & TDE & \citet{jiang_2021_MidinfraredOutburstsNearby, wang_2022_MidinfraredOutburstsNearby}\\
39627829549207044 & TDE 2018hyz & TDE & \citet{arcavi_2018_TransientClassificationReporta, short_2020_tidaldisruptionevent, gomez_2020_TidalDisruptionEvent} \\
39632976878768365 & TDE 2021gje & TDE & \citet{hammerstein_2021_ZTFTransientClassification} \\
39632941722109945 & TDE 2024lhc & TDE & \citet{yao_2024_Classification2024lhcTidal} \\ \hline
\end{tabular}
\end{table*}

\subsection{SDSS crossmatches}
\label{sec:Appendix_SDSS_Crossmatches}

Here we describe the full results of the crossmatches conducted between the \sleipnir\ Input Galaxy Sample and the samples of CrL galaxies constructed from the previous uses of \sleipnir\ on SDSS data.

As described in Section~\ref{sec:SDSS_Crossmatch}, 18 crossmatches were obtained between the \sleipnir\ Input Galaxy Sample and the CrL object sample of \citet{callow_2024_rateextremecoronal}, five of which are in the EDR CrL Object Sample. Two of these objects failed to meet the CrL strength cut relative to \fspectralline{O}{iii}{5007} used in the \citet{callow_2024_rateextremecoronal} analysis (20 per cent) to trigger visual inspection. The remaining three were rejected as false detections following visual inspection of the SDSS spectra primarily owing to the weakness of the CrL features. The higher SNR of the DESI spectra of these objects allowed the CrL features to be reassessed as real in following the classification process in this work.

Of the remaining 13 objects that were not flagged in the initial DESI EDR analysis (i.e., not part of the EDR CrL Object Sample), ten failed to meet the requirements for visual inspection in the search of the SDSS Legacy sample, with the remaining three being discarded as false-positive detections following visual inspection. After visual inspection of the DESI spectra of all 13 galaxies, 11 were confirmed to be non-CrL hosting galaxies, with the remaining 2 being likely AGNs with weak \Fevii\ emission that was insufficient to trigger automatic \sleipnir\ flagging alone (DESI~39628465170813469, DESI~39633362041702102). Of these, DESI~39628465170813469 failed to meet the CrL strength requirements in the SDSS Legacy search whilst DESI~39633362041702102 was discarded as a false positive following visual inspection by \citet{callow_2025_rateextremecoronal}.

The full crossmatch to the BOSS LOWZ CrL sample from \citet{callow_2025_rateextremecoronal} returned seven objects. Of these, six were identified as being sky-line contaminated false-positive detections with the final object (SDSS~J113104.36+531338.8) not meeting the CrL strength threshold to trigger visual inspection. Visual inspection of both the DESI spectrum (which was not flagged by \sleipnir) and the BOSS LOWZ spectrum reveal this to be a low-SNR false positive. No objects within the DESI EDR CrL sample were crossmatched to the BOSS LOWZ CrL sample -- unsurprising given the small number of matches overall.

\begin{table*}
\caption{Crossmatch results between the full \sleipnir\ Input Galaxy Sample and the sample of galaxies identified as showing potential CrLs by an earlier version of \sleipnir\ in the search for ECLEs in the SDSS Legacy DR17 survey by  \citet{callow_2024_rateextremecoronal}.}
\label{tab:Appendix_Full_Legacy_Crossmatch}
\begin{tabular}{lccc}
\hline
\textbf{DESI ID} & \textbf{DESI EDR Classification} & \textbf{SDSS Short Name} & \textbf{SDSS Legacy Classification} \\ \hline
39628449278591796 & CrL-AGN & SDSS J1300+2824 & Rejected ${^1}$ \\
39633066712369198 & CrL-AGN & SDSS J0731+3926 & Weak CrLs ${^2}$ \\
39633236212582085 & CrL-AGN & SDSS J1134+4912 & Rejected \\
39633304827202451 & CrL-AGN & SDSS J1606+5340 & Weak CrLs \\
39633416353743637 & CrL-AGN & SDSS J1243+6225 & Rejected \\
 & \multicolumn{1}{l}{} & \multicolumn{1}{l}{} & \multicolumn{1}{l}{} \\
39627775576903981 & - & SDSS J1138-0035 & Weak CrLs \\
39627788335973998 & - & SDSS J1420-0004 & Weak CrLs \\
39627799132112775 & - & SDSS J0914+0023 & Rejected \\
39627818576909309 & - & SDSS J1430+0118 & Weak CrLs \\
39627823480048542 & - & SDSS J0959+0126 & Weak CrLs \\
39627836692105274 & - & SDSS J1429+0154 & Weak CrLs \\
39628465170813469 & - & SDSS J1301+2918 & Weak CrLs \\
39628487006358229 & - & SDSS J1648+3022 & Weak CrLs \\
39633274808568349 & - & SDSS J1148+5145 & Weak CrLs \\
39633278537304499 & - & SDSS J1143+5153 & Weak CrLs \\
39633335630169483 & - & SDSS J1204+5602 & Weak CrLs \\
39633338981418326 & - & SDSS J1151+5613 & Weak CrLs \\
39633362041702102 & - & SDSS J1153+5806 & Rejected \\ \hline
\end{tabular}
\begin{flushleft}
\textit{Notes:} ${^1}$ SDSS spectrum was manually inspected and determined to be a false-positive detection.\\
${^2}$ In cases of ``Weak CrLs'', the SDSS spectrum was not manually inspected as no CrLs exceeded the strength threshold relative to \fspectralline{O}{iii}{5007} (20 per cent) used by \citet{callow_2024_rateextremecoronal}.\\
\end{flushleft}
\end{table*}

\begin{table*}
\caption{Crossmatch results between the full \sleipnir\ Input Galaxy Sample and the sample of galaxies identified as showing potential CrLs by an earlier version of \sleipnir\ in the search for ECLEs in the SDSS BOSS LOWZ survey by \citet{callow_2025_rateextremecoronal}.}
\label{tab:Appendix_Full_BOSS_Crossmatch}
\begin{tabular}{lccc}
\hline
\textbf{DESI ID} & \textbf{DESI EDR Classification} & \textbf{SDSS Short Name} & \textbf{SDSS BOSS LOWZ Classification} \\ \hline
39632966892126678 & - & SDSS J1645+3418 & Skyline ${^1}$ \\
39633153945503422 & - & SDSS J1546+4415 & Skyline \\
39633296920938570 & - & SDSS J1131+5313 & Weak CrLs ${^2}$\\
39633300582564759 & - & SDSS J1147+5323 & Skyline \\
39633300582564766 & - & SDSS J1147+5323 & Skyline \\
39633304143528511 & - & SDSS J1131+5339 & Skyline \\
39633311940741336 & - & SDSS J1603+5409 & Skyline \\
\hline
\end{tabular}
\begin{flushleft}
\textit{Notes:} ${^1}$ SDSS spectrum was manually inspected and determined to be a false-positive detection as a result of sky-line contamination of one or more CrLs.\\
${^2}$ In cases of ``Weak CrLs'', the SDSS spectrum was not manually inspected as no CrLs exceeded the strength threshold relative to \fspectralline{O}{iii}{5007} (20 per cent) used by \citet{callow_2025_rateextremecoronal} search.\\
\end{flushleft}
\end{table*}

\subsection{MILLIQUAS crossmatch}
\label{sec:Appendix_MILLIQUAS_Crossmatch}

Table~\ref{tab:Appendix_Full_MILLIQUAS_Crossmatch} outlines the full results of the crossmatch between the between the \sleipnir\ Input Galaxy Sample and the MILLIQUAS AGN database \citep{flesch_2023_MillionQuasarsMilliquas}. We remind the reader that some MILLIQUAS classifications have been made using DESI spectra, so these datasets are not fully independent. A total of \MILLIQUASGalaxyMatches\ crossmatches were returned. Of the subset flagged by \sleipnir\ for visual inspection, \MILLIQUASCrLMatches\ have been determined to show real CrL emission, with the remaining identified as being the result of false-positive contamination.

\begin{table*}
\caption{The results of the MILLIQUAS v8 crossmatch to the \sleipnir\ Input Galaxy Sample. A small portion of the table is included here for guidance of form. The full machine-readable version is included in the supplementary material.}
\label{tab:Appendix_Full_MILLIQUAS_Crossmatch}
\begin{tabular}{lccc}
\hline
\textbf{DESI ID} & \textbf{DESI EDR Classification} & \textbf{MILLIQUAS ID} & \textbf{MILLIQUAS Classification} ${^1}$ \\ \hline
39627628482659947 & CrL-AGN & SDSS J020852.34-063403.4 & AX \\
39627671155511332 & CrL-AGN & DESI 39627671155511332 & A \\
39627694769445378 & CrL-AGN & SDSS J022659.82-035015.0 & A \\
39627758170540645 & CrL-AGN & SDSS J142827.38-011845.9 & Q \\
39627758191511075 & CrL-AGN & PGC 3096363 & AX \\
... & ... & ... & ... \\
\hline
\end{tabular}
\begin{flushleft}
\textit{Note:} ${^1}$ Classification description available through the MILLIQUAS documentation: \url{https://quasars.org/Milliquas-ReadMe.txt} \\
\end{flushleft}
\end{table*}

\subsection{Ding et al.~(2025) crossmatch}
\label{sec:Appendix_Ding_Crossmatch}

Table~\ref{tab:Appendix_Full_Ding_Crossmatch} outlines the full results of the crossmatch between the \sleipnir\ Input Galaxy Sample and the independent sample of CrL objects within DESI EDR identified by \citet{ding_2025_ExploringLinkExtreme}.

\begin{table*}
\caption{The results of the \citet{ding_2025_ExploringLinkExtreme} crossmatch to the \sleipnir\ Input Galaxy Sample. A small portion of the table is included here for guidance of form. The full machine-readable version is included in the supplementary material.}
\label{tab:Appendix_Full_Ding_Crossmatch}
\begin{tabular}{lccc}
\hline
\textbf{DESI ID} & \textbf{DESI EDR Classification} & \textbf{Ding ID} & \textbf{Ding Classification}${^1}$ \\ \hline
39627887837451274 & CrL-AGN & DESI J027.9525+04.1951 & FeX6376 \\
39627902945332471 & CrL-AGN & DESI J210.4959+04.6738 & NeV3347  NeV3427  FeVII6088 \\
39627956879886559 & CrL-AGN & DESI J199.4801+06.9125 & NeV3427 \\
39628346308430103 & CrL-AGN & DESI J208.7671+23.7873 & NeV3427  FeVII6088 \\
39628357146511265 & CrL-AGN & DESI J194.5984+24.3064 & NeV3427\\
... & ... & ... & ... \\
\hline
\end{tabular}
\begin{flushleft}
\textit{Note:} ${^1}$ Classification indicates which CrLs were detected in the \citet{ding_2025_ExploringLinkExtreme} analysis.\\
\end{flushleft}
\end{table*}

\section{Coronal Line AGNs}
\label{sec:Appendix_CrL_AGN_Sample_Information}

In addition to the six objects initially flagged as potential TDE-ECLE candidates, the \sleipnir\ classification analysis identified a further 201 objects as being AGNs displaying real CrL emission features. This assessment is based on initial BPT diagnostics, visual inspection of their optical spectra and MIR light curves, and with a comparison of their MIR colours to the previously described \citet{stern_2012_MIDINFRAREDSELECTIONACTIVE} colour cut. Whilst not the primary focus of this work, given their role as significant contaminants in TDE searches, identifying information for each of these objects is provided in Table~\ref{tab:Appendix_CrL_AGN_Info}, and we draw attention to several of the most unusual galaxies within this sample.

\begin{table}
\caption{Identified information for sources identified as CrL-AGN in the DESI EDR sample. A small portion of the table is included here for guidance of form. The full machine-readable version is included in the supplementary material.}
\label{tab:Appendix_CrL_AGN_Info}
\begin{adjustbox}{width=0.95\columnwidth}
\begin{tabular}{llllll}
\hline
\textbf{DESI ID} & \textbf{RA (J2000)} & \textbf{Dec. (J2000)}  & \textbf{$z$} & \textbf{$z$ err} & \textbf{$E(B-V)$ (mag)} \\ \hline
39627628482659947 & 32.21810 & -6.56766 & 0.238 & 1.15E-05 & 0.025 \\
39627630114245131 & 130.07619 & -6.40654 & 0.095 & 1.15E-06 & 0.025 \\
39627640474179036 & 31.02454 & -6.05817 & 0.298 & 5.62E-06 & 0.027 \\
39627671155511332 & 66.39076 & -4.77105 & 0.359 & 7.21E-06 & 0.036 \\
39627694769445378 & 36.74930 & -3.8375 & 0.292 & 5.34E-06 & 0.030\\
... & ... & ... & ... & ... & ... \\ \hline
\end{tabular}
\end{adjustbox}
\end{table}

\begin{figure*}
    \centering
    \includegraphics[width=\textwidth]{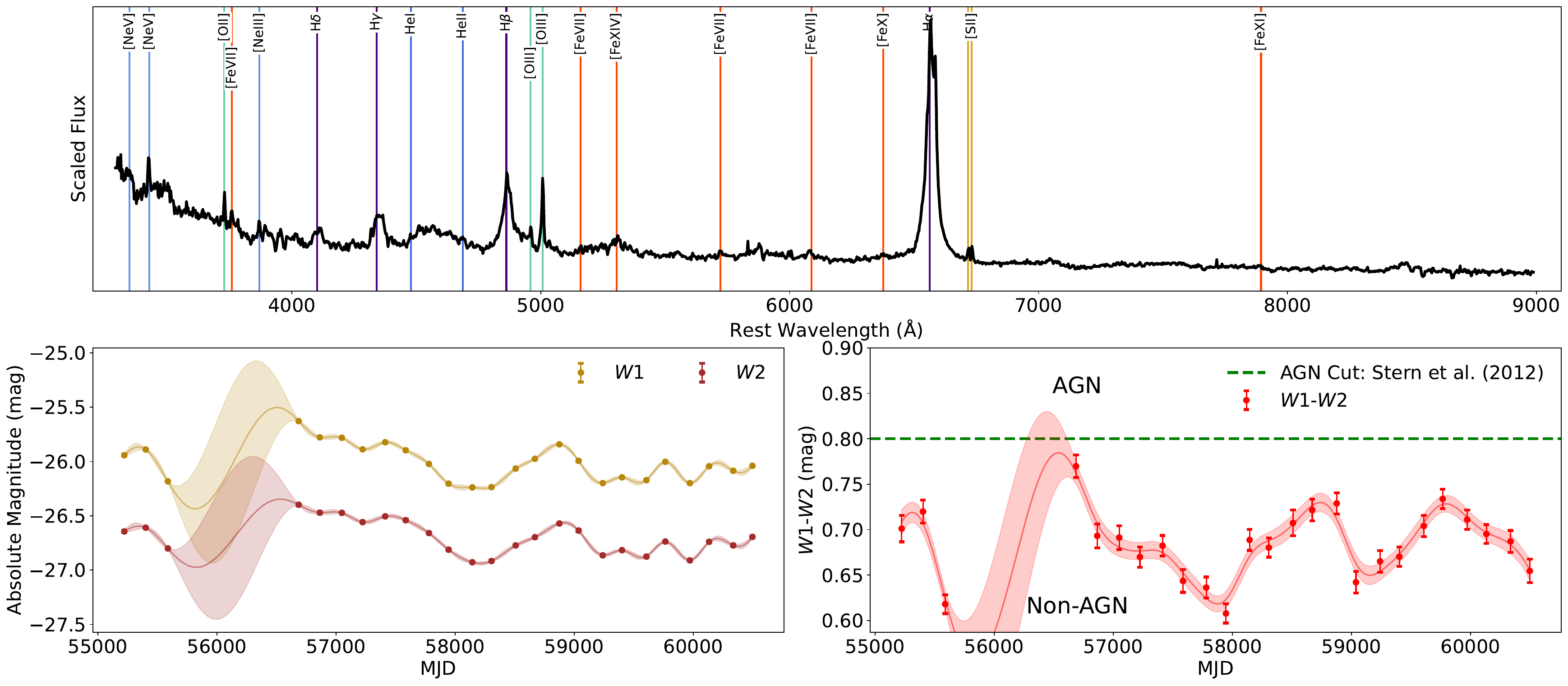}
    \caption{\textit{Top:} Spectrum of the unusual CrL-AGN DESI~39627758191511075. Prominent emission lines, including the broader than usual Fe CrLs, are indicated by the vertical lines. \textit{Bottom left:} MIR evolution of DESI~39627758191511075 showing repeated variations of $\sim$~0.1~mag, as expected of an AGN. \textit{Bottom right:} MIR \WOneminusWTwo\ colour evolution of DESI~39627758191511075, which whilst showing variation, remains below the \citet{stern_2012_MIDINFRAREDSELECTIONACTIVE} selection cut for typical AGNs. In both MIR plots, fits displayed are obtained through Gaussian processes, with the shaded regions indicating the 1$\sigma$ fitting uncertainties.}
    \label{fig:X1}
\end{figure*}

DESI~39627758191511075 (see Fig.~\ref{fig:X1}) displays broader than usual CrL \fspectralline{Fe}{vii}{6088} emission, accompanied by strong and broad H$\alpha$ emission typical of broad-line AGNs and evidence for an underlying blue AGN continuum. In the MIR, DESI~39627758191511075, whilst not having the typical AGN \WOneminusWTwo\ colour above the \citet{stern_2012_MIDINFRAREDSELECTIONACTIVE} cut for selecting AGNs, does display erratic evolution with repeated episodes of brightening and fading at the level of several tenths of a magnitude. Additionally, DESI~39627758191511075 is associated with the unclassified TNS AT~2025ils \citep{hall_2025_DESIRTTransientDiscovery}.

\begin{figure*}
    \centering
    \includegraphics[width=\textwidth]{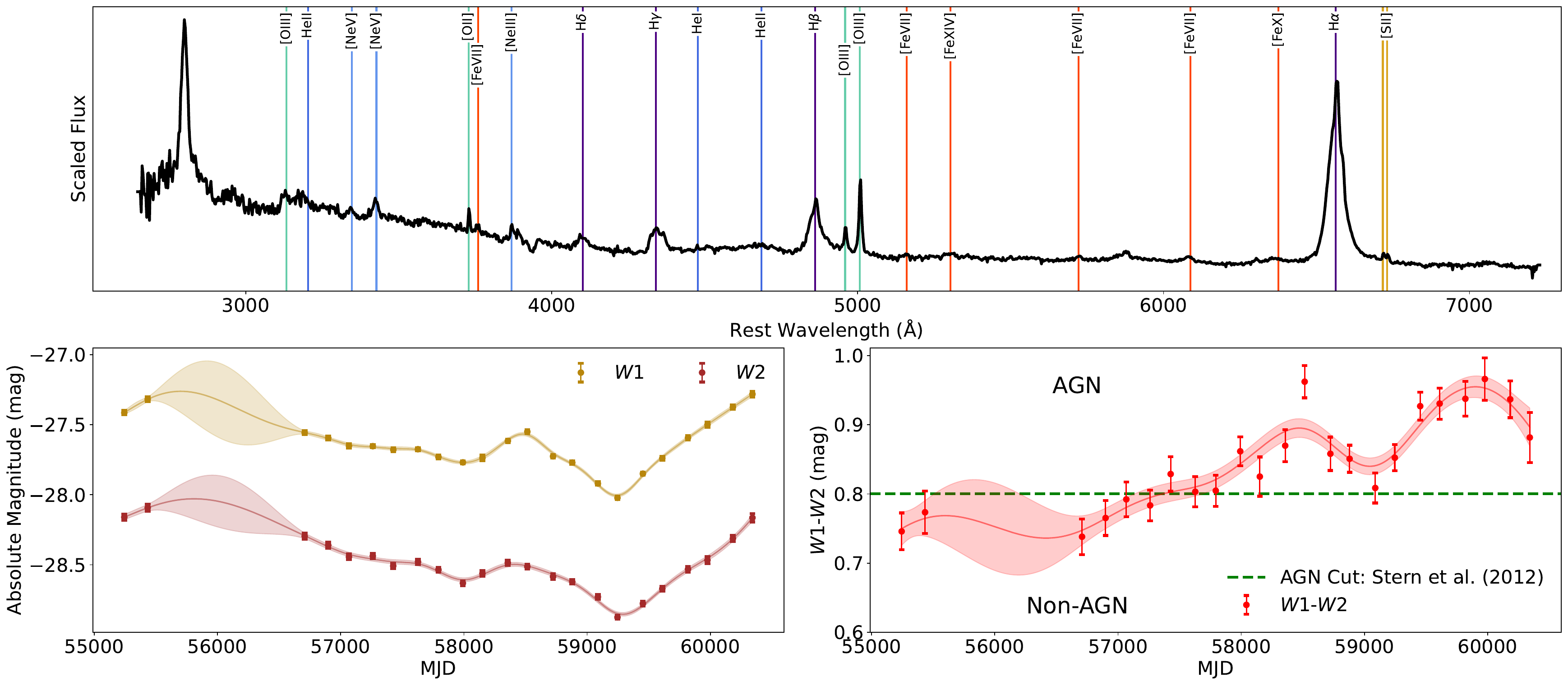}
    \caption{\textit{Top:} Spectrum of the unusual CrL-AGN DESI~39627671155511332. Prominent emission lines, including the broader than usual Fe CrLs, are indicated by the vertical lines. \textit{Bottom left:} MIR evolution of DESI~39627671155511332 showing several phases of both fading and brightening, most recently brightening by $\sim$~0.7~mag in both bands. \textit{Bottom right:} MIR \WOneminusWTwo\ colour evolution of DESI~39627671155511332, which
    shows repeated variation and a long-term reddening trend moving the measured colour above the \citet{stern_2012_MIDINFRAREDSELECTIONACTIVE} selection cut for typical AGNs. In both MIR plots, fits displayed are obtained through Gaussian processes, with the shaded regions indicating the 1$\sigma$ fitting uncertainties.}
    \label{fig:X2}
\end{figure*}

DESI~39627671155511332 (see Fig.~\ref{fig:X2}) also displays broader than usual CrL emission features and a multipeaked MIR light curve with overall changes in observed brightness $\sim$~0.7~mag, larger than the typical expected variability of AGNs, though with a \WOneminusWTwo\ colour more typical of AGNs. 

\section{MIR outburst peak analysis}
\label{Appendix:MIR_Outbust_Fit}

In Table~\ref{tab:Appendix_Fit_Stats_Report}, we include the full fitting parameters and statistical test results for the MIR outburst-peak analysis as described in Section~\ref{sec:MIR_Comp} and Fig.~\ref{fig:MIR_Delta_Fig}.

\begin{table*}
\caption{Fitting parameters obtained in the $\Delta$\ value analysis. Data and fits shown in Fig.~\ref{fig:MIR_Delta_Fig}}
\label{tab:Appendix_Fit_Stats_Report}
\begin{tabular}{llccccc}
\hline
 & Model & Parameter & Value & $t$-statistic & $p$-value & $\sigma$ \\ \hline
\textbf{$\Delta$\textit{W2} vs. $\Delta$\textit{W1}} & Quadratic $^*$ &  &  &  &  &  \\
 &  & $a$ & $-0.20 \pm 0.08$ & $-2.59$ & 0.0487 & 1.97 \\
 &  & $b$ & $0.66 \pm 0.21$ & 3.07 & 0.0279 & 2.20 \\
 &  & $c$ & $-0.47 \pm 0.11$ & 4.36 & 0.00732 & 2.68 \\
 &  &  &  &  &  &  \\
\textbf{$\Delta$(\textit{W1$-$W2}) vs. $\Delta$\textit{W2}} & Linear \textsuperscript{\textsquare} &  &  &  &  &  \\
 &  & $m$ & $-0.27 \pm 0.05$ & $-5.49$ & 0.00537 & 2.78 \\
 &  & $c$ & $0.15 \pm 0.07$ & 2.21 & 0.0918 & 1.69 \\ \hline
\end{tabular}
\begin{flushleft}
$^*$ Selected through maximum-likelihood analysis and AIC value comparison between a fixed constant, a linear model, and a quadratic model\\
\textsuperscript{\textsquare}Selected through maximum-likelihood analysis and AIC value comparison between a fixed constant and a linear model. A quadratic model was not included in this comparison owing to the small number of data points (six) available for inclusion. \\
\end{flushleft}
\end{table*}
\raggedbottom


\bsp	
\label{lastpage}
\end{document}